\documentclass[12pt]{article}
\usepackage{epsfig,color,amsmath,amssymb,euscript,eufrak,mathrsfs,multirow,enumerate,mathabx,algorithm,algpseudocode,subcaption,amsthm,enumitem,setspace}
\usepackage{makecell}
\allowdisplaybreaks[4]
\RequirePackage[colorlinks,citecolor=blue,urlcolor=blue]{hyperref}
\usepackage[round,authoryear]{natbib}
\usepackage{amsthm}
\usepackage{authblk}
\def\spacingset#1{\renewcommand{\baselinestretch}%
	{#1}\small\normalsize} \spacingset{1}
\providecommand{\keywords}[1]{\textbf{\textit{Keywords: }} #1}
\newcommand{\blind}{1}
\makeatletter

\newcommand{\inner}[2]{ \mbox{$ \langle #1, #2 \rangle $} }

\newtheorem{definition}{Definition}
\newtheorem{theorem}{Theorem}
\newtheorem{lemma}{Lemma}

\newtheorem{proposition}{Proposition}
\newtheorem{inequality}{Inequality}
\newtheorem{equality}{Equality}
\theoremstyle{definition}
\newtheorem{condition}{Condition}
\newtheorem{remark}{Remark} 
\newtheorem{example}{Example} 

\def\singlespace{\def\baselinestretch{1}\@normalsize}

\def\wh{\widehat}

\newcommand{\cov}{{\rm cov}}

\newcommand{\var}{{\rm Var}}

\def\la{\lambda}

\newcommand{\tF}{\text{F}}

\newcommand{\bA}{{\mathbf A}}
\newcommand{\bB}{{\mathbf B}}

\newcommand{\bE}{{\mathbf E}}

\newcommand{\bH}{{\mathbf H}}
\newcommand{\bI}{{\mathbf I}}
\newcommand{\bK}{{\mathbf K}}

\newcommand{\bM}{{\mathbf M}}
\newcommand{\bQ}{{\mathbf Q}}
\newcommand{\bP}{{\mathbf P}}
\newcommand{\bR}{{\mathbf R}}

\newcommand{\bU}{{\mathbf U}}
\newcommand{\bV}{{\mathbf V}}
\newcommand{\bW}{{\mathbf W}}
\newcommand{\bX}{{\mathbf X}}
\newcommand{\bY}{{\mathbf Y}}
\newcommand{\bZ}{{\mathbf Z}}

\newcommand{\bb}{{\mathbf b}}

\newcommand{\be}{{\mathbf e}}

\newcommand{\bg}{{\boldsymbol g}}

\newcommand{\bs}{{\mathbf s}}

\newcommand{\bbf}{{\boldsymbol f}}

\newcommand{\bbeta}  {\boldsymbol{\beta}}
\newcommand{\bfeta}  {\boldsymbol{\eta}}
\newcommand{\bvarphi}  {\boldsymbol{\varphi}}
\newcommand{\bvartheta}  {\boldsymbol{\vartheta}}

\newcommand{\bPi}  {\boldsymbol{\Pi}}
\newcommand{\bphi}  {\boldsymbol{\phi}}
\newcommand{\bkappa}  {\boldsymbol{\kappa}}

\newcommand{\beps}{\boldsymbol \epsilon}
\newcommand{\bOmega}{\boldsymbol{\Omega}}

\newcommand{\bSigma}{\boldsymbol \Sigma}
\newcommand{\bDelta}{\boldsymbol{\Delta}}
\newcommand{\bgamma}{\boldsymbol{\gamma}}

\newcommand{\bvarepsilon}{\boldsymbol \varepsilon}
\newcommand{\bve}{\mbox{\boldmath$\varepsilon$}}

\newcommand{\bTheta} {\boldsymbol{\Theta}}
\newcommand{\bPhi} {\boldsymbol{\Phi}}
\newcommand{\bPsi} {\boldsymbol{\Psi}}

\newcommand{\btheta} {\boldsymbol{\theta}}
\newcommand{\bxi} {\boldsymbol{\xi}}
\newcommand{\bpsi} {\boldsymbol{\psi}}
\newcommand{\bmu} {\boldsymbol{\mu}}
\newcommand{\bzeta} {\boldsymbol{\zeta}}
\newcommand{\bvar} {\boldsymbol{\varepsilon}}
\newcommand{\bGamma} {\boldsymbol{\Gamma}}
\newcommand{\bLambda} {\boldsymbol{\Lambda}}

\newcommand{\bD}{{\mathbf D}}
\newcommand{\bzero}{{\mathbf 0}}

\newcommand{\calL}{{\mathcal L}}

\newcommand{\cH}{\mathbb{H}}
\newcommand{\cE}{\mathbb{E}}

\newcommand{\calU}{{\mathcal U}}

\newcommand{\calM}{{\mathcal M}}

\newcommand{\calS}{{\mathcal S}}
\newcommand{\calT}{{\mathcal T}}

\newcommand{\calC}{{\mathcal C}}

\newcommand{\ttS}{{\scriptscriptstyle\textup{S}}}

\def\6bullets{\bullet\bullet\bullet\bullet\bullet\bullet}

\newcommand{\eulC}{\EuScript C}

\DeclareMathAlphabet\EuScriptBF{U}{eus}{b}{n}

\newcommand{\eulbC}{\EuScriptBF C}

\DeclareSymbolFont{rmsymbols}{OMX}{mdbch}{m}{n}
\DeclareMathSymbol{\rmintop}{\mathop}{rmsymbols}{82}
\newcommand{\rmint}{\rmintop\nolimits} 

\DeclareMathAlphabet\EuScriptBF{U}{eus}{b}{n}

\def\T{{ \mathrm{\scriptscriptstyle \top} }}

\newcommand{\Date}[1]{\def\@Date{#1}}
\def\today{\number\day~\ifcase\month\or
 January\or February\or March\or April\or May\or June\or
 July\or August\or September\or October\or November\or December\fi~\number\year}

\begin{document}
\if1\blind
{
\spacingset{1.25}
  \title{\bf \Large On the Modelling and Prediction of High-Dimensional Functional Time Series}
\author[1,2,3]{Jinyuan Chang}
\author[4]{Qin Fang}
\author[5]{Xinghao Qiao}
\author[6]{Qiwei Yao}

\affil[1]{\it \small Joint Laboratory of Data Science and Business Intelligence, Southwestern University of Finance and Economics, Chengdu, Sichuan, China}
\affil[2]{\it \small Big Data Laboratory on Financial Security and Behavior (MOE Philosophy and Social Sciences Laboratory), Southwestern University of Finance and Economics, Chengdu, Sichuan, China}
\affil[3]{\it \small Academy of Mathematics and Systems Science, Chinese Academy of Sciences, Beijing, China}
\affil[4]{\it \small Business School, University of Sydney, Sydney, Australia}
\affil[5]{\it \small Faculty of Business and Economics, The University of Hong Kong, Hong Kong
}
\affil[6]{\it \small Department of Statistics, London School of Economics, London, U.K.
}

		\setcounter{Maxaffil}{0}
		
		\renewcommand\Affilfont{\itshape\small}
		\date{\vspace{-5ex}}
		\maketitle
	} \fi
	\if0\blind
	{
		\bigskip
		\bigskip
		\bigskip
		\begin{center}
			{
			\Large \bf On the Modelling and Prediction of High-Dimensional Functional Time Series
			}
		\end{center}
		\medskip
	} \fi
\spacingset{1.5}
\begin{abstract}
We propose a two-step procedure to model and predict high-dimensional functional time series, where the number of function-valued time series $p$ is large in relation to the length of time series $n$. Our first step
performs an eigenanalysis of a positive definite matrix, which leads to a one-to-one linear transformation for the original high-dimensional functional time series, and the transformed curve series can be segmented into several groups such that any two subseries from any two different groups 
are uncorrelated both contemporaneously and serially. Consequently in our second step those groups are handled separately without the information loss on the overall linear dynamic structure. The second step is devoted to establishing a finite-dimensional dynamical structure for all the transformed
functional time series within each group. Furthermore the finite-dimensional structure is represented by that of a vector time series. Modelling and forecasting for the original high-dimensional functional time series are realized via those for the vector time series in all the groups. We investigate the theoretical properties of our proposed methods, and illustrate the finite-sample performance through both extensive simulation and two real datasets.
\end{abstract}

\noindent\keywords{Dimension reduction; Eigenanalysis; Functional thresholding; Hilbert--Schmidt norm; Permutation; Segmentation transformation.}

\spacingset{1.69}
\setlength{\abovedisplayskip}{0.2\baselineskip}
\setlength{\belowdisplayskip}{0.2\baselineskip}
\setlength{\abovedisplayshortskip}{0.2\baselineskip}
\setlength{\belowdisplayshortskip}{0.2\baselineskip}

\section{Introduction}
Functional time series typically refers to continuous-time records that are naturally divided into consecutive time intervals, such as days, months or years. 
With recent advances in data collection technology, multivariate or even high-dimensional functional time series arise ubiquitously in many applications, including 
daily pollution concentration curves 
over different locations,
annual temperature curves  
at different stations, annual age-specific mortality rates 
for different countries, and intraday energy consumption trajectories from different
households. 
Those data can be represented as a 
$p$-dimensional functional time series $\bY_t(u)=\{Y_{t1}(u), \ldots, Y_{tp}(u)\}^{\T}$  
defined on a compact set $u\in\calU$, and we observe $\bY_t(\cdot)$ for $t=1, \ldots, n$.
In this paper we tackle the high-dimensional settings when the dimension $p$ is comparable to, or even greater than, the sample size $n$, which poses new challenges in modelling and forecasting $\bY_{t}(\cdot)$.

By assuming $\bY_t(\cdot)$ is stationary, a conventional approach is first to extract features by performing dimension reduction for each component series $Y_{tj}(\cdot)$ separately via, e.g.  functional principal component analysis (FPCA) or dynamic FPCA \citep{Bathia2010,hormann2015}, and then to model $p$ vector time series 
by, e.g., regularized vector autoregressions \cite[]{guo2021} or factor model 
\cite[]{Gao2019}. 
However, 
more effective dimension-reduction can be achieved by pulling together the information from different component series in the first place.
This is in the same spirit of 
multivariate FPCA \cite[]{Chiou2014,Happ2018} 
and sparse FPCA \cite[]{hu2021}, though those approaches make no use of the information on the serial dependence which is the most relevant for future prediction.
To achieve  more effective dimension reduction and better predictive performance, 
we propose in this paper 
a two-step approach. 
Our first step is a segmentation transformation step in which we seek for 
a linear transformation $\bY_t(\cdot)=\bA \bZ_t(\cdot)$,
where $\bA$ is a $p\times p$ invertible constant matrix,
such that the transformed series $\bZ_t(\cdot) = \{\bZ_{t}^{(1)}(\cdot)^\T, \dots,\bZ_{t}^{(q)}(\cdot)^\T\}^\T$ can be  segmented into $q$ groups $\bZ_{t}^{(1)}(\cdot), \dots,\bZ_{t}^{(q)}(\cdot)$, and curve subseries $\bZ_t^{(i)}(\cdot)$ and
$\bZ_t^{(j)}(\cdot)$ are uncorrelated at all time lags for any $i \neq j $, i.e.,
\[
\cov\{ \bZ_t^{(i)}(u), \bZ_{t+k}^{(j)}(v) \} =\bzero\,, \quad  (u,v) \in \calU^2 \;\; {\rm and}\;\; k=0, \pm 1, \pm 2, \ldots\,.
\]
Hence each $\bZ_t^{(i)}$ can be modelled and forecasted separately
as far as the linear dynamics is concerned.
Under the stationarity assumption,
the estimation of the transformation matrix $\bA$ boils down to the eigenanalysis of a positive definite matrix defined by the double integral 
of quadratic forms in the autocovariance functions of $\bY_t(\cdot)$.
An additional 
permutation on the components of $\bZ_t(\cdot)$  will be
specified in order to identify the latent group structure.

Our second step is to identify a finite-dimensional dynamic structure
for each transformed subseries $\bZ_{t}^{(l)}(\cdot)$ separately, which is based 
 on a latent decomposition
\begin{equation}
    \label{Y.err}
    \bZ^{(l)}_t(u) = \bX^{(l)}_t(u) + \bvar^{(l)}_{t}(u)\,, ~~~~u \in \calU \,, 
\end{equation}
where $\bX^{(l)}_t(\cdot)$ represents the dynamics of $\bZ^{(l)}_t(\cdot)$,
$\bvar_t^{(l)}(\cdot) $ 
is white noise with
$\cE \{ \bvar^{(l)}_t(u) \} =\bzero$ and $\cE\{\bvar^{(l)}_t(u)\bvar^{(l)}_s(v)^\T\}={\bf0}$ for any $(u,v)\in \calU^2$ and $t\ne s$, and  $\{\bX_t^{(l)}(\cdot)\}_{t=1}^n$ are uncorrelated with  $\{\bvar_t^{(l)}(\cdot)\}_{t=1}^n$. 
Furthermore we assume that the dynamic structure of $\bX_t^{(l)}(\cdot)$ admits a vector time series presentation via
a variational multivariate FPCA. 
For given $\{\bZ_t^{(l)}(\cdot)\}_{t=1}^n$, the standard multivariate FPCA performs dimension reduction based on the eigenanalysis of the sample covariance function of $\bZ_{t}^{(l)}(\cdot),$ which cannot be used to identify the finite-dimensional dynamic structure of $\bX_{t}^{(l)}(\cdot)$ due to the contamination of $\bvar^{(l)}_{t}(\cdot).$
Inspired by the fact that the lag-$k$ ($k \ne 0$)  autocovariance function of $\bZ_{t}^{(l)}(\cdot)$ automatically filters out the white noise, our variational multivariate FPCA is 
based on the eigenanalysis of a positive-definite matrix defined in terms of its nonzero lagged autocovariance functions; leading to
a low-dimensional vector time series which bears all the dynamic 
structure of $\bX_t^{(l)}(\cdot)$, and consequently, also that of
$\bZ_{t}^{(l)}(\cdot)$. This is possible as the number of components in each $\bZ_{t}^{(l)}(\cdot)$ is usually small in practice. 
Finally, owing to the one-to-one linear transformation in the segmentation step, the good predictive performance of $\bZ_t(\cdot)$ 
can be easily carried back to $\bY_t(\cdot).$

Our paper makes useful contributions on multiple fronts. 
Firstly, the segmentation transformation in the first step transforms the serial
correlations across different series into the autocorrelations within
each of the identified $q$ subseries. This not only avoids the direct modelling
of the $p$ functional time series together, but also makes each of
those transformed subseries more serially correlated and, hence, more predictable. As the serial correlations across different series are valuable for future prediction, the segmentation provides an effective way to use the information.
Note that
the prediction directly based on a multivariate ARMA-type model with even a moderately large dimension is not recommendable, as the gain from
using the autocorrelations across different component series is often
cancelled off by the errors in estimating too many parameters. 
Furthermore, even in the special case with $q = 1$,
    our decorrelation transformation can effectively push the cross-autocorrelations that are previously spread over $p$ components into a block-diagonally dominate structure, where the cross-autocorrelations along the block diagonal are significantly stronger than those off the diagonal. This still leads to reasonably good segmentation by retaining the strong within-group cross-autocorrelations while ignoring the weak between-group cross-autocorrelations and, as evidenced by simulations in Section~\ref{sec.sim.nogroup}, results in more accurate future predictions than those based on models without transformation. Therefore, the proposed transformation can always be used as an initial step in modelling high-dimensional functional time series.

Secondly, though the segmentation transformation is motivated from the decorrelation idea of \cite{chang2018} for vector time series, its adaption to the functional setting introduces
additional methodological and theoretical complexities and requires innovative advancements in both methodology and theory due to the intrinsic infinite-dimensionality of functional data.
A simple extension of \cite{chang2018} would be to apply their method to the $p$-dimensional vector $\bY_t(u)$ on each evaluation grid value $u$ followed by aggregation, which
fails to account for the smoothness and continuity of the functional nature of observed data.
In contrast, our proposal on $\bY_t(\cdot)$ implements novel integral-based normalization 
and utilizes double integral over $(u,v) \in \calU^2$ to fully leverage the autocovariance information, thus leading to more efficient estimation.
Moreover, when performing permutation on the components of the transformed series, our method relies on Hilbert--Schmidt norm to measure the magnitude of bivariate functions, which introduces extra theoretical complexities compared to the absolute value measure used in \cite{chang2018}. Finally, we develop a novel functional thresholding procedure,
which guarantees the consistency of our estimation under high-dimensional scaling.
Its theoretical analysis involves establishing novel inequalities between functional versions of matrix norms.

Thirdly, the nonzero lagged autocovariance-based dimension reduction approach in the second step makes the good use of the serial dependence information in our estimation, which is most relevant in time series prediction. 
On the method side, our proposed variational multivariate FPCA extends the univariate method of \cite{Bathia2010} by incorporating the cross-autocovariance. 
This extension addresses a crucial gap in dimension-reduction techniques, enabling us to accommodate multivariate functional time series. Importantly, when $p$ is fixed or moderately large, such method can be directly applied to the observed curve series $\bY_t(\cdot)$ for dimension reduction and forecasting purposes.
On the theory side, we demonstrate that our proposal exhibits appealing convergence properties despite the additional transformation and estimation errors arisen from the first step, which are not involved in \cite{Bathia2010}. By comparison, standard (multivariate) FPCA methods under (\ref{Y.err}) suffer from inconsistent estimation and less efficient dimension reduction. 

Existing research on functional time series has mainly focused on adapting the univariate or low-dimensional multivariate time series methods to the functional domain. An incomplete list of the relevant references includes
\cite{Bathia2010}, 
\cite{cho2013}, \cite{aue2015}, 
\cite{hormann2015}, \cite{aue2018} and  
\cite{li2019}. 
Following the recent emergence of high-dimensional functional time series data, there has been a wave of significant advancements aimed at addressing its complexities. Notable developments include functional factor models \citep{Gao2019,Tavakoli2023}, functional dependence analysis \citep{guo2021}, functional clustering \citep{tang2022}, statistical inference for mean functions \citep{zhou2023}, sparse vector functional autoregressions \citep{chang2024}, 
and graphical PCA \citep{tan2024}.

The rest of the paper is organized as follows. In Section~\ref{sec:seg}, we develop the methods employed in the first step, i.e. the segmentation transformation, the permutation and the functional thresholding. Section~\ref{sec:dr} specifies the variational multivariate FPCA method used in the second dimension reduction step.
We investigate the associated theoretical properties of the proposed methods in Section~\ref{sec.th}. 
The finite-sample performance of our methods is examined through extensive simulations in Section~\ref{sec.sim}. Section~\ref{sec.real} applies our proposal to two real datasets, revealing its superior predictive performance over most frequently used competitors. 

{\it Notation.} Denote by $I(\cdot)$ the indicator function. For a positive integer $m$, write $[m] = \{1,\dots, m\}$ and denote by $\bI_m$ the identity matrix of size $m \times m$. For $x, y \in \mathbb{R},$ we use $x \vee y = \max(x,y).$ For two positive sequences $\{a_n\}$ and $\{b_n\}$, we write $a_n\ll b_n$ or $b_n\gg a_n$ if  $\lim \sup_{n \to \infty} a_n/b_n = 0$. For a $p \times q$ real matrix $\bE$, denote by $\bE^{\T}$ its transpose, and write   $\bE^{\otimes2} = \bE \bE^{\T}$ and  $\|\bE\|_{2}= \lambda_{\max}^{1/2}(\bE^{\T}\bE),$  where $\lambda_{\max}(\bM)$ denotes the largest  eigenvalue of the matrix $\bM.$
Let $L_2(\calU)$ be the Hilbert space of square integrable functions defined on $\calU$ and equipped with the inner product $\langle f,g \rangle=\int_{\calU} f(u)g(u)\,{\rm d}u$ for $f,g \in L_2(\calU)$ and the induced norm $\|\cdot\|=\langle \cdot,\cdot \rangle^{1/2}.$ For any $B$ in ${\mathbb S}\equiv L_2(\calU \times \calU), $ we denote the Hilbert--Schmidt norm by $\|B\|_{\calS} = \{\int_{\calU} \int_{\calU} B^2(u,v)\, {\rm d}u {\rm d}v\}^{1/2}.$

\vspace{-0.2cm}
\section{Segmentation transformation} 
\label{sec:seg}


\vspace{-0.2cm}
\subsection{Linear decomposition of $\bY_t(u)$}
We consider the following linear decomposition of 
$\bY_t(u)$: 
\begin{equation} \label{c1}
\bY_t(u)  = \bA \bZ_t(u) = \bA_1\bZ_{t}^{(1)}(u)+\cdots+\bA_q\bZ_{t}^{(q)}(u)\,, ~~~~u \in \calU\,,
\end{equation}
where $q\in[p]$ is an unknown positive integer, $\bA=(\bA_1,\ldots,\bA_q)$ is a $p\times p$ unknown loading matrix, and $\bZ_t(u)=\{\bZ_{t}^{(1)}(u)^\T,\ldots\bZ_t^{(q)}(u)^\T\}^\T$ is a latent $p$-dimensional functional time series such that $\cov\{\bZ_{t}^{(l)}(u), \bZ_{s}^{(l')}(v)\} = \boldsymbol{0}$ for all $t,s\in[n]$,  $l \neq l'$ and $(u,v)\in \calU^2$.  Such linear decomposition possesses three key properties:  

\begin{itemize}
\item For any $p$-dimensional functional time series $\bY_t(u)$, its linear decomposition \eqref{c1} always exists by setting $q=1$ and choosing $(\bA, \bZ_t(u))=(\bH,\bH^{-1}\bY_t(u))$ for some invertible matrix $\bH$.

\item The linear decomposition \eqref{c1} is not uniquely determined.  Alternative segmentations of $\bZ_t(u)$ can be obtained by merging multiple uncorrelated groups  into a single group.

\item For a given segmentation, $\bA_l$ for  $l \in[q]$ cannot be uniquely identified, as within-group rotations will not distort the uncorrelated group structure. In fact, only the linear spaces spanned by the columns of $\bA_l$, denoted by $\calC(\bA_l)$,  $l \in[q]$, are uniquely defined.
\end{itemize}

Our goal is then to find a linear decomposition \eqref{c1} for $\bY_t(\cdot)$, where each group $\bZ_t^{(l)}(\cdot)$ for $l \in [q]$ cannot be further divided into smaller uncorrelated subgroups. This allows us to model each $\bZ_t^{(l)}(\cdot)$ separately, as there are no cross-correlations among them at all time lags.  We formalize the inseparability for each $\bZ_t^{(l)}(\cdot)$ as Condition~\ref{con.inseparable} in Section~\ref{sec.th}, which in turn defines
the number of groups $q$ and the segmentation of $\bZ_t(\cdot)$ in \eqref{c1}.
In Section~\ref{sec:seg.est}, we will present the estimation of the number of groups $q$, the linear spaces $\calC(\bA_l)$ and the associated transformed subseries $\bZ_t^{(l)}(\cdot)$ of group size $p_l$ for $l \in [q]$.
Before that, let us firstly illustrate the validity and benefit of the linear decomposition \eqref{c1},  i.e. segmentation transformation, in predicting multivariate functional time series with a real-life example. As we will demonstrate, such a decomposition \eqref{c1} is commonly achieved with a relatively large $q$ in practice. This effectively reduces the modelling burden while retaining the full linear dynamics of the original curve series $\mathbf{Y}_t(\cdot),$ thus leading to more accurate predictions.


\begin{example}
\label{exp:motivation.model}
We consider the global age-specific mortality dataset
analyzed in \cite{tang2022}.
To simplify the presentation, we examine only the female mortality curve series $\bY_t(\cdot)$ with $p=8$ randomly selected countries (Australia, Canada, Switzerland, Denmark, Finland, Great Britain, Japan and Portugal) and the transformed curve series $\bZ_t(\cdot)$ in \eqref{c1}, which are obtained by the proposed method in Section \ref{sec:seg.est}. Let
$$\hat \sigma_{y,k,ij}(u,v)  = \frac{1}{n-k}\sum_{t=1}^{n-k}\{ Y_{ti}(u) - \bar Y_i(u) \}
\{ Y_{(t+k)j}(v) - \bar Y_j(v) \}$$
with $\bar{Y}_i(u)=n^{-1}\sum_{t=1}^nY_{ti}(u)$. 
We use $$\hat \varpi_{y,k,ij} = \frac{\|\hat \sigma_{y,k,ij}\|_\calS}{\{\int{\hat \sigma_{y,0,ii}^2(u,u)\, {\rm{d}} u} \int{\hat \sigma_{y,0,jj}^2(u,u) \,{\rm{d}} u} \}^{1/2}}\, ,$$ as proposed by \cite{Rice2019}, to measure the functional cross-autocorrelation between $Y_{ti}(\cdot)$ and $Y_{tj}(\cdot)$  at lag $k$.
Figure~\ref{plot.mortality} displays $\hat \varpi_{y,k,ij}$ and $\hat \varpi_{z,k,ij}$ for $-5\leq k \leq 5$, where $\hat \varpi_{z,k,ij}$ is defined by substituting each $Y_{ti}(\cdot)$ in $\hat \varpi_{y,k,ij}$ with $Z_{ti}(\cdot).$ It is evident that the transformation effectively channels the strong cross-autocorrelations  over different time lags among all 8 countries into significant autocorrelations within each of the 6 groups of $\bZ_t(\cdot)$, i.e., $\{1,2,3\}, \{4\}, \{5\},\{6\}, \{7\}$ and $\{8\}$, while the cross-autocorrelations among
these six groups are identified as weak and statistically insignificant across
all time lags at the $5\%$ significance level. 

\begin{figure}[tbp]
\centering
\begin{subfigure}{1\linewidth}
  \centering\includegraphics[width=16cm,height=10cm]{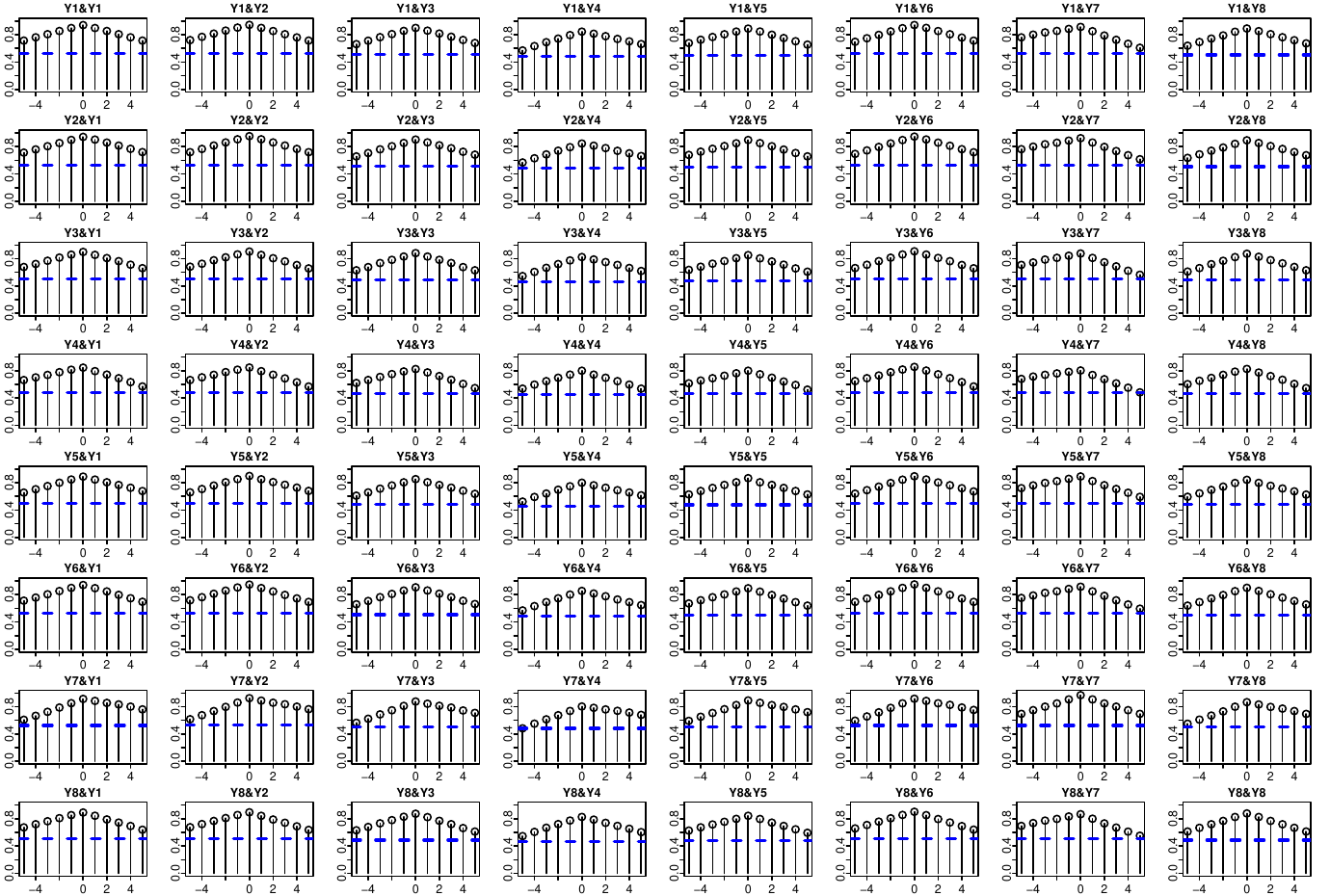}
  
  \caption{Functional cross-autocorrelations of the 8 original curve series.} 
 \end{subfigure}

\vspace{0.1cm}
\begin{subfigure}{1\linewidth}
  \centering\includegraphics[width=16cm,height=10cm]{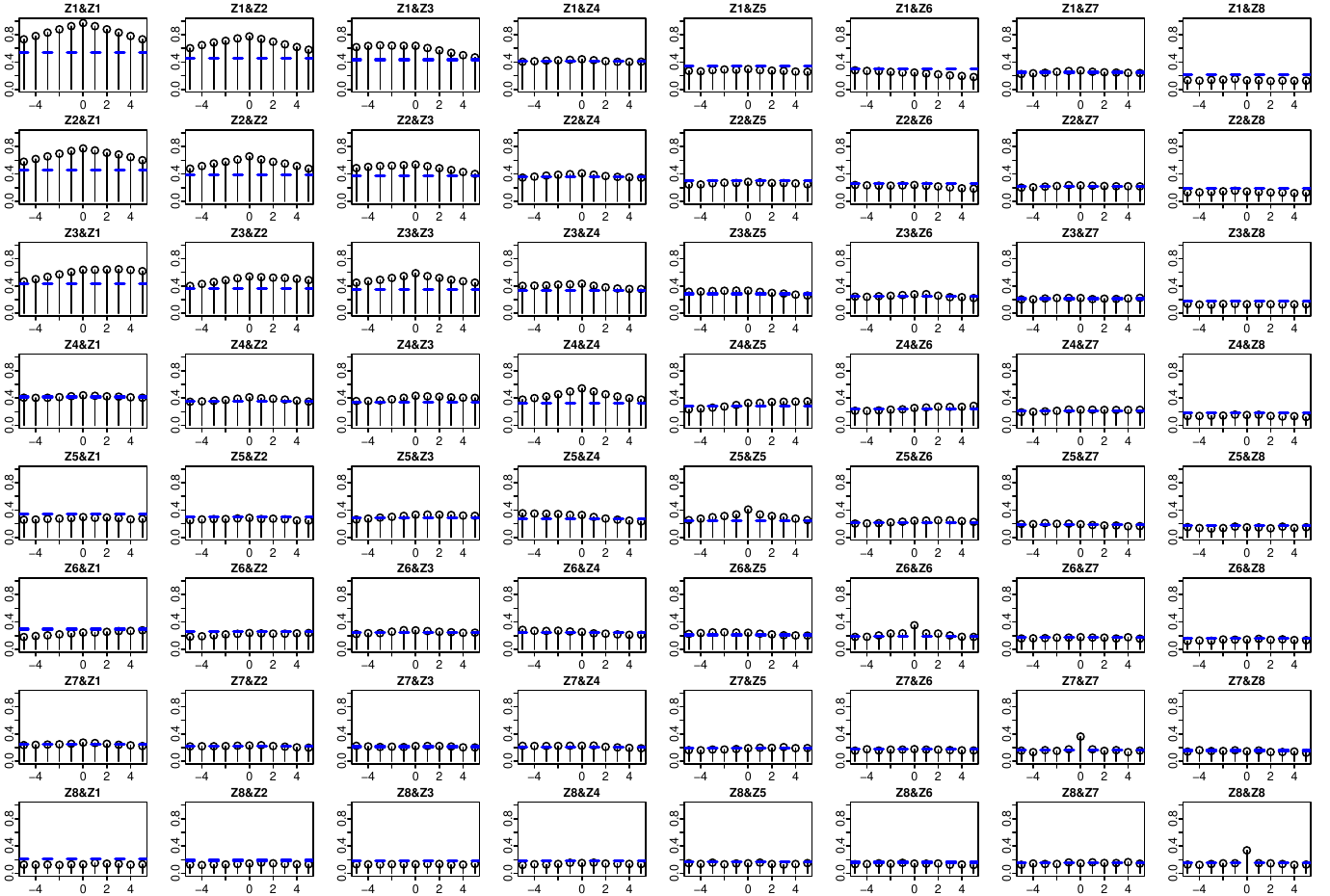}
  \caption{Functional cross-autocorrelations of the 8  transformed curve series.} 
 \end{subfigure}

\centering
\caption{\label{plot.mortality}{Functional cross-autocorrelation of mortality (original and transformed) curve series versus size $0.95$ upper confidence bound 
(blue dotted line).
}}
\end{figure}

We then implement 
two prediction methods on  $\bY_t(\cdot)$ and $\bZ_t(\cdot)$, respectively, to demonstrate that 
forecasting $\bY_t(\cdot)$  through the forecasting of the transformed series \(\bZ_t(\cdot)\) can yield more accurate predictive performance than directly forecasting \(\bY_t(\cdot)\):
\begin{itemize}
    \item (Joint prediction)  We treat  the $p$ components of $\bY_t(\cdot)$ as one group and perform the \textbf{V}ariational-\textbf{m}ultivariate-FPCA-and-\textbf{V}AR-based procedure (VmV), i.e. Step \ref{step 2.sim} of our proposed Algorithm~\ref{alg1} in Section~\ref{sec.sim.moderate}, on $\bY_t(\cdot)$ directly.  Based on the identified group structure by Figure~\ref{plot.mortality}, we implement SegV 
on $\bZ_{t}(\cdot)$, which performs VmV on each of the 6 groups of $\bZ_{t}(\cdot)$ separately.
    \item (Marginal prediction) We implement 
    UniV and  Uni.SegV, which respectively perform VmV
on each component of  $\bY_t(\cdot)$ and $\bZ_t(\cdot)$ separately.
\end{itemize}
Note that the difference in each prediction method comes solely from the transformation.
See details of these methods in Sections~\ref{sec.sim} and \ref{sec.real}.
Table~\ref{table.mortality.small} reports one-step ahead
mean absolute prediction errors (MAPE)
and mean squared prediction errors (MSPE) defined as \eqref{MSP} in Section~\ref{sec.real}, with a test size of $15$. As expected, 
methods that employ the transformation, namely SegV and Uni.SegV, significantly outperform their counterparts VmV and UniV without any transformation.
This highlights the benefit of integrating the transformation as an initial step in modelling multivariate functional time series. 



\begin{table}[!t]
\vspace{-0.2cm}
	\caption{\label{table.mortality.small}
	MAPEs and MSPEs for four competing methods on the female mortality curves. All numbers are multiplied by $10.$ The lowest values are in bold font.}
	\begin{center}
		\vspace{-0.5cm}
		\resizebox{3in}{!}{
\begin{tabular}{c|cccc}
\hline
Method & SegV           & VmV     & Uni.SegV & UniV  \\ \hline
MAPE   & \textbf{1.205} & 1.890  & 1.454    & 1.765 \\
MSPE   & \textbf{0.296} & 0.662 & 0.386    & 0.616 \\ \hline
\end{tabular}
}	
	\end{center}
	\vspace{-0.5cm}
\end{table}

\vspace{-0.2cm}
\end{example}

\vspace{0.5cm}

\vspace{-0.7cm}
\subsection{Estimation procedure}
\label{sec:seg.est}
\vspace{-0.1cm}
We now consider how to find the segmentation transformation under \eqref{c1}. Assume that $\max_{i \in [p]}\int_{\calU} \mathbb{E}\{Z_{ti}^2(u)\}\,{\rm d}u = O(1)$. 
Define
$\bSigma_{y,k}(u, v) = \cov\{\bY_{t}(u), \bY_{t+k}(v)\}$ and $\bSigma_{z,k}(u, v) = \cov\{\bZ_{t}(u),  \bZ_{t+k}(v)\}.$
Without loss of generality, we focus on 
the orthogonal transformations only, i.e., $\bA^{\T}\bA = \bA\bA^{\T}=\bI_p$, as we can replace $(\bY_t, \bZ_t)$ in (\ref{c1}) by
$(\bV_y^{-1/2} \bY_t, \bV_z^{-1/2} \bZ_t)$ with $\bV_y = \int_{\calU}\bSigma_{y,0} (u,u)\,{\rm d}u$ and $\bV_z = \int_{\calU}\bSigma_{z,0} (u,u)\,{\rm d}u$. Then $\bA$ is replaced by
$\bV_y^{-1/2} \bA \bV_z^{1/2}$ which is an orthogonal matrix as 
\begin{equation} \label{eq.normal}    \bI_p=\int_{\calU}\var\{\bV_y^{-1/2}\bY_t(u)\}\,{\rm d}u = \int_{\calU}\var\{\bV_z^{-1/2}\bZ_t(u)\}\,{\rm d}u\,.
\end{equation}
Due to the unobservability of $\bZ_t$, we can take $\bV_z^{-1/2} \bZ_t$ as $\bZ_t$ since they share the same block structure. In practice, we can replace observations $\bY_t$
by $\wh \bV_y^{-1/2} \bY_t$, where 
$\wh \bV_y $ is a consistent estimator of $\bV_y$.



For a given integer $k_0\ge 1,$ let
\begin{align} \label{c3}
\bW_z=\sum_{k=0}^{k_0}\int_{\calU}\int_{\calU} \bSigma_{z,k}(u,v)^{\otimes2}\,{\rm d}u{\rm d}v~~\textrm{and}~~\bW_y=\sum_{k=0}^{k_0}\int_{\calU}\int_{\calU} \bSigma_{y,k}(u,v)^{\otimes2}\,{\rm d}u{\rm d}v\,.
\end{align}
Then both $\bW_y$ and $\bW_z$ are non-negative definite. According to \eqref{c1}, it holds that $
\bSigma_{y,k}(u, v) 
=\bA
\bSigma_{z,k}(u, v)\bA^\T\,,$
where $\bSigma_{z,k}(u,v)$ is block-diagonal with blocks on
the main diagonal of sizes $p_1\times p_1, \ldots, p_q\times p_q$. 
Due to $\bA \bA^\T= \bI_p$,
by (\ref{c3}),
\begin{align} \label{c4}
\bW_z= \bA^\T\bW_y \bA\,.
\end{align}
%
As all $\bSigma_{z,k}(u, v)$ for $k\ge 0$ and $(u, v)\in \calU^2$ are block-diagonal matrices of the same sizes, so is $\bW_z$.
Perform the eigenanalysis for each of $q$ blocks on the main diagonal of $\bW_z$ separately, 
leading to $q$ orthogonal matrices of sizes $p_l \times p_l$ for $l\in [q]$. The columns of each of those orthogonal matrices are the $p_l$ orthonormal eigenvectors from the corresponding eigenanalysis. We form a
$p\times p$ block diagonal orthogonal matrix $\bGamma_z$ with those $q$ orthogonal matrices along the  main block diagonal. Then the columns of
$\bGamma_z$ are the orthonormal eigenvectors of $\bW_z$, i.e.,  
\begin{equation} \label{c5}
\bW_z \bGamma_z = \bGamma_z \bD\,,
\end{equation}
where $\bD$ is a diagonal matrix consisting of the $p$ eigenvalues.
Then by (\ref{c4}) and \eqref{c5},
$
\bW_y \bA \bGamma_z = \bA \bW_z \bGamma_z =  \bA \bGamma_z \bD$. 
Thus the columns of $ \bGamma_y\equiv\bA \bGamma_z$ are the orthonormal
eigenvectors of $\bW_y$.
Combining this with (\ref{c1}) yields that
$
\bGamma_y^\T\bY_t(\cdot)  = \bGamma_z^\T\bA^\T\bY_t(\cdot) = \bGamma_z^\T\bZ_t(\cdot)\,.
$
Since $\bGamma_z$ is a block-diagonal orthogonal matrix with $q$ blocks,
$\bGamma_z^\T\bZ_t(\cdot)$ effectively applies orthogonal transformation within each of the $q$ groups of $\bZ_t(\cdot)$. Thus $\bGamma_z^\T\bZ_t(\cdot)$ is of the same segmentation structure of $\bZ_t(\cdot)$, i.e. knowing
$\bGamma_z^\T\bZ_t(\cdot)$ is as good as knowing the latent segmentation of $\bZ_t(\cdot).$ 
By \eqref{c1}, we have $\bZ_t(\cdot) = \bA^{\T} \bY_t(\cdot)$.
Hence $\bGamma_y$ can be taken as the required transformation matrix $\bA$.

Let $\wh \bSigma_{y,k}(u,v)$ be some consistent estimator of $\bSigma_{y,k}(u,v)$ for $k \in \{0\}\cup [k_0]$, to be specified in Section \ref{sec:ft}. We define an estimator of $\bW_y$ as
	\begin{equation}
	\label{Wy.est}
	  	\wh \bW_y = \sum_{k=0}^{k_0}\int_{\calU}\int_{\calU} \wh \bSigma_{y,k}(u,v)^{\otimes2}\,{\rm d}u{\rm d}v\,,
	\end{equation}
and calculate  
	its orthonormal eigenvectors $\wh \bfeta_1,\dots,\wh \bfeta_p$.
Let $\wh \bGamma_y=(\widehat{\bfeta}_1,\ldots,\widehat{\bfeta}_p)$. 
Then the required transformation matrix $\bA$ can be estimated by
a (latent)
column-permutation of $\wh \bGamma_y$.
More specifically, put 
\begin{align}\label{eq:hatZ}
\widehat{\bZ}_t(\cdot)\equiv\{\wh Z_{t1}(\cdot),\ldots,\wh Z_{tp}(\cdot)\}^\T= \wh \bGamma_y^\T \bY_t(\cdot)\,.
\end{align}
We propose below a data-driven procedure to divide the $p$ components of $\widehat{\bZ}_t(\cdot)$ into $\hat q$ uncorrelated groups. 


Recall
$
\bZ_t(\cdot)=\{Z_{t1}(\cdot),\ldots,Z_{tp}(\cdot)\}^\T$ with $\bSigma_{z,k}(\cdot,\cdot)=\{\Sigma_{z,k,ij}(\cdot,\cdot)\}_{i,j\in[p]}.$ 
For two curve series $Z_{ti}(\cdot)$ and $ Z_{tj}(\cdot)$ within the same group, one would expect that their lag-$k$ cross-autocovariance function
$\Sigma_{z,k,ij}(u,v)$ to be significantly different from zero for some integer $k$ and $(u,v)\in \calU^2$, thus leading to at least one large $\|\Sigma_{z,k,ij}\|_{\calS}$ for some integer $k$.
Based on $\widehat{\bZ}_t(\cdot)$ defined as \eqref{eq:hatZ}, we let
$
\widehat{\bSigma}_{z,k}(u,v)\equiv\{\widehat{\Sigma}_{z,k,ij}(u,v)\}_{i,j\in[p]}=\wh \bGamma_y^\T \widehat{\bSigma}_{y,k}(u,v)\wh \bGamma_y$
for any $(u,v)\in\calU^2$. 
Given a fixed integer $m\geq0$, we define the maximum cross-autocovariance over the lags between prespecified $-m$ and $m$ as 
\begin{equation}
\label{eq_T}
    \wh T_{ij}=\max_{ |k|\leq m }\|\wh\Sigma_{z,k,ij}\|_\calS
\end{equation}
for any pair $(i,j)\in[p]^2$ such that $i < j$,  
and regard $\wh Z_{ti}(\cdot)$ and $\wh Z_{tj}(\cdot)$ from the same group  if $\wh T_{ij}$ takes some large value.
To be specific, we  rearrange $\aleph=p(p-1)/2$ values of $\wh T_{ij}$ ($1 \leq i < j \leq p$) in the descending order $\wh T_{(1)} \geq \cdots \geq \wh T_{(\aleph)}$ and compute
\begin{equation} \label{eq.r}
    \hat  \varrho = \arg\max_{ j\in [\aleph]} \frac{\wh T_{(j)}+\delta_{n}}{\wh T_{(j+1)}+\delta_{n}}
\end{equation}
for some $\delta_{n}>0$. Corresponding to $\wh T_{(1)},\dots, \wh T_{(\hat \varrho)},$ we identify $\hat \varrho$ pairs of cross-correlated curves. To divide the $p$ components of $\widehat{\bZ}_t(\cdot)$ into several uncorrelated groups, we can first start with $p$ groups with each $\wh Z_{tj}(\cdot)$ in one group and then repeatedly merge two groups if two cross-correlated curves are split over the two groups. The iteration is terminated until all the cross-correlated pairs are within one group. Hence we obtain the estimated group structure of $\wh \bZ_t(\cdot)$ with the number of the final groups $\hat q$ being the estimated value for $q$. Denote by $\wh \bZ_t^{(l)}(\cdot)$ the estimated $l$-th group for $l 
\in [\hat q]$.  The estimated transformation matrix $ \wh \bA =  (\wh \bA_1, \ldots, \wh \bA_{\hat q})$ can then be  found by reorganizing the order of 
$(\widehat{\bfeta}_1,\ldots,\widehat{\bfeta}_p)$ such that 
\begin{equation} \label{est.trans.curve}
    \wh \bZ_{t}^{( l)}(\cdot) ={\wh \bA_l}^{\T} \bY_t(\cdot)\,,~~~~l \in [\hat q]\,.
\end{equation}

\begin{remark}
(i) 
We include a small term $\delta_n>0$ in (\ref{eq.r}) to stabilise the 
estimates for `0/0'. Given a suitable order of  $\delta_n$, we can  establish the group recovery consistency. See Theorem~\ref{thm.seg.msc} in Section \ref{sec.th}. A common practice is to set $\delta_n =0$ and
replace $\aleph$ by $c_\varrho\aleph$  in (\ref{eq.r}) for some constant $c_\varrho \in (0,1)$, see \cite{lam2012} and \cite{ahn2013}.

(ii)
All integrated terms in $\bW_y$ are non-negative definite. Hence there is no information cancellation over different lags.  Therefore
the estimation is insensitive to the choice of $k_0.$ In practice a small $k_0$ (such as $k_0 \le 5$) is often sufficient, 
while further enlarging $k_0$ tends to add more noise to $\bW_y$. 

\end{remark}


\vspace{-0.5cm}
\subsection{Selection of $\widehat{\bSigma}_{y,k}(u,v)$}
\label{sec:ft}
\vspace{-0.1cm}

The estimate $\widehat{\bSigma}_{y,k}(u,v)$ plays a key role in 
Section \ref{sec:seg.est}. 
Let $\bar{\bY}(u)=n^{-1}\sum_{t=1}^n\bY_t(u).$ A natural candidate for $\widehat{\bSigma}_{y,k}(u,v)$ is the sample version of $\bSigma_{y,k}(u,v)$ defined as
\begin{equation} 
\label{eq_sample}
     \wh \bSigma{}^{\ttS}_{y,k}(u,v)  = \frac{1}{n-k}\sum_{t=1}^{n-k}\{ \bY_{t}(u) - \bar \bY(u) \}
\{ \bY_{t+k}(v) - \bar \bY(v) \}^\T\,, ~~~~k\in\{0\}\cup[k_0]\,.
\end{equation}
When $p^2/n\rightarrow0$, $\wh \bSigma{}^{\ttS}_{y,k}(u,v)$ is a valid estimator for $\bSigma_{y,k}(u,v)$. However when $p$ grows faster than $n^{1/2}$,
it does not always hold that $\|\wh \bSigma{}^{\ttS}_{y,k}(u,v)-\bSigma_{y,k}(u,v)\|_2 \to 0$ in probability.
Under the high-dimensional scenario, the orthogonality of $\bA$ naturally results in 
the magnitude of   
many of its entries being small, leading to certain sparsity on $\bA$ which  will then pass onto the autocovariance functions $\bSigma_{y,k}(\cdot,\cdot)$,
as $\bSigma_{y,k}(\cdot, \cdot) = \bA \bSigma_{z,k}(\cdot, \cdot)\bA^{\T}.$  

Inspired by the spirit of threshold estimator for large covariance matrix \cite[]{bickel2008}, we apply the functional thresholding rule, which combines the functional generalizations of hard thresholding and shrinkage with the aid of the Hilbert--Schmidt norm of functions, on the entries of the sample autocovariance function $\wh \bSigma{}^{\ttS}_{y,k}(u,v)=\{\widehat\Sigma_{y,k,ij}^{\ttS}(u,v)\}_{i,j\in[p]} $ in (\ref{eq_sample}). 
This leads to the estimator
\begin{equation}
\label{th-est}
    \calT_{\omega_k}(\wh \bSigma{}^{\ttS}_{y,k})(u,v) =   \big[\wh{\Sigma}^{\ttS}_{y,k,ij}(u,v) I\{\|\wh{\Sigma}^{\ttS}_{y,k,ij}\|_{\calS}\geq \omega_k\}\big]_{i,j \in [p]}\,, ~~~~(u,v)\in \calU^2\,,
\end{equation}
where $\omega_k \geq 0$ is the thresholding parameter at lag $k.$ Taking $\wh \bSigma_{y,k}$ in (\ref{Wy.est}) as $\calT_{\omega_k}(\wh \bSigma{}^{\ttS}_{y,k})$ yields 
\begin{equation}\label{Wy.th.est}
    \wh \bW_y = \sum_{k=0}^{k_0}\int_{\calU} \int_{\calU} \calT_{\omega_k}(\widehat \bSigma{}^{\ttS}_{y,k})(u,v)^{\otimes2}\,{\rm d} u{\rm d}v\,.
\end{equation}

\begin{remark}
\label{rmk:fth}
The thresholding parameter $\omega_k$ for each $k\in\{0\}\cup[k_0]$ can be selected using an  $L$-fold cross-validation approach. Specifically, we sequentially divide the set $[n]$ into $L$ validation sets $V_1, \dots, V_L$ of approximately equal size. For each $l\in[L]$, let $\wh \bSigma{}^{\ttS,(l)}_{y,k}(u,v)=\{\wh\Sigma{}^{\ttS,(l)}_{y,k,ij}(u,v)\}_{i,j\in[p]}$ and $\wh \bSigma{}^{\ttS,(-l)}_{y,k}(u,v)=\{\wh\Sigma{}^{\ttS,(-l)}_{y,k,ij}(u,v)\}_{i,j\in[p]}$ be the sample lag-$k$ autocovariance functions based on the $l$-th validation set $\{\bY_{t}(\cdot): t \in V_l\}$ and
the remaining $L-1$ sets $\{\bY_{t}(\cdot): t \in [n] \setminus  V_l\},$ respectively.
We select the optimal $\hat \omega_k$ by minimizing 
\begin{equation*} 
\label{eq.cv}
     \text{Error}(\omega_k) = \frac{1}{L}\sum_{l=1}^L\sum_{i,j=1}^p \big\|
\calT_{\omega_k}(\wh \Sigma{}^{\ttS,(l)}_{y,k,ij}) - \wh \Sigma{}^{\ttS,(-l)}_{y,k,ij}
\big\|_{\calS}^2\,,
\end{equation*}
where $\calT_{\omega_k}(\wh \Sigma{}^{\ttS,(l)}_{y,k,ij})(u,v)=\wh{\Sigma}^{\ttS,(l)}_{y,k,ij}(u,v) I\{\|\wh{\Sigma}^{\ttS,(l)}_{y,k,ij}\|_{\calS}\geq \omega_k\}$.
\end{remark}

\vspace{-0.3cm}
\section{Variational multivariate FPCA}
\label{sec:dr}
\vspace{-0.2cm}
Our second step is to represent (linear) dynamic structure of each $\bZ_t^{(l)}(\cdot)$ in terms of a
vector time series via representation (\ref{Y.err}). The key
idea is to identify the finite decomposition
for $\bX_t^{(l)}(\cdot)$. 
For $(u, v) \in \calU^2$ and $k\ge 0$, let $\bmu^{(l)}(u)= \cE\{\bX_t^{(l)}(u)\}$ and
\begin{equation} \nonumber
\bM_k^{(l)}(u, v) =\cE[ \{\bX_{t}^{(l)}(u) - \bmu^{(l)}(u)\} \{ \bX_{t+k}^{(l)}(v) - \bmu^{(l)}(v)\}^\T]\,.
\end{equation}
Then the multivariate Karhunen-Lo\`eve decomposition
for $\bX^{(l)}_t(\cdot)$ serving as the foundation of multivariate FPCA \citep{Chiou2014, Happ2018} admits the form
\begin{equation}
    \label{kl-decomp}
    \bM_0^{(l)}(u,v) = \sum_{j=1}^{\infty} \lambda_j^{(l)} \bvarphi_j^{(l)}(u) \bvarphi_j^{(l)}(v)^\T\,,
    \quad\quad
    \bX_t^{(l)}(u) - \bmu^{(l)}(u) = \sum_{j=1}^{\infty} \xi^{(l)}_{tj} \bvarphi_j^{(l)}(u)\,,
\end{equation}
where $\la_1^{(l)} \ge \la_2 ^{(l)}\ge \dots \ge 0$ are the ordered eigenvalues of $\bM_0^{(l)}(\cdot,\cdot)$, $\bvarphi_1^{(l)}(\cdot), \bvarphi_2^{(l)}(\cdot), \dots $ are the corresponding orthonormal eigenfunctions satisfying $\int_{\calU} \bvarphi_{j}^{(l)}(u)^\T \bvarphi_{k}^{(l)}(u) \,{\rm d}u = I(j=k)$, and
$
\xi_{tj}^{(l)} = \int_\calU \bvarphi_j^{(l)}(u) ^\T\{ \bX_t^{(l)}(u) - \bmu^{(l)}(u) \}\,{\rm d}u
$
with $\cE\{\xi_{tj}^{(l)}\}=0$ 
and $\cov\{\xi_{tj}^{(l)},\xi_{tk}^{(l)}\} = \la_j^{(l)}I(j=k).$ 

When $\bX_t^{(l)}(\cdot)$ is $r_l$-dimensional
in the sense that $\la_{r_l}^{(l)}>0$ and $\la_{r_l+1}^{(l)}=0,$ 
 the dynamics of $\bX^{(l)}_t(\cdot)$ is entirely determined by that of $r_l$-vector time series $\bxi_t^{(l)}
= \{\xi_{t1}^{(l)}, \ldots, \xi_{tr_l}^{(l)}\}^\T$.
Unfortunately, under the latent decomposition \eqref{Y.err}, i.e.,
\begin{equation}
    \label{z.kl.err}
    \bZ^{(l)}_t(u) = \bX^{(l)}_t(u) + \bvar^{(l)}_{t}(u)= \bmu^{(l)}(u) + \sum_{j=1}^{r_l} \xi^{(l)}_{tj} \bvarphi_j^{(l)}(u) + \bvar^{(l)}_{t}(u)\,, ~~~~u \in \calU \,, 
\end{equation}
the standard multivariate FPCA based on (\ref{kl-decomp}) is inappropriate as $\bX_t^{(l)}(\cdot)$ is unobservable and we cannot provide a consistent estimator for $\bM_0^{(l)}(u, v)$ based on $\bZ_t^{(l)}(\cdot)$ due to the fact $\cov\{\bZ_{t}^{(l)}(u),\bZ_{t}^{(l)}(v)\}=\bM_0^{(l)}(u, v) + \cov\{\bve_t^{(l)}(u),\bve_t^{(l)}(v)\}.$

Now we introduce the variational multivariate FPCA based on a variational multivariate Karhunen-Lo\`eve decomposition
for $\bX_t^{(l)}(\cdot)$.
Motivated from the fact $\cov\{\bZ_{t}^{(l)}(u),\bZ_{t+k}^{(l)}(v)\}=\bM_k^{(l)}(u, v)$ for any $k \geq 1,$ 
for a prespecified small integer $k_0 \ge 1$, we define
\begin{equation} \label{b5}
\bK^{(l)} (u, v) = \sum_{k=1}^{k_0} \int_\calU \bM_k^{(l)}(u, w) \bM_k^{(l)}(v, w)^\T
\,{\rm d}w\,.
\end{equation}
Similar to $\bM_0^{(l)}$, $\bK^{(l)}$ is also non-negative definite and admits a spectral decomposition
\begin{equation} \nonumber
\bK^{(l)} (u, v) = \sum_{j=1}^{\infty} \theta_j^{(l)} \bpsi_j^{(l)}(u) \bpsi_j^{(l)}(v) ^\T\,,
\end{equation}
where $\theta_1^{(l)} \ge \theta_2^{(l)} \ge \cdots \ge 0$ are the eigenvalues of $\bK^{(l)}$, and $\bpsi_1^{(l)}(\cdot), \bpsi_2^{(l)}(\cdot), \ldots $ are the corresponding orthonormal eigenfunctions. 

\begin{proposition} \label{lm.space}
    Let $\bOmega_k^{(l)} = \cE[\bxi_{t}^{(l)} \{\bxi_{t+k}^{(l)}\}^\T]$ be a full-ranked matrix for some $k\in [k_0]$. Then it holds that {\rm(i)} $\theta_{r_l}^{(l)}>0$ and  $\theta_{r_l+1}^{(l)} = 0$; 
    {\rm(ii)} ${\rm span}\{\bvarphi_1^{(l)}(\cdot), \dots, \bvarphi_{r_l}^{(l)}(\cdot)\}={\rm span}\{\bpsi_1^{(l)}(\cdot), \dots, \bpsi_{r_l}^{(l)}(\cdot)\}$.
\end{proposition} 

Proposition~\ref{lm.space} shows that, under the expansion (\ref{z.kl.err}), 
$\bK^{(l)}$ has exactly $r_l$ nonzero eigenvalues
, and the dynamic space spanned by $\{\bpsi_1^{(l)}(\cdot), \dots, \bpsi_{r_l}^{(l)}(\cdot)\}$ remains the same as that spanned by $\{\bvarphi_1^{(l)}(\cdot), \dots, \bvarphi_{r_l}^{(l)}(\cdot)\}.$
Therefore, $\bX_t^{(l)}(\cdot)$ can be expanded using $r_l$ basis functions  $\bpsi_1^{(l)}(\cdot), \dots, \bpsi_{r_l}^{(l)}(\cdot),$ i.e.,  \begin{equation}  \label{eq.zeta}
\bX_t^{(l)}(u) - \bmu^{(l)}(u) = \sum_{j=1}^{r_l} \zeta_{tj}^{(l)} \bpsi_j^{(l)}(u)\,, ~~~~u \in \calU\,,
\end{equation}
where the basis coefficients
$\zeta^{({l})}_{tj} = \int_\calU \bpsi_j^{(l)}(u)^\T \{ \bX_t^{(l)}(u) - \bmu^{(l)}(u) \}\, {\rm d}u$. 
Note that we take the sum in defining $\bK^{(l)}$ in (\ref{b5}) to accumulate the information from different lags, and there is no information cancellation as each term in the sum is non-negative definite. 
An additional advantage for using the nonzero lagged autocovariance-based decomposition is that the identified directions 
$\bpsi_1^{(l)}(\cdot),\ldots,\bpsi_{r_l}^{(l)}(\cdot)$ catch the most significant serial dependence, which leads to the most efficient dimension reduction and is thus advantageous for prediction.

Noting that $\bZ_t^{(l)}(\cdot)$ is not directly observable, we can only estimate
$\bM_k^{(l)}$ and $\bK^{(l)}$ based on 
$\hat p_l$-vector of estimated transformed curve subseries  $\wh \bZ_t^{(l)}(\cdot)= \{\wh Z_{t1}^{(l)}(\cdot), \dots, \wh Z_{t\hat p_l}^{(l)}(\cdot)\}^{\T}$  obtained in the segmentation transformation step.
With the aid of \eqref{est.trans.curve},
for $k \in \{0\}\cup[k_0]$, put
\begin{equation} \label{M.gp.est}
\wh \bM_k^{(l)}(u, v) = \wh \bA_l^{\T} \wh \bSigma_{y,k}(u,v)\wh\bA_l\,.
\end{equation}
It is easy to see from (\ref{Y.err}) that $\wh \bM_k^{(l)}(u, v)$ is a reasonable
estimator for $\bM_k^{(l)}(u, v)$ when $k \geq1$, as it filters out white noise $\bve_t^{(l)}(\cdot)$ automatically. 
It is noteworthy that (\ref{M.gp.est}) requires the consistent estimators for $\bSigma_{y,k}(u,v)$. 
Its implementation under the high-dimensional setting can thus be done by setting $\wh\bSigma_{y,k}(u,v)=\calT_{\omega_k}(\wh \bSigma{}^{\ttS}_{y,k})(u,v)$ defined in \eqref{th-est}. 


To estimate $\bpsi_j^{(l)}(\cdot)$ and $\zeta^{(l)}_{tj}$ in (\ref{eq.zeta}), we perform eigenanalysis of the estimator for $\bK^{(l)},$
\begin{align} \label{K.gp.est}
\wh \bK^{(l)} (u, v)& = \sum_{k=1}^{k_0} \int_\calU \wh\bM_k^{(l)}(u, w)
 \wh\bM_k^{(l)}(v, w)^\T\,{\rm d}w\,,
\end{align}
leading to the eigenvalues $\hat \theta_1^{(l)} \ge \hat \theta_2^{(l)} \ge  \cdots \ge 0$, and the corresponding orthonormal eigenfunctions
$\wh \bpsi_1^{(l)}(\cdot), \wh\bpsi_2^{(l)}(\cdot), \dots$.  
To estimate $r_l$ (i.e. the number of nonzero
eigenvalues), we take the commonly-adopted ratio-based estimator for $r_l$ as:
\begin{equation} \label{eq.r.2}
    \hat r_l = \arg\max_{j\in [n-k_0]}
    \frac{\hat \theta_{j}^{(l)}+\tilde \delta_{n}}{\hat \theta_{j+1}^{(l)}+\tilde\delta_{n}} 
\end{equation}
for some $\tilde\delta_{n}>0$. Under some regularity conditions, such defined $\hat r_l$ is a consistent
estimator for $r_l$; 
see Theorem~\ref{thm.dr.dim} in Section \ref{sec.th}.
In practice, since $\tilde \delta_n$ is usually unknown, we instead adopt 
$
\hat r_l = \arg \max_{j \in [c_r(n-k_0)]} \hat \theta_j^{(l)} / \hat \theta_{j+1}^{(l)}$,
where $c_r \in (0, 1)$ is a prescribed constant aiming to avoid fluctuations due to the ratios of extreme small values.

Let 
$\hat \zeta_{tj}^{(l)}=\int_{\calU} \wh \bpsi_j^{(l)}(u)^\T \{\wh \bZ_t^{(l)}(u)- \widebar\bZ^{(l)}(u)\}\, {\rm d}u$ for $t\in[n]$, $j\in[\hat r_l]$ and $l \in [\hat q]$. We can fit a model for the $\hat r_l$-dimensional vector time series $\widehat\bzeta_{t}^{(l)}=\{\hat \zeta_{t1}^{(l)}, \ldots, \hat  \zeta_{t \hat r_l}^{(l)}\}^\T$ with $t\in[n]$ to obtain its $h$-step ahead prediction $\mathring \bzeta_{n+h}^{(l)}$ and then recover the $h$-step ahead functional prediction as
\begin{equation} \nonumber
    \mathring \bZ_{n+h}^{(l)}(u) = \widebar \bZ^{(l)}(u) + \sum_{j = 1}^{\hat r_l} \mathring \zeta_{(n+h)j}^{(l)} \widehat \bpsi_j^{(l)}(u)\,, ~~~h \geq 1\,.
\end{equation}
We finally obtain the $h$-step ahead prediction $ \wh \bA \mathring \bZ_{n+h}(\cdot)$ for original functional time series,
 where $\wh \bA = (\wh \bA_1, \ldots, \wh \bA_{\hat q})$ and   $\mathring \bZ_{n+h} (\cdot)=\{\mathring\bZ_{n+h}^{(1)} (\cdot)^{\T}, \dots, \mathring \bZ_{n+h}^{(\hat q)}(\cdot)^{\T}\}^{\T}.$

\vspace{-0.3cm}
\section{Theoretical properties} 
\label{sec.th}
\vspace{-0.2cm}
This section presents theoretical analysis of our two-step estimation procedure. 
To ease presentation, we focus on the high-dimensional scenario  and develop the theoretical results based on the estimator $ \calT_{\omega_k}(\wh \bSigma{}^{\ttS}_{y,k})(u,v)$ in (\ref{th-est}). 
To simplify notation, we use $B  $ to denote the linear  operator induced from the kernel function
$B \in {\mathbb S}, $ i.e., for any $f \in L_2(\calU),$ $B(f)(\cdot)=\int_{\calU}B(\cdot\,,v)f(v)\,{\rm d}v \in L_2(\calU).$ Denote the $p$-fold Cartesian product $\mathbb{H}=L_2(\calU) \times \dots \times L_2(\calU).$
For any $\bbf, \bg\in \mathbb{H},$ we denote the inner product by
$
\inner{\bbf}{\bg}=\int_{\calU}\bbf(u)^{\T}\bg(u)\,{\rm d}u 
$
with the induced norm $\|\cdot\| =\inner{\cdot}{\cdot}^{1/2},$
and use $\bB  $ to denote the linear  operator induced from the kernel matrix function
$\bB = (B_{ij})_{m_1 \times m_2}$ with each $B_{ij} \in {\mathbb S},$ i.e., for any $\bbf \in \mathbb{H},$  $\bB(\bbf)(\cdot) = \int_\calU
\bB(\cdot\,, v) \bbf(v)\,{\rm d} v \in \mathbb{H}$.  We write $ \|\bB\|_{\calS,\infty} = \max_{i\in [m_1]} \sum_{j = 1}^{m_2}\|B_{ij}\|_{\calS}.$
Before imposing the regularity conditions, we firstly define the functional version of sub-Gaussianity that facilities the development of non-asymptotic results for Hilbert space-valued random elements.

\begin{definition}
\label{def_subGaussian}
Let $\Upsilon_t(\cdot)$ be a mean zero random variable in $ L_2(\calU)$ and $\Sigma_0:  L_2(\calU) \to  L_2(\calU)$ be a covariance operator. We call  $\Upsilon_t(\cdot)$ a sub-Gaussian process if there exists a constant $c>0$ such that
$\mathbb{E}[\exp\{\langle f, \Upsilon_t - \mathbb{E}(\Upsilon_t)\rangle\}] \leq \exp\{2^{-1}c^2\langle f, \Sigma_{0}(f)\rangle\}$ for all $f \in  L_2(\calU).$
\end{definition}

\begin{condition} 
\label{con_subG}
(i) $\{\bY_t(\cdot)\}$ is a sequence of multivariate functional linear processes with sub-Gaussian errors, i.e., $ \bY_t(\cdot) =  \sum_{l=0}^{\infty}\bPsi_l(\beps_{t-l}),$ where $\bPsi_l=(\Psi_{l,ij})_{p\times p}$ with each $\Psi_{l,ij} \in {\mathbb S}$ and  $\beps_{t}(\cdot) = \{\epsilon_{t1}(\cdot),\dots,\epsilon_{tp}(\cdot)\}^{\T}$ with independent components of mean-zero sub-Gaussian processes satisfying Definition~\ref{def_subGaussian};
(ii) The coefficient functions satisfy
$\sum_{l=0}^{\infty}\|\bPsi_l\|_{\calS,\infty} = O(1);$ 
(iii) $\max_{j\in[p]}\int_{\calU} \cov\{\epsilon_{tj}(u),\epsilon_{tj}(u)\}\,{\rm d}u =O(1).$
\end{condition}

\begin{condition} 
\label{con_stability}
For $\{\bY_t(\cdot)\},$ its spectral density operator $\bLambda_{y,\theta}= (2\pi)^{-1}\sum_{k\in \mathbb{Z}}\bSigma_{y,k}\mathrm{exp}(-k\theta\sqrt{-1})$ for $\theta \in [-\pi,\pi]$ exists and the functional stability measure 
\begin{equation} 
\label{eq_M_X}
\calM_y = 2\pi \mathop{\text{ess sup}}\limits_{\theta \in [-\pi,\pi],\bPhi\in \mathbb{H}_0}\frac{\langle\bPhi,\bLambda_{y,\theta}(\bPhi)\rangle}{\langle\bPhi,\bSigma_{y,0}(\bPhi)\rangle} < \infty\,,
\end{equation}
where $\mathbb{H}_0 = \{\bPhi \in \mathbb{H}:\langle\bPhi,\bSigma_{y,0}(\bPhi)\rangle \in (0, \infty)\}.$
\end{condition}


Write $ \bSigma{}_{y,k}(u,v)=\{\Sigma_{y,k,ij}(u,v)\}_{i,j\in[p]}$. Conditions \ref{con_subG}(ii) and \ref{con_subG}(iii) guarantee the covariance-stationarity of $\{\bY_t(\cdot)\}$ and imply that $\max_{j\in[p]}\rmint_{\calU} \Sigma_{y,0,jj}(u,u)\,{\rm d}u = O(1).$ 
Condition~\ref{con_stability} places a finite upper bound on the functional stability measure, which characterizes the effect of small decaying eigenvalues of $\bSigma_{y,0}$ on the numerator of (\ref{eq_M_X}), thus being able to handle infinite-dimensional functional objects $Y_{tj}(\cdot).$ See its detailed discussion in \cite{fang2021a,guo2021}.  
Conditions \ref{con_subG} and \ref{con_stability} are essential to derive 
$\max_{i,j\in[p]}\| {\widehat\Sigma{}^{\ttS}_{y,k,ij}  - \Sigma_{y,k,ij}}\|_\calS =  O_{\rm p}\{\mathcal{M}_y (n^{-1}{\log p})^{1/2}\}$  for $\widehat\Sigma{}^{\ttS}_{y,k,ij}$ 
involved in \eqref{th-est},
which plays a crucial rule in our theoretical analysis.

\begin{condition}
\label{cond.sA}
For $\bA=(A_{ij})_{p \times p},$ $\max_{i\in[p]}\sum_{j=1}^p |A_{ij}|^{\alpha} \leq s_1$ and $\max_{j\in[p]}\sum_{i=1}^p |A_{ij}|^{\alpha} \leq s_2$ for some constant $\alpha \in [0,1).$
\end{condition}

The parameters $s_1$ and $s_2$ determine the row and column sparsity levels of $\bA,$ respectively. 
The row sparsity with small $s_1$ entails that each component of $\bY_t(\cdot)$ is a linear combination of a small number of components in $\bZ_t(\cdot),$ while the column sparsity with small $s_2$
corresponds to the case that each $Z_{tj}(\cdot)$ has impact on only a few components of $\bY_t(\cdot).$ The parameter $\alpha$ also controls the sparsity level of $\bA$ with a smaller value yielding a sparser $\bA$. 
Write  
\begin{equation} \label{p.dagger}
    p_{\dagger} = \max_{l\in[q]} p_l\,.
\end{equation}
Lemma~\ref{lemma.Sigma.k} in the supplementary material reveals that the functional sparsity structures in columns/rows of $\bSigma_{y,k}(\cdot,\cdot)$ are determined by $s_1s_2p_{\dagger}$ with smaller values of $s_1$, $s_2$ and $p_{\dagger}$ yielding functional sparser $\bSigma_{y,k}(\cdot,\cdot)$.  

Recall that $\bW_z=\textrm{diag}(\bW_{z,1},\ldots,\bW_{z,q})$ in (\ref{c4}) is a block-diagonal matrix, where $\bW_{z,l}$ is a $p_l\times p_l$ matrix. We  further define 
\begin{equation} \label{eq.rho}
\rho=\min_{j\neq l}\min_{\lambda\in\Lambda(\bW_{z,l}),\,\tilde \lambda\in\Lambda(\bW_{z,j})}| \lambda-\tilde \lambda|\,,
\end{equation}
where $\Lambda(\cdot)$ denotes the set of eigenvalues of the matrix, and 
assume  $\rho>0$.

We first establish the group recovery consistency of the segmentation step. To do this, we
reformulate the permutation in Section~\ref{sec:seg.est} in an equivalent graph representation way. 
Recall $\bGamma_y = \bA \bGamma_z$ and $\bGamma_z$ is a block-diagonal orthogonal matrix with the main block sizes $p_1,\dots, p_q$. Write 
$
    \bGamma_z = \text{diag}(\bGamma_{z,1}, \dots, \bGamma_{z,q})$. 
Since $\bA = (\bA_1,\dots, \bA_q),$ we have $\bGamma_y\equiv(\bfeta_1,\ldots,\bfeta_p) = (\bA_1\bGamma_{z,1}, \ldots, \bA_q\bGamma_{z,q})$. The columns of $\bGamma_y$ are naturally partitioned in to $q$ groups $G_1, \dots, G_q$, where $ G_l = \{\bfeta_{\sum_{l'=0}^{l-1} p_{l'}+1}, \dots, \bfeta_{\sum_{l'=0}^{l} p_{l'}}\}$ with $p_0=0$. To simplify the notation, we just write 
\begin{equation} \label{G.true}
    G_l = \bigg\{\sum_{l'=0}^{l-1} p_{l'}+1, \dots, \sum_{l'=0}^{l} p_{l'}\bigg\}\,,~~~~l\in[q]\,.
\end{equation}

 Recall that the columns of such defined $\bGamma_y$ are the eigenvectors of $\bW_y$.  For $\rho$ defined in \eqref{eq.rho}, if $\|\widehat{\bW}_y-\bW_y\|_2 \leq \rho/5$,  by Lemma~\ref{la:space} in the supplementary material, there exists an orthogonal matrix $\bH = \text{diag} (\bH_1, \dots, \bH_q)$  with $\bH_l\in\mathbb{R}^{p_l\times p_l}$ for each $l \in [q]$ and  a column permutation matrix $\bR$ for $\wh \bGamma_y$,
such that $ \wh \bGamma_y \bR  \equiv (\wh \bPi_1, \dots, \wh \bPi_q)$ with $\wh \bPi_l \in \mathbb{R}^{p\times {p_l}}$, and 
\begin{equation} \label{eq.th}
    \|\wh \bPi_l  - \bA_l \bGamma_{z,l} \bH_l \|_2 \leq8 \rho^{-1}\|\widehat{\bW}_y-\bW_y\|_2\,.
\end{equation}
If the $p$ eigenvalues of $\bW_y$ are distinct, $\bH$ is a diagonal matrix with elements in the diagonal being $1$ or $-1$. Write $\bGamma_y \bH = (\bA_1 \bGamma_{z,1} \bH_1, \dots, \bA_q \bGamma_{z,q} \bH_q)\equiv (\bgamma_1,\dots, \bgamma_p)$. For each $l\in[q]$, we can define a graph $(G_l,E_l)$ such that $(i,j)\in E_l$ if and only if $\max_{|k| \leq m}\|\bgamma_{i}^{\T}\bSigma_{y,k}\bgamma_{j}\|_\calS\neq 0$.


\begin{condition}
\label{con.inseparable}
	There exists some $\varsigma>0$ such that
	$
	\inf_{(i,j)\in E_l}\max_{|k| \leq m}\|\bgamma_{i}^{\T}\bSigma_{y,k}\bgamma_{j}\|_\calS\geq \varsigma
	$
for each $l \in [q]$, where $m$ is specified in \eqref{eq_T}.	
\end{condition}

Condition~\ref{con.inseparable}  ensures that the group $G_l$ is inseparable at the minimal signal level $\varsigma$ given the  transformation $\bA_l \bGamma_{z,l} \bH_l$ for each $l\in[q]$, and facilitates the specifications of  the true number of groups $q$ and the associated segmentation structure under \eqref{c1}.
Define
 $ T_{ij}
 =\max_{|k| \leq m}\|{\bgamma}_{i}^\T  \bSigma_{y,k}{\bgamma}_{j}\|_{\cal S}$ and $\varrho=\sum_{l=1}^q|E_l|$. Rearrange $\aleph=p(p-1)/2$ values of $ T_{ij}$ ($1 \leq i < j \leq p$) in the descending order, $ T_{(1)} \geq \cdots \geq  T_{(\aleph)}. $
We then have  $T_{(i)} \geq \varsigma$ for $i \in [\varrho]$ and $T_{(i)} = 0$ for $i  \geq \varrho+1$. Denote by $E = \{(i,j): T_{ij} \geq T_{(\varrho)}, 1 \leq i < j \leq p\}$
the  edge set of $G = [p]$ under the transformation $\bGamma_y \bH.$  The true segmentation $\{G_1, \dots, G_q\}$  in (\ref{G.true}) can then be identified by splitting $(G,E)$ into $q$ isolated subgraphs $(G_1, E_1),\ldots,(G_q,E_q)$, where $q$ represents the true number of uncorrelated groups.

Recall that with the aid of $\wh \bGamma_y$,  the estimated segmentation is obtained via the ratio-based estimator
$\hat \varrho$ as defined in (\ref{eq.r}). To be specific, we  build an estimated graph $(G, \widetilde E)$ with vertex set $G=[p]$ and edge set 
$
    \widetilde E = \{(i,j): \wh T_{ij} \geq \wh T_{( \hat \varrho)}, 1 \leq i < j \leq p\}
$,
and split it into $\hat{q}$ isolated subgraphs $(\widetilde G_{1}, \widetilde E_{1}),\ldots,(\widetilde{G}_{\hat{q}},\widetilde{E}_{\hat{q}})$. Note that $p$ columns of $\wh \bGamma_y=(\widehat{\bfeta}_1,\ldots,\widehat{\bfeta}_p) $ correspond to the ordered eigenvalues $\lambda_1(\wh \bW_y)\geq\cdots\geq\lambda_p(\wh \bW_y)$.  Write 
$ \wh \bGamma_y \bR \equiv (\wh \bgamma_1,\dots, \wh \bgamma_p)$ and 
let $\pi:[p] \to [p]$ denote the permutation associated with $\bR$, i.e., 
$\wh \bgamma_i =\wh \bfeta_{\pi(i)}$. Based on the permutation mapping $\pi$, we let
$
    \wh G_{l} = \{\pi^{-1}(i): i \in \widetilde G_{l}\}$ for $l\in [\hat q]$. 

\begin{theorem}
\label{thm.seg.msc}
Let Conditions {\rm\ref{con_subG}--\ref{con.inseparable}} hold.
For each $|k|\leq k_0 \vee m $, select  $\omega_k = c_k \mathcal{M}_y (n^{-1}{\log p})^{1/2}$ in \eqref{th-est} for some sufficiently large constant $c_k>0$. Assume 
   $(\rho^{-1}s_1^2s_2^2 p_{\dagger}^{3-\alpha})^{2/(1-\alpha)} \calM_y^2 \log p = o(n)$
   and 
   $\delta_{n}$ in 
 \eqref{eq.r} satisfies  
  $\rho^{-1}s_1^3 s_2^3 p_{\dagger}^{5-2\alpha}\calM_y^{1-\alpha}(n^{-1}\log p)^{(1-\alpha)/2}\ll \delta_n\ll\varsigma^2T_{(1)}^{-1}$, where $p_\dagger$ and $\rho$ are specified in \eqref{p.dagger} and \eqref{eq.rho}, respectively. As $n\rightarrow\infty$, it holds that {\rm (i)} 
$\mathbb{P}(\hat q  = q) \to 1$ and 
{\rm (ii)}  there exists a permutation $\tilde\pi:[q] \to [q]$ such that 
$\mathbb{P}[\bigcap_{l= 1}^q \{\widehat{G}_{\tilde\pi(l)} =  G_l\}\, |\, \hat q 
 = q ]\rightarrow 1$. 
\end{theorem}

Theorem~\ref{thm.seg.msc} gives the group recovery consistency of our segmentation step. 
We next evaluate the errors in estimating ${\cal C}(\bA_l)$ for $l\in[q]$. Based on the estimated groups $\{\wh G_1, \dots, \wh G_{\hat q}\}$, we 
reorganize the order of 
$(\widehat{\bgamma}_1,\ldots,\widehat{\bgamma}_p) = (\wh \bfeta_{\pi(1)}, \dots, \wh \bfeta_{\pi(p)})$ and
define $\wh \bA_l$ in \eqref{est.trans.curve} as $\wh \bA_l = (\wh \bgamma_i)_{i \in \wh G_l}$ for $l \in [\hat q]$.
We consider a general discrepancy measure \cite[]{chang2015,chang2018} 
between two linear spaces $\calC(\bE_1)$ and $\calC(\bE_2)$ spanned by the columns of $\bE_1 \in \mathbb{R}^{p\times \tilde p_1}$ and $\bE_2 \in \mathbb{R}^{p\times \tilde p_2}$, respectively, 
with $\bE_i^{\T} \bE_i =\bI_{\tilde p_i}$  for $i \in[2]$ as
\begin{equation} \label{eq.dis.gr}
      D\{\calC( \bE_1),\calC( \bE_2)\} =  \sqrt{1 - \frac{\text{tr}(\bE_1\bE_1^{\T}\bE_2\bE_2^{\T})}{\max(\tilde p_1,\tilde p_2)}}  \in [0,1]\,.
\end{equation}
Then  $ D\{\calC( \bE_1),\calC( \bE_2)\}$ is equal to 0 if and only if $\calC(\bE_1) \subset \calC(\bE_2)$ or $\calC(\bE_2) \subset \calC(\bE_1),$ and to 1 if and only if the two spaces are orthogonal. 

\begin{theorem}
\label{thm.seg.distance}
Let conditions for Theorem {\rm\ref{thm.seg.msc}} hold. As $n \to \infty$, it holds that  
\begin{equation} 
    \nonumber
    \max_{l \in [q]}  \min_{j\in[\hat q]}   D\{\calC( \bA_l),\calC( \widehat \bA_j)\} = O_{\rm p}\big\{\rho^{-1} s_1^2 s_2^2 p_{\dagger}^{3-\alpha}\mathcal{M}_y^{1-\alpha}(n^{-1}{\log p})^{(1-\alpha)/{2}} \big\}\,.
\end{equation}

\end{theorem}


Theorem~\ref{thm.seg.distance} presents the uniform convergence rate for $\min_{j\in[\hat q]}   D\{\calC( \bA_l),\calC( \widehat \bA_j)\}$ over $l\in [q]$. For given $\alpha$, the rate is faster for smaller values of $\{s_1, s_2, p_\dagger, {\cal M}_y\},$ while enlarging the minimum eigen-gap between different blocks (i.e., larger $\rho$) reduces the difficulty of estimating each $\calC(
\bA_l).$

Supported by Theorems~\ref{thm.seg.msc} and \ref{thm.seg.distance}, our subsequent theoretical results are developed by assuming that the group structure of $\bZ_t(\cdot)$ is correctly identified or known, i.e., $\hat{q}=q$ and $\widehat{G}_l=G_l$ for each $l$. 
We now turn to investigate the theoretical properties of the dimension reduction step. 
Inherited from the segmentation step, $\bZ_{t}^{(1)}(\cdot),\ldots,\bZ_t^{(q)}(\cdot)$ rely on the specific form of $\bA = (\bA_1, \dots, \bA_q),$ and thus are not uniquely defined.  Yet intuitively, we only require a certain transformation matrix 
to make our subsequent analysis related to  $\wh \bfeta_1,\ldots,\wh \bfeta_p$ mathematically tractable. 
Based on (\ref{eq.th}), we define $\bPi_{l} = \bA_l \bGamma_{z,l} \bH_l$ and it holds that $\calC(\bPi_{l}) = \calC(\bA_l)$ for each $l\in[q]$. Let $\bZ_{t}^{(l)}(\cdot) =\bPi_{l}^{\T}\bY_t(\cdot)$. Recall \eqref{Y.err} and \eqref{eq.zeta}.
The primary goal of the second dimension reduction step is to identify each $r_l$ and to estimate the associated dynamic space 
$\calC_l =\text{span}\{\bpsi_{1}^{(l)}(\cdot), \dots, \bpsi_{r_l}^{(l)}(\cdot)\}$.
Recall that $\{\hat \theta_{j}^{(l)},\wh\bpsi_{j}^{(l)}(\cdot)\}_{j \geq 1}$ are the eigenvalue/eigenfunction pairs of $\wh \bK^{(l)}(\cdot,\cdot)$ defined in (\ref{K.gp.est}) with  $\wh \bA_l = (\wh \bgamma_i)_{i \in  G_l}$ 
and the dimension $r_l$ is fixed for all $l \in [q]$. 
Our asymptotic results are based on the following regularity condition:


\begin{condition}
    \label{cond_eigenvalue}
    For each $l\in[q]$, all $r_l$ nonzero eigenvalues of $\bK^{(l)}(\cdot,\cdot)$ are different, i.e., $\theta_{1}^{(l)} > \cdots >\theta_{r_l}^{(l)}>0=\theta_{r_l+1}^{(l)}=\cdots.$
\end{condition}



\begin{theorem}
    \label{thm.dr.dim}
    Let Conditions~{\rm\ref{con_subG}--\ref{cond.sA}} and {\rm\ref{cond_eigenvalue}} hold. 
    Assume   $(\rho^{-1}s_1^3 s_2^3 p_{\dagger}^{5-2\alpha})^{2/(1-\alpha)} \calM_y^2 \log p = o(n)$ and $\tilde \delta_n$ in \eqref{eq.r.2}
    satisfies  
  $\rho^{-1}s_1^3 s_2^3 p_{\dagger}^{7-2\alpha}\calM_y^{1-\alpha}(n^{-1}\log p)^{(1-\alpha)/2}\ll \tilde \delta_n \ll \min_{l \in [q]}\{\theta_{r_l}^{(l)}\}^2/\max_{l \in [q]}\theta_{1}^{(l)}$, where $p_\dagger$ and $\rho$ are specified in \eqref{p.dagger} and \eqref{eq.rho}, respectively. 
   As $n\rightarrow\infty$, it holds that
    $
    \mathbb{P}[\bigcap_{l= 1}^q \{\hat r_l = r_l\}] \to 1.$ 
   
\end{theorem}

Theorem~\ref{thm.dr.dim} shows that $r_l$ can be correctly identified  with probability tending to one uniformly over $l \in [q].$ 
Let $\widehat{\calC}_l = \text{span}\{\wh \bpsi_{1}^{(l)}(\cdot),\dots, \wh \bpsi_{\hat r_l}^{(l)}(\cdot)\}$ be the dynamic space spanned by $\hat r_l$ estimated eigenfunctions.
To measure the discrepancy between $\calC_l$ and $\wh \calC_l,$ we introduce the following metric. For two  subspaces
$\calC(\bb_1) = \text{span} \{\bb_{11}(\cdot), \dots,\bb_{1\tilde r_1}(\cdot) \}$ and $\calC(\bb_2)= \text{span} \{\bb_{21}(\cdot), \dots,\bb_{2\tilde r_2}(\cdot) \}$ satisfying $\langle \bb_{ij}, \bb_{ik}\rangle=I(j=k)$ for each $i \in [2],$ the discrepancy measure between  $\calC(\bb_1)$ and $\calC(\bb_2)$ is defined as
$$
\widetilde D\{\calC(\bb_1),\calC(\bb_2)\} = \sqrt{1-\frac{1}{\max(\tilde r_1,\tilde r_2)}\sum_{j=1}^{\tilde r_1} \sum_{k=1}^{\tilde r_2}\langle \bb_{1j}, \bb_{2k}\rangle^2}\, \in [0,1]\,,
$$ which equals 0 if and only if $\calC(\bb_1)\subset\calC(\bb_2)$ 
 or $\calC(\bb_2)\subset\calC(\bb_1)$ and 1 if and only if two spaces are orthogonal. 

\begin{theorem}
    \label{thm.dr.space}
 Let conditions for Theorem {\rm\ref{thm.dr.dim}} hold. Assume
 $( \Delta^{-1}\rho^{-1}s_1^3 s_2^3 p_{\dagger}^{7-2\alpha})^{2/(1-\alpha)} \calM_y^2 \log p = o(n)$ with $\Delta = \min_{l \in [q], j \in [r_l]}\{\theta_{j}^{(l)} - \theta_{j+1}^{(l)}\}$.  As $n \to \infty$, it holds that
    \begin{equation}\nonumber
        \max_{l \in [q]}\widetilde D ( \widehat{\calC}_l,\calC_l)  =O_{\rm p }\big\{\Delta^{-1}\rho^{-1} s_1^3 s_2^3 p_{\dagger}^{7-2\alpha}\mathcal{M}_y^{1-\alpha}(n^{-1}{\log p})^{(1-\alpha)/{2}} \big\} \,.
\end{equation}
\end{theorem}

\vspace{-0.3cm}
\section{Simulation studies} 
\label{sec.sim}
\vspace{-0.2cm}
We conduct a series of simulations to illustrate the finite sample performance of the proposed methods. 
To simplify the data-generating process, we consider
a relaxed form of (\ref{c1}) as 
\begin{equation} \label{eq.relax}
   \widecheck \bY_t(u)  = \widecheck \bA \widecheck \bZ_t(u) 
 = \widecheck \bA \{\widecheck \bZ_{t}^{(1)}(u)^\T, \dots,\widecheck \bZ_{t}^{(q)}(u)^\T \}^\T \,, ~~~~u \in \calU=[0,1]\,,
\end{equation}
with no orthonormality restriction on the transformation matrix $\widecheck \bA = (\widecheck \bA_1,\dots,\widecheck \bA_q)$.
The $p$-dimensional functional time series
$\widecheck \bZ_t(\cdot)$ is formed by  $q$ uncorrelated groups $\{\widecheck \bZ_t^{(l)}(\cdot): l \in [q]\}$, where each $\widecheck \bZ_t^{(l)}(\cdot)$ arises as the sum of dynamics $ \widecheck\bX_{t}^{(l)}(\cdot)$ and white noise 
$  \check\bvarepsilon_{t}^{(l)}(\cdot).$ 
Based on \eqref{eq.normal} in Section \ref{sec:seg.est}, 
(\ref{eq.relax}) can then be easily reformulated as (\ref{c1}) by setting
\begin{equation} \label{eq.to2}
    \bY_t(\cdot) = \bV_{\check y}^{-1/2}\widecheck \bY_t(\cdot)\,, ~~ \bA=\bV_{\check y}^{-1/2}\widecheck\bA\bV_{\check z}^{1/2} ~~\text{and}~~\bZ_t(\cdot) = \bV_{\check z}^{-1/2}\widecheck \bZ_t(\cdot)\,,
\end{equation}
where 
 $\bV_{\check y} = \int_{\calU}\cov\{\widecheck\bY_{t}(u), \widecheck\bY_{t} (u)\}\,{\rm d}u$ and    $\bV_{\check z} = \int_{\calU}\cov\{\widecheck\bZ_{t}(u), \widecheck\bZ_{t} (u)\}\,{\rm d}u$. Then the orthonormality of $\bA$ is satisfied. 

Write  $\check\bvarepsilon_{t}(\cdot)= \{\check\bvarepsilon_{t}^{(1)}(\cdot)^\T, \dots,\check\bvarepsilon_{t}^{(q)}(\cdot)^\T \}^\T \equiv \{\check \varepsilon_{t1}(\cdot), \dots, \check \varepsilon_{tp}(\cdot)\}^{\T}.$
We generate each curve component of $\check\bvar_t(\cdot)$  independently  by  $\check\varepsilon_{tj}(\cdot) = \sum_{l=1}^{10} {2^{-(l-1)}}e_{tjl} \psi_{l}(\cdot)$ for $j \in [p],$ where $e_{tjl}$'s are sampled independently from $\mathcal{N}(0,1)$ and $\{\psi_l(\cdot)\}_{l=1}^{10}$ is a 10-dimensional Fourier basis function. 
The  finite-dimensional dynamics $\widecheck\bX_t(\cdot)= \{\widecheck\bX_{t}^{(1)}(\cdot)^\T, \dots,\widecheck\bX_{t}^{(q)}(\cdot)^\T \}^\T$ with  prescribed group structure is generated based on some $5$-dimensional curve dynamics  $\vartheta_{tg}(\cdot) = \sum_{l = 1}^5\kappa_{tgl} \psi_{l}(\cdot)$ for $g\in[20]$.
The basis coefficients $\bkappa_{tg}=(\kappa_{tg1}, \dots, \kappa_{tg5})^{\T}$ are generated from a stationary VAR model $\bkappa_{tg}=\bU_g\bkappa_{(t-1)g}+ \be_{tg}$ for each $g$.  
To guarantee the stationarity of $\bkappa_{tg},$ we generate $\bU_g = \iota\widecheck \bU_g/\rho(\widecheck \bU_g)$ with $\iota \sim$ $\text{Uniform}[0.5,1]$ and $\rho(\widecheck \bU_g)$ being the spectral radius of $\widecheck \bU_g \in {\mathbb R}^{5 \times 5},$ the entries of which are sampled independently from $\text{Uniform}[-3,3]$. The components of the innovation $\be_{tg}$ are sampled independently from ${\cal N}(0,1).$ 
We will specify  the exact forms of  $\widecheck\bX_t(\cdot)$ under the fixed and large $p$ 
scenarios in Sections~\ref{sec.sim.moderate}
and \ref{sec.sim.large}, respectively.
The white noise sequence $\check \bvar_t(\cdot)$ ensures that $\widecheck \bZ_t(\cdot)$ as well as $\bZ_t(\cdot)$ share the same group structure as $\widecheck \bX_t(\cdot).$ 
Unless otherwise stated, we set $k_0  = m =  5$ and $c_r = c_\varrho =0.75$ in our procedure, 
as our simulation results suggest that our procedure is robust to the choices of  these parameters.

\vspace{-0.2cm}
\subsection{Cases with fixed $p$} 
\label{sec.sim.moderate}
We consider the following three examples of $\widecheck\bX_t(\cdot)=\{\check X_{t1}(\cdot), \dots, \check X_{tp}(\cdot)\}^{\T}$ with different group structures for $p\in\{6, 10, 15\}$ based on independent $\vartheta_{t1}(\cdot), \dots, \vartheta_{t5}(\cdot).$ 

\newcounter{bean}
\setcounter{bean}{0}
\begin{center}
	\begin{list}
		{\textsc{Example} \arabic{bean}.}{\usecounter{bean}}
		\item \label{exp1} $\check X_{t1}(\cdot) = \vartheta_{t1}(\cdot),$ $\check X_{tj}(\cdot) = \vartheta_{(t+j-2)2}(\cdot)$ for $j\in\{2, 3\}$ and $\check X_{tj}(\cdot) = \vartheta_{(t+j-4)3}(\cdot)$ for $j \in\{4, 5, 6\}$.
		\item \label{exp2} $\check X_{tj}(\cdot)$ for $j\in[6]$ are the same as those in Example~\ref{exp1} and $\check X_{tj}(\cdot) = \vartheta_{(t+j-7)4}(\cdot)$ for $j\in\{7, \dots, 10\}$.  
        \item \label{exp3} $\check X_{tj}(\cdot)$ for $j\in[10]$ are the same as those in Example~\ref{exp2} and $\check X_{tj}(\cdot) = \vartheta_{(t+j-11)5}(\cdot)$ for $j \in\{11, \dots, 15\}$. 
	\end{list}
\end{center}
Therefore, $\widecheck \bX_t(\cdot)$ consists of $q=3, 4$ and $5$ uncorrelated groups of curve subseries in Examples~\ref{exp1}, \ref{exp2} and \ref{exp3}, respectively, where the number of component curves per group is $p_l=l$ for $l\in [q].$
The $p$-dimensional observed functional time series $\widecheck \bY_t(\cdot) = \{\check Y_{t1}(\cdot), \dots, \check Y_{tp}(\cdot)\}^{\T}$ for $t\in [n]$  is then generated by (\ref{eq.relax}) with the entries of $\widecheck \bA$ sampled independently  from $\text{Uniform}[-3,3]$. To obtain $h$-step ahead prediction of $\widecheck \bY_t(\cdot),$ we integrate  the segmentation and dimension reduction steps respectively in Sections~\ref{sec:seg} and \ref{sec:dr}
into the  VAR estimation as outlined in Algorithm~\ref{alg1}. 
For each of the three examples introduced above,
we select 
\begin{equation}
    \label{eq.V.est}
    \wh \bV_{\check y}^{(h)}=\frac{1}{n-h}\sum_{t=1}^{n-h}\int_{\calU} \bigg\{\widecheck\bY_{t}(u)-\frac{1}{n-h}\sum_{t=1}^{n-h} \widecheck \bY_t(u)\bigg\}^{\otimes2}\,{\rm d}u\,,
\end{equation}
\begin{equation} 
\label{eq.sigma.sim}
     \wh \bSigma_{\tilde{y},k}^{(h)}(u,v)  = \frac{1}{n-h-k}\sum_{t=1}^{n-h-k}\{ \widetilde \bY_{t}(u) - \bar \bY_*(u) \}
\{ \widetilde \bY_{t+k}(v) - \bar \bY_*(v) \}^\T\,,
\end{equation}
with  $ \bar \bY_*(\cdot) = (n-h-k)^{-1}\sum_{t=1}^{n-h-k} \widetilde \bY_{t}(\cdot)$, 
for the quantities  involved in 
Step~\ref{step 1.sim} of Algorithm~\ref{alg1}. 
We refer to the \textbf{seg}mentation-(\textbf{V}ariational-\textbf{m}ultivariate-FPCA)-and-\textbf{V}AR-based Algorithm~\ref{alg1}
with selections of $\wh \bV_{\check y}^{(h)}$ in (\ref{eq.V.est}) and $\wh \bSigma_{\tilde{y},k}^{(h)}(u,v) $ in (\ref{eq.sigma.sim}) as SegV hereafter.
\begin{center}
	\begin{algorithm}[!t] \caption{\label{alg1}{General prediction procedure for multivariate functional time series}}
		\begin{enumerate}[label=(\roman*)]
			\item\label{step 1.sim}    Treat the first $n-h$ observations as training data, adopt the normalization step  to obtain  $\widetilde \bY_t(\cdot)= \{\wh\bV_{\check y}^{(h)}\}^{-1/2}\widecheck \bY_t(\cdot),$ where  
$\wh \bV_{\check y}^{(h)}$ is the consistent estimator of $ \bV_{\check y}$ in (\ref{eq.to2}),
and implement the procedure in Section~\ref{sec:seg.est} on $\{\widetilde \bY_t(\cdot)\}_{t=1}^{n-h}$  to obtain estimated  
transformation matrix $\widehat \bA = (\widehat \bA_1, \dots, \widehat \bA_{\hat q})$ and transformed curve subseries $\{\wh \bZ_{t}^{(l)}(\cdot): l \in [\hat q]\}$.
\vspace{-0.6cm}
			\item \label{step 2.sim} 
    Apply the procedure in Section~\ref{sec:dr} on each $\{\wh \bZ_{t}^{(l)}(\cdot)\}_{t=1}^{n-h}$ to achieve the $h$-step ahead prediction denoted as $\mathring \bZ_n^{(l)}(\cdot)$ for $l \in [\hat q].$ In particular, for each $l,$ select the best VAR model that best fits each vector time series $\{\widehat \bzeta_{t}^{(l)}\}_{t=1}^{n-h}$
    according to the AIC criterion.
    \vspace{-0.1cm}
			\item\label{step 3.sim}
    Obtain the $h$-step ahead prediction $\wh \bA \mathring \bZ_n(\cdot)$ for the normalized curves $\widetilde \bY_{n}(\cdot)$  with  $\mathring \bZ_{n} (\cdot)=\{\mathring \bZ_{n}^{(1)} (\cdot)^{\T}, \dots, \mathring \bZ_{n}^{(\hat q)}(\cdot)^{\T}\}^{\T}$. Then the $h$-step ahead prediction  for  the original curves $\widecheck \bY_n(\cdot)$ is given by $\wh \bY_n(\cdot)\equiv\{\hat Y_{n1}(\cdot), \dots, \hat Y_{np}(\cdot)\}^{\T}=\{\wh\bV_{\check y}^{(h)}\}^{1/2}\wh \bA \mathring \bZ_n(\cdot).$ 
		\end{enumerate}
	\end{algorithm}  \vspace{-0.5cm}		
\end{center}

The performance of our two-step proposal is examined in terms of linear space estimation, group identification and post-sample prediction. For $\bA=(\bA_1,\dots, \bA_q)$ specified in (\ref{eq.to2}),
with the aid of \eqref{eq.dis.gr},
 define
$
f(l) =  \arg\min_{j\in[\hat q]}   D^2\{\calC( \bA_l),\calC( \widehat \bA_j)\}$
for each $l\in[q]$.
We  then call $\wh \bA = (\wh \bA_1,\dots, \wh \bA_{\hat q})$ an effective segmentation of $\bA$ if (i) $1<\hat q \leq q$, and (ii) 
$\text{rank}(\widehat \bA_{l'})=\sum_{l\in[q]:\,f(l) = l' }\text{rank}(\bA_l) $ for each $l'  \in [\hat{q}]$.  
The intuition is as follows. The effective segmentation implies that
each identified group in $\widehat \bZ_t(\cdot)$ contains at least one, but not all, groups in $ \bZ_t(\cdot).$  Since our main target is to forecast $\widecheck \bY_t(\cdot)$ based on  the cross-serial  dependence in $\{\bZ_t^{(l)}(\cdot):l \in [q]\},$ this segmentation result is effective in the sense that
the linear dynamics in $\bZ_t(\cdot)$ is well kept in $\{\widehat \bZ_t^{(l)}(\cdot): l \in [\hat q]\}$ without any contamination or damage
and a mild dimension reduction is achieved with $\hat q>1.$
For the special case of complete segmentation ($\hat q =  q$), we use the maximum and averaged estimation errors for $(\wh \bA_1,\dots, \wh \bA_{\hat q})$, respectively, defined as
$
{\rm MaxE} = \max_{l \in [q]}  D^2\{\calC( \bA_l),\calC( \wh \bA_{f(l)})\}$ and ${\rm AvgE} = q^{-1}\sum_{l = 1}^q  D^2\{\calC( \bA_l),\calC( \wh \bA_{f(l)})\}
$
 to assess the ability of our method in fully
recovering the spanned spaces $\calC(\bA_1), \dots, \calC(\bA_q).$ 
Note that $\bA$ in (\ref{eq.to2}) can not be easily computed, as the true $\bV_{\check y}$ and $\bV_{\check z}$ are  hard to find even for simulated examples.
For $\widecheck \bA $ specified in (\ref{eq.relax}), let 
 $\widetilde \bA =\bV_{\check y}^{-1/2} \widecheck\bA \equiv (\widetilde\bA_1,\dots, \widetilde\bA_q)$ with $\widetilde \bA_l = \bV_{\check y}^{-1/2} \widecheck\bA_l$.  Since  $\bV_{\check z}$ is a 
 block-diagonal matrix,  then 
$\calC(\widetilde \bA_l) = \calC(\bA_l)$ for $l \in [q]$. 
Hence, we can  replace $\calC(\bA_l)$ by $\calC( \{\widehat\bV_{\check y}^{(h)}\}^{-1/2}\widecheck \bA_l)$ to obtain the approximations of 
 MaxE  and AvgE in our simulations.

To evaluate the post-sample predictive accuracy, we define the mean squared prediction error (MSPE) as
    \begin{equation} 
    \label{df_MSPE_11}
         \text{MSPE} =\frac{1}{pN}\sum_{j=1}^p\sum_{i=1}^{N} \{\hat Y_{nj}(v_i) -\check Y_{nj}(v_i)\}^2
    \end{equation} 
with $v_1, \dots, v_N$ being equally-spaced points in $[0,1],$ and
compute the relative prediction error as the ratio of MSPE  in (\ref{df_MSPE_11}) to that under the `oracle' case. In the oracle case, we apply the procedure in Section~\ref{sec:dr} directly on each true $\{\widecheck \bZ_{t}^{(l)}(\cdot)\}_{t=1}^{n-h}$ to achieve the $h$-step ahead prediction for  $\{\widecheck \bZ_{n}^{(l)}(\cdot) :l \in [ q]\},$ denoted by $\{\breve \bZ_{n}^{(l)}(\cdot) :l \in [ q]\},$ and further obtain the $h$-step ahead prediction $\widecheck \bA \{\breve \bZ_{n}^{(1)} (\cdot)^{\T}, \dots, \breve \bZ_{n}^{( q)}(\cdot)^{\T}\}^{\T}$ for  the original curves $\widecheck \bY_t(\cdot).$
By comparison, we also implement an {\bf uni}variate functional prediction method on each $\check Y_{tj}(\cdot)$ separately by performing {uni}variate dimension reduction \cite[]{Bathia2010}, then predicting vector time series based on the best fitted {\bf V}AR model and finally recovering functional prediction (denoted as UniV).

We generate $n\in \{200, 400, 800, 1600\}$  observations with $N=30$ for each example and replicate each simulation 500 times. Table~\ref{table.low} provides numerical summaries, including the relative frequencies of the effective segmentation with $\hat q=q$ and $\hat q \geq q-1,$ and the estimation errors for $\wh \bA=(\wh \bA_1,\dots, \wh \bA_{\hat q})$ under the complete segmentation case.
As one would expect, the proposed method provides higher proportions of effective segmentation and lower estimation errors as $n$ increases, and performs fairly well for reasonably large $n$ as $p$ increases. 
For $(p,n)=(6,200),$ we observe $62.6 \%$ complete segmentation with ${\rm AvgE}$ as low as $0.079.$ 
Furthermore, the proportions of effective segmentation with $\hat q \geq q-1$ are above $93\%$ for $n \geq 200.$ Similar results can be found for cases of $(p,n) = (10,800+)$ and $(15, 1600),$ whose proportions of effective segmentation with $\hat q \geq q-1$ remain higher than $87.4\%$ and $83.2\%,$ respectively.  
Table~\ref{table.low} also reports the relative one-step ahead prediction errors. It is evident that SegV significantly outperforms UniV in all settings, demonstrating the effectiveness of our proposed segmentation transformation and dimension reduction in predicting future values.
Although the proportions of complete segmentation are not high when $p=15$, the corresponding proportions of $\hat q \geq q-1$ become satisfactorily higher, and SegV performs similarly to the oracle case with its relative prediction errors being  closer to 1 as $n$ increases. 

\begin{table}[!t]
\vspace{-0.2cm}
	\caption{\label{table.low}  The relative frequencies of effective segmentation with respect to $\hat q=q$ and $\hat q \geq q-1,$ and the means (standard deviations) of {\rm MaxE}, {\rm AvgE}, and relative MSPEs (rMSPE) over 500 simulation runs.}
	\begin{center}
		\vspace{-0.5cm}
		\resizebox{4.5in}{!}{
			\begin{tabular}{cccccc}
			\hline
	&		 &	$n$ = 200	&	$n$ = 400	&	$n$ = 800	&	$n$ = 1600		\\	\hline
\multirow{6}{*}{\begin{tabular}[c]{@{}c@{}}Example 1\\ ($p$ = 6)\end{tabular}}	&	$\hat q =q$	&	0.626	&	0.722	&	0.772	&	0.880		\\	
	&	$\hat q \ge q-1$	&	0.930	&	0.988	&	0.998	&	1.000		\\	
	&	MaxE	&	0.128(0.088)	&	0.089(0.066)	&	0.053(0.048)	&	0.035(0.037)		\\	
	&	AvgE	&	0.079(0.052)	&	0.053(0.038)	&	0.030(0.025)	&	0.019(0.019)	\\	
	&	rMSPE - SegV	&	1.081(0.172)	&	1.048(0.105)	&	1.026(0.065)	&	1.014(0.048)		\\	
	&	rMSPE - UniV	&	1.584(0.453)	&	1.598(0.423)	&	1.596(0.379)	&	1.651(0.443)		\\	\hline

\multirow{6}{*}{\begin{tabular}[c]{@{}c@{}}Example 2\\ ($p$ = 10)\end{tabular}}	&	$\hat q =q$	&	0.324	&	0.444	&	0.644	&	0.806		\\	
	&	$\hat q \ge q-1$	&	0.490	&	0.688	&	0.874	&	0.972		\\	
	&	MaxE	&	0.301(0.108)	&	0.193(0.090)	&	0.117(0.064)	&	0.072(0.049)	\\	
	&	AvgE	&	0.183(0.059)	&	0.115(0.047)	&	0.069(0.035)	&	0.041(0.024)	\\	
	&	rMSPE - SegV	&	1.291(0.271)	&	1.174(0.215)	&	1.089(0.143)	&	1.059(0.091)		\\	
	&	rMSPE - UniV	&	1.708(0.404)	&	1.836(0.410)	&	1.841(0.436)	&	1.862(0.392)	\\	\hline

\multirow{6}{*}{\begin{tabular}[c]{@{}c@{}}Example 3\\ ($p$ = 15)\end{tabular}}	&	$\hat q =q$	&	0.032	&	0.178	&	0.410	&	0.622		\\	
	&	$\hat q \ge q-1$	&	0.086	&	0.344	&	0.616	&	0.832		\\	
	&	MaxE	&	0.426(0.091)	&	0.347(0.121)	&	0.241(0.113)	&	0.157(0.091)		\\	
	&	AvgE	&	0.273(0.054)	&	0.195(0.050)	&	0.128(0.042)	&	0.077(0.033)	\\	
	&	rMSPE - SegV	&	1.477(0.313)	&	1.363(0.277)	&	1.166(0.156)	&	1.091(0.098)		\\	
	&	rMSPE - UniV	&	1.805(0.370)	&	1.967(0.394)	&	2.033(0.394)	&	2.001(0.384)		\\	\hline

			\end{tabular}
		}	
	\end{center}
	\vspace{-0.5cm}
\end{table}

\vspace{-0.2cm}
\subsection{Cases with large $p$}
\label{sec.sim.large}

Under a large $p$ scenario, a natural question to ask is whether the segmentation method based on the classical estimation for autocovariance functions of $\widetilde \bY_t(\cdot)$
(denoted as NonT) as \eqref{eq.sigma.sim} in Section~\ref{sec.sim.moderate}  still performs well, and if not, whether a satisfactory improvement is attainable via the functional-thresholding estimation 
 (denoted as FunT) developed in Section~\ref{sec:ft}. To this end, we generate $\widecheck \bY_t(\cdot)$ from  
 (\ref{eq.relax}) with $p\in\{30, 60\}$ and $n\in\{200, 400\}$. Specifically, we let
$\check X_{t(3l-2)}(\cdot)=\vartheta_{tl}(\cdot),$
$\check X_{t(3l-1)}(\cdot)=\vartheta_{(t+1)l}(\cdot),$
$\check X_{t(3l)}(\cdot)=\vartheta_{(t+2)l}(\cdot)$
for $l\in [q].$ This setting ensures $q$ uncorrelated groups of curve subseries in $\widecheck \bX_t(\cdot)$ with $p_l=3$ component curves per group and hence $q=10$ and $20$ correspond to $p=30$ and $60,$ respectively.
Let the $p \times p$ transformation matrix $\widecheck \bA = \bDelta_1 + \delta \bDelta_2.$
Here $\bDelta_1 = \text{diag} \{\bDelta_{11},\dots, \bDelta_{1 (p/6)}\}$ with elements of each $\bDelta_{1i} \in \mathbb{R}^{6 \times 6}$ being sampled independently from $\text{Uniform}[-3,3]$ for $i \in [p/6],$ and $\bDelta_2$ is a matrix with two randomly selected nonzero elements from $\text{Uniform}[-1,1]$ each row. We set $\delta\in\{0.1, 0.5\}$. It is notable that our setting results in a very high-dimensional learning task in the sense that the intrinsic dimension $30\times 5=150$ or $60 \times 5=300$ is large relative to the sample size $n=200$ or $400.$

We assess the performance of  NonT and FunT in discovering the group structure. The optimal thresholding parameters $\hat\omega_k$ in FunT are selected by the five-fold cross-validation 
(see Remark~\ref{rmk:fth}), 
and  $ \bV_{\check y}$ in the normalization step is estimated by $\wh \bV_{\check y}^{(0)}$ given in  (\ref{eq.V.est}), as the threshold version of $\wh \bV_{\check y}^{(0)}$ might not be positive definite.
In practice, when $p$ is large,  FunT may  lead to segmentation  with a small $\hat q,$ indicating that some groups of $\{\wh \bZ_{t}^{(l)}(\cdot): l \in [\hat q]\}$   contain multiple groups in $\{ \bZ_{t}^{(l)}(\cdot): l \in [ q]\}.$ 
To ease the modelling burden of complex VAR process, we may consider performing further segmentation transformation on the estimated groups by repeating FunT $R$ 
times. 
To be precise, the $i$-th round of 
segmentation transformation via FunT is performed within each group discovered in the $(i-1)$-th round with $c_\varrho = 1$  for $i \in [R],$ and hence $(\wh \bA_1, \dots, \wh \bA_{\hat q})$ is updated after each iteration. 
Table \ref{table.high.run} reports the relative frequencies of the effective segmentation for NonT and FunT with  $R\in\{1,5,10\}$. 
Finally, we apply FunT-based SegV (denoted as FTSegV) combined with the $R$-round segmentation transformation
for $R\in\{1,5,10\}$ in Step~\ref{step 1.sim} of Algorithm~\ref{alg1}, and compare their one-step ahead predictive performance with UniV and SegV. 
Table~\ref{table.high.error} summarizes the relative prediction errors for all five comparison methods. 

\begin{table}[!t]
\vspace{-0.2cm}
	\caption{ \label{table.high.run} The relative frequencies of effective segmentation over 500 simulation runs.}
	\begin{center}
		\vspace{-0.5cm}
		\resizebox{4.8in}{!}{
\begin{tabular}{c|cc|cccccc}
\hline
\multirow{3}{*}{$(p,\delta)$} & \multicolumn{2}{c|}{NonT}                                                                      & \multicolumn{6}{c}{FunT}                                         
\\ \cline{4-9} 
& \multicolumn{1}{c}{\multirow{2}{*}{$n = 200$}} & \multicolumn{1}{c|}{\multirow{2}{*}{$n = 400$}} &  \multicolumn{2}{c|}{$R = 1$}  
& \multicolumn{2}{c|}{$R = 5$}               & \multicolumn{2}{c}{$R = 10$} \\
& \multicolumn{1}{c}{}                           & \multicolumn{1}{c|}{}                           &  $n = 200$ & \multicolumn{1}{c|}{$n = 400$}   & $n = 200$ & \multicolumn{1}{c|}{$n = 400$} & $n = 200$     & $n = 400$    \\ \hline
(30, 0.1)                           & 0                                              & 0                                               & 0.706      & \multicolumn{1}{c|}{1.000}     & 0.556     & \multicolumn{1}{c|}{1.000}      & 0.546         & 1.000        \\
(30, 0.5)                           & 0                                              & 0                                               & 0.588     & \multicolumn{1}{c|}{1.000  }     & 0.436     & \multicolumn{1}{c|}{1.000  }      & 0.420       & 1.000           \\
(60, 0.1)                           & 0                                              & 0                                               & 0.298     & \multicolumn{1}{c|}{1.000  }     & 0.148     & \multicolumn{1}{c|}{1.000  }     & 0.144         & 1.000           \\
(60, 0.5)                           & 0                                              & 0                                               & 0.194     & \multicolumn{1}{c|}{0.996}     & 0.078     & \multicolumn{1}{c|}{0.990}      & 0.072         & 0.990        \\ \hline
\end{tabular}
		}	
	\end{center}
	\vspace{-0.5cm}
\end{table}

\begin{table}[!t]
\vspace{-0.2cm}
	\caption{\label{table.high.error} Means (standard deviations) of relative MSPEs over 500 simulation runs.}
	\begin{center}
		\vspace{-0.5cm}
		\resizebox{5.2in}{!}{
			\begin{tabular}{c|c|cc|c|cc}
			\hline
{Method}	&	$(p,\delta)$	&	$n$ =200	&	$n$ =400	&	$(p,\delta)$	&			$n$ =200	&	$n$ =400		\\	\hline
FTSegV ($R=1$)	&	\multirow{5}{*}{(30, 0.1)}	&	1.243(0.162)	&	1.095(0.105)	&	\multirow{5}{*}{(60, 0.1)}	&			1.249(0.122)	&	1.110(0.073)		\\	
FTSegV ($R=5$)	&		&	1.225(0.153)	&	1.091(0.101)	&		&			1.250(0.123)	&	1.104(0.071)		\\	
FTSegV ($R=10$)	&		&	1.222(0.151)	&	1.087(0.099)	&		&			1.249(0.122)	&	1.099(0.071)		\\	
SegV	&		&	1.814(0.376)	&	1.901(0.368)	&		&			1.813(0.271)	&	1.907(0.265)		\\	
UniV	&		&	1.631(0.313)	&	1.735(0.317)	&		&			1.599(0.214)	&	1.682(0.210)		\\	\hline
FTSegV ($R=1$)	&	\multirow{5}{*}{(30, 0.5)}	&	1.268(0.176)	&	1.134(0.134)	&	\multirow{5}{*}{(60, 0.5)}	&			1.285(0.134)	&	1.149(0.101)		\\	
FTSegV ($R=5$)	&		&	1.255(0.171)	&	1.128(0.130)	&		&			1.282(0.136)	&	1.142(0.098)		\\	
FTSegV ($R=10$)	&		&	1.250(0.168)	&	1.128(0.127)	&		&			1.281(0.136)	&	1.141(0.099)		\\	
SegV	&		&	1.815(0.377)	&	1.903(0.369)	&		&			1.813(0.271)	&	1.905(0.264)		\\	
UniV	&		&	1.635(0.315)	&	1.740(0.317)	&		&			1.603(0.215)	&	1.684(0.209)		\\	\hline

			\end{tabular}
		}	
	\end{center}
	\vspace{-0.5cm}
\end{table}

Several conclusions can be drawn from Tables~\ref{table.high.run} and \ref{table.high.error}.
Firstly, the performance of SegV severely deteriorates under the high-dimensional setting, as this procedure fails to detect any effective segmentation, resulting in elevated prediction errors. 
By comparison, FTSegV 
exhibits superior predictive ability over SegV and UniV. In particular, for large $n,$ e.g., $n = 400$,  
FTSegV does a reasonably good job in recovering the group structure of $\bZ_t(\cdot)$ and performs comparably well to the oracle method with the relative prediction errors lower than $1.149$ in all scenarios.
Secondly, comparing the results for $n = 200$ among different $R,$ we observe an interesting phenomenon that even though the relative frequencies of effective segmentation for FunT drop as $R$ increases, implying that some groups in $\{\wh \bZ_{t}^{(l)}(\cdot): l \in [\hat q]\}$  are split incorrectly before forecasting,  the prediction errors stay low and slightly decrease as shown in Table~\ref{table.high.error}.
This is not surprising, since further segmentation based on FunT
yields fewer parameters to be estimated in VAR models and  
thus  benefits the forecasting accuracy even if a few small but significant cross-covariances of $\bZ_t(\cdot)$  are ignored. Such finding highlights the success of FTSegV and its $R$-round segmentation in the sense that although FTSegV may not be able to  accurately recover the group structure in $\bZ_t(\cdot)$ for a small $n$, it achieves
an appropriate dimension reduction to provide significant improvement in high-dimensional
functional prediction.

\vspace{-0.2cm}
\subsection{General data-generating cases}
\label{sec.sim.nogroup}
To further illustrate the advantage of our proposed segmentation transformation 
in predicting high-dimensional functional time series, 
we simulate data from a more generalized functional time series framework instead of strictly adhering to \eqref{c1}.  Specifically,  we consider the
vector functional autoregressive (VFAR) model of order 1, 
\begin{equation} \label{VFAR1}
    \bY_t(u) = \int_\calU \bQ(u,v) \bY_{t-1}(v) \, {\rm d} v + \beps_t(u)\,, ~~~ u \in \calU\,,  ~ t \in [200]\,,
\end{equation}
where $\beps_t(\cdot) = \{\epsilon_{t1}(\cdot), \dots, \epsilon_{tp}(\cdot) \}^{\T}$ are independently sampled from a $p$-dimensional vector of mean zero Gaussian processes, independent of $\bY_{t-1}(\cdot)$, and $\bQ = (Q_{ij})_{i,j \in [p]}$ is the functional transition matrix with each $Q_{ij} \in \mathbb{S}$.  
See Section~\ref{supp.sec_nogroup} of the supplementary material for the detailed data-generating process.


We compare the predictive performance of three competing methods. The first VFAR method is developed by knowing the true data-generating process through VFAR model. 
We relegate the detailed prediction procedure to Section~\ref{supp.sec_nogroup} of the supplementary material.
We next consider two segmentation-based prediction methods:
\begin{itemize}
    \item (Seg+Y method)  
    For the original curve series $\{ \bY_{t}(\cdot)\}_{t \in [200]}$,
    we compute the sample estimates $ \{\widehat{\Sigma}_{ y,k,ij}(u,v)\}_{i,j\in[p]}$ for $k \in \{0\}\cup [5]$ as in Section \ref{sec:ft}.
Let $\mathring T_{y,ij}=\max_{ |k|\leq 5 }\|\wh\Sigma_{ y,k,ij}\|_\calS$ and 
sort $\mathring T_{y,ij}$'s for $1 \leq i < j \leq p$ in descending order. 
We recognize $ Y_{ti}(\cdot)$ and $ Y_{tj}(\cdot)$ as belonging to the same group if $\mathring T_{y,ij}$ is ranked among 10\% of all $p(p-1)/2$ sorted values. 
We then segment the $p$ component series $ Y_{tj}(\cdot)$'s into several non-overlapping groups and apply VFAR to each identified group to obtain its one-step ahead prediction.
\item (Seg+Z method) Consider the transformed curve series 
$\widehat \bZ_{t}(u) = \widehat \bA^{\T} \{\wh\bV_{ y}^{(0)}\}^{-1/2} \bY_t(u)$, where $\widehat \bA$  is obtained by implementing the procedure in Section \ref{sec:seg.est} on the normalized process $\{[\wh\bV_{ y}^{(0)}]^{-1/2} \bY_t(\cdot)\}_{t \in [200]}$.  We perform  the same segmentation procedure as in Seg+Y to $\{\widehat \bZ_{t}(\cdot)\}_{t \in [200]}$,
 apply VFAR to each of the identified groups of  $\{\widehat \bZ_{t}(\cdot)\}_{t \in [200]}$  to obtain the one-step ahead prediction $\breve \bZ_{201}(\cdot),$ 
and finally obtain $\{\wh\bV_{ y}^{(0)}\}^{1/2}\widehat \bA \breve \bZ_{201}(\cdot)$ as the one-step ahead prediction for the original curve series. 
\end{itemize}

\begin{table}[!t]
\vspace{-0.2cm}
	\caption{\label{table.nogroup} The  mean of MSPEs over 500 simulation runs.}
	\begin{center}
		\vspace{-0.5cm}
		\resizebox{3.8in}{!}{
		\begin{tabular}{ccccccc}
\hline
Method & $p=10$ & $p=15$ & $p=20$ & $p=25$ & $p=30$ & $p=35$ \\ \hline
VFAR & 6.709  & 7.974  & 10.506 & 13.439 & 20.149 & 40.552 \\
Seg+Y     & 6.314  & 6.691  & 8.931  & 11.313 & 17.001 & 32.846 \\
Seg+Z   & 6.324  & 6.682  & 8.267  & 8.998  & 12.605 & 17.275 \\ \hline
\end{tabular}
		}	
	\end{center}
	\vspace{-0.5cm}
\end{table}

Table~\ref{table.nogroup} reports 
one-step ahead MSPEs for three methods with different values of $p.$ 
As anticipated, the performance of VFAR deteriorates severely as $p$ increases, demonstrating that the joint model suffers from the high-dimensionality, even when the true model is known. Meanwhile,
both segmentation-based prediction methods exhibit improved predictive performance, with Seg+Z notably outperforming  Seg+Y, particularly in scenarios with large $p.$ It is crucial to emphasize that the improvement of  Seg+Z over Seg+Y is attributed to the decorrelation transformation.
Table~\ref{table.nogroup.A} in the supplementary material provides further insights into the impact of transformation, where $\mathring q_y$ and $\mathring q_z$ denote the numbers of the identified groups using Seg+Y and Seg+Z, respectively. Interestingly,  Seg+Z yields more groups than Seg+Y while retaining the same amount of strongly connected pairs. This observation indicates that the decorrelation transformation effectively pushes the cross-autocorrelations that were previously spread over $p$ components into a block-diagonally dominate structure, where the cross-autocorrelations along the block diagonal are significantly stronger than those off the diagonal.
Such enhancement of within-group autocorrelations, along with the reduction of cross-autocorrelations between the groups, leads to reasonably good
segmentation by only retaining the strong within-group cross-autocorrelations while ignoring the weak between-group cross-autocorrelations, and thus yields more accurate future predictions.

\vspace{-0.3cm}
\section{Real data analysis} 
\label{sec.real}
\vspace{-0.2cm}
In this section, we apply our proposed SegV and FTSegV to two real data examples arising from different fields. 
Our main goal is to evaluate the post-sample predictive accuracy of both methods. By comparison, we also implement componentwise univariate prediction method (UniV) and the multivariate prediction method of \cite{Gao2019} (denoted as GSY) to jointly predict $p$ component series by fitting a factor model to estimated scores obtained via eigenanalysis of the long-run covariance function \cite[]{hormann2015}. It is worth mentioning that the joint prediction model VmV (see Example~\ref{exp:motivation.model}) completely fail due to high dimensionality, so we do not report their results here. 
To evaluate the effectiveness of the segmentation transformation and its impact on prediction,  we forge two other segmentation cases, namely \textbf{under}-segmentation and \textbf{uni}-segmentation,  for both SegV and FTSegV (denoted as Under.SegV, Uni.SegV, Under.FTSegV and Uni.FTSegV, respectively). Denote by $\{\widehat G_l:l \in [\hat q]\}$ the segmented groups of $\{\wh \bZ_{t}^{(l)}(\cdot): l \in [\hat q]\}$ discovered in Step~\ref{step 1.sim} of Algorithm \ref{alg1}  (seen also as correct-segmentation). The under-segmentation updates $\{\widehat G_l:l \in [\hat q]\}$  by merging two groups $\widehat{G}_{l_1}$ and $\widehat{G}_{l_1'}$ together  before subsequent analysis, where $\arg\max_{(i,j):\,i\in\widehat{G}_l,\,j\in\widehat{G}_{l'},\, 1 \leq l\neq l' \leq \hat q}\widehat{T}_{ij} \in \widehat{G}_{l_1}\times \widehat{G}_{l_1'}$ with 
$\widehat{T}_{ij}$ defined in (\ref{eq_T}).
The uni-segmentation, on the other hand, 
regards each curve component of  $\{\wh \bZ_{t}^{(l)}(\cdot): l \in [\hat q]\}$ as an individual group and then applies UniV componentwisely.
For a fair comparison, the orders of VAR models adopted in all SegV/FTSegV-related methods and UniV are determined by the AIC criterion, 
while GSY is implemented using the R package \verb"ftsa".

To examine the predictive performance, we apply an expanding window approach to the observed data $\check Y_{tj}(v_i)$ for $t\in[n], j\in[p], i\in[N]$. 
We first split the dataset into a training set and a test set respectively consisting of the first $n_1$ and the remaining $n_2$ observations. For any positive integer $h,$ we implement each comparison method on the training set $\{\check Y_{tj}(v_i): t\in[n_1], j\in[p], i\in[N]\}$ and obtain its $h$-step ahead prediction, denoted as  $\hat Y_{(n_1+h)j}^{(h)}(v_i)$, 
based on the fitted model. We then 
increase the training size by one, i.e. $\{\check Y_{tj}(v_i):t\in[n_1+1], j\in[p], i\in[N]\},$ refit the model and   compute the next $h$-step ahead prediction $\hat Y_{(n_1+1+h)j}^{(h)}(v_i)$ for $ j\in[p], i\in[N].$
Repeat the above procedure
until the last $h$-step ahead prediction
 $\hat Y_{nj}^{(h)}(v_i)$ is produced. Finally, 
we compute the $h$-step ahead 
MAPE 
and MSPE 
as
\begin{equation}
\label{MSP}
\begin{split}
    \text{MAPE}(h) =&\,\,\frac{1}{(n_2+1-h)pN} \sum_{t=n_1+h}^{n}\sum_{j=1}^p\sum_{i=1}^{N}|\hat Y_{tj}^{(h)}(v_i)-\check Y_{tj}(v_i)|\,,\\
    \text{MSPE}(h) =&\,\,\frac{1}{(n_2+1-h)pN} \sum_{t=n_1+h}^{n}\sum_{j=1}^p\sum_{i=1}^{N}\{\hat Y_{tj}^{(h)}(v_i)-\check Y_{tj}(v_i)\}^2\,.
\end{split}
\end{equation}

\vspace{-0.2cm}
\subsection{Age-specific mortality data}
\label{sec.real.mortality}
The first dataset, 
analyzed in \cite{tang2022}, 
contains age-specific and gender-specific mortality rates for developed countries during 1965 to 2013 ($n=49$).
See Table~\ref{tab.country} in the supplementary material for the list of $p = 29$ countries after removing certain countries with missing data.
Following the proposal of \cite{tang2022}, we model the log transformation of the mortality rate of people aged $v_i=i-1$ 
living in the $j$-th country during year $1964+t$ as a random curve $\check Y_{tj}(v_i)$ ($t\in[49],$ $j\in [29],$ $i \in [101]$) and perform smoothing for observed mortality curves via smoothing splines. 
We divide the smoothed dataset into the training set of size $n_1=34$ and the test set of size $n_2=15.$ Since the smoothed curve series exhibit  weak autocorrelations when lags are beyond 3 and the training size is relatively small, we use $k_0=m=3$ in our procedure for this example.

Table~\ref{table.mortality} reports the MAPEs and MSPEs for females and males. Several obvious patterns are observable.
Firstly, our proposed methods,  SegV and FTSegV, provide the best predictive performance uniformly for both
females and males, and all $h$. This demonstrates the effectiveness of reducing the number of parameters via the segmentation transformation in predicting high-dimensional functional time series. 
Secondly, although the cases of under- and uni-segmentation are inferior to the correct-segmentation case, they significantly outperform UniV and GSY.  
Note that the improvement of Uni.SegV over UniV reveals the capability of the transformation matrix $\widehat \bA$  to effectively decorrelate the original curves, thereby leading to more accurate predictions.
One may also notice that, Uni.SegV does not perform as well as SegV and Under.SegV. In most cases, the transformed curve series exhibits $\hat{q}=26$ groups, 
with $25$ groups of size 1 and one large group of size 4; see Figures~\ref{plot21}--\ref{plot22} in the supplementary material. The limitation of Uni.SegV thus becomes apparent as it fails to account for the cross-serial dependence within the large group, resulting in less accurate predictions.   
This finding again confirms the effectiveness of our procedure, in particular, the within-group cross-autocorrelations  is also valuable in forecasting future values.

\begin{table}[!t]
\vspace{-0.2cm}
	\caption{\label{table.mortality}
	MAPEs and MSPEs for eight competing methods on the female and male mortality curves for $h \in\{1,2,3\}$. All numbers are multiplied by $10.$ }
	\begin{center}
		\vspace{-0.5cm}
		\resizebox{6.6in}{!}{

\begin{tabular}{c|c|ccc|ccc|c|ccc|ccc}
\hline
\multirow{2}{*}{Method} & \multirow{2}{*}{}       & \multicolumn{3}{c|}{MAPE}                        & \multicolumn{3}{c|}{MSPE}                        & \multirow{2}{*}{}     & \multicolumn{3}{c|}{MAPE}                        & \multicolumn{3}{c}{MSPE}                         \\ \cline{3-8} \cline{10-15} 
                        &                         & $h=1$          & $h=2$          & $h=3$          & $h=1$          & $h=2$          & $h=3$          &                       & $h=1$          & $h=2$          & $h=3$          & $h=1$          & $h=2$          & $h=3$          \\ \hline
SegV                    & \multirow{8}{*}{Female} & \textbf{1.157} & 1.461          & 1.806          & \textbf{0.291} & \textbf{0.401} & \textbf{0.566} & \multirow{8}{*}{Male} & 1.104          & 1.391          & \textbf{1.727} & \textbf{0.251} & 0.354          & \textbf{0.499}          \\
Under.SegV              &                         & 1.201          & 1.510          & 1.874          & 0.304          & 0.417          & 0.593          &                       & 1.123          & 1.425          & 1.751          & 0.251          & 0.358          & 0.500          \\
Uni.SegV               &                         & 1.526          & 1.821          & 2.154          & 0.441          & 0.579          & 0.767          &                       & 1.302          & 1.573          & 1.892          & 0.324          & 0.443          & 0.598          \\
FTSegV                  &                         & 1.175          & \textbf{1.458} & \textbf{1.801} & 0.301          & 0.405          & 0.569          &                       & \textbf{1.101} & \textbf{1.391} & 1.732          & 0.251          & \textbf{0.353} & 0.502          \\
Under.FTSegV            &                         & 1.206          & 1.510          & 1.876          & 0.309          & 0.421          & 0.598          &                       & 1.118          & 1.418          & 1.743          & 0.251          & 0.356          & 0.499 \\
Uni.FTSegV             &                         & 1.560          & 1.838          & 2.173          & 0.457          & 0.585          & 0.776          &                       & 1.300          & 1.573          & 1.897          & 0.324          & 0.444          & 0.602          \\
UniV                    &                         & 1.761          & 2.032          & 2.325          & 0.603          & 0.749          & 0.925          &                       & 1.561          & 1.825          & 2.127          & 0.467          & 0.596          & 0.759          \\
GSY                     &                         & 2.476          & 2.515          & 2.577          & 1.434          & 1.447          & 1.451          &                       & 2.144          & 2.110          & 2.201          & 1.112          & 1.023          & 1.043          \\ \hline
\end{tabular}

}	
	\end{center}
	\vspace{-0.5cm}
\end{table}

\vspace{-0.2cm}
\subsection{Energy consumption data}
\vspace{-0.1cm}
Our second dataset contains energy consumption readings (in kWh) taken at half hourly intervals for thousands of London households, and is available at \url{https://data.london.gov.uk/dataset/smartmeter-energy-use-data-in-london-households}. In our study, we select households with flat energy prices during the period between December 2012 and May 2013 ($n=182$) after removing samples with too many missing records, and hence construct $4000$ samples of daily energy consumption curves observed at $N=48$ equally spaced time points following the proposal of \cite{cho2013}. To alleviate the impact of randomness from individual curves, we randomly split the data into $p$ groups of equal size, then take the sample average of curves within each group and finally smooth the averaged curves based on a 15-dimensional Fourier basis. 
We target to evaluate the $h$-day ahead predictive accuracy for the $p$-dimensional  intraday energy consumption averaged curves in May 2013 based on the training data from December 2012 to the previous day. The eight comparison methods are built in the same manner as Section~\ref{sec.real.mortality} with $k_0= m = 5.$


Table~\ref{tabel.edf} presents the mean prediction errors for $h\in\{1,2,3\}$ and $p\in\{40, 80\}.$ A few trends are apparent. 
Firstly, the prediction errors for $p=80$ are higher than those for $p=40$ as higher dimensionality poses more challenges in prediction.
Secondly, likewise in previous examples, SegV and FTSegV attain the lowest prediction errors in comparison to five competing methods under all scenarios. All segmentation-based methods consistently outperform UniV and GSY by a large margin. Thirdly, despite being developed for high-dimensional functional time series prediction, GSY provides the worst result in this example.


\begin{table}[t]
\vspace{-0.2cm}
	\caption{\label{tabel.edf}
	MAPEs and MSPEs for eight competing methods on the energy consumption curves for $h\in\{1, 2, 3\}$ and $p\in \{40, 80\}.$ All numbers are multiplied by $10^2.$} 
	\begin{center}
		\vspace{-0.5cm}
		\resizebox{6.6in}{!}{
\begin{tabular}{c|c|ccc|ccc|c|ccc|ccc}
\hline
\multirow{2}{*}{Method} & \multirow{2}{*}{}       & \multicolumn{3}{c|}{MAPE}                        & \multicolumn{3}{c|}{MSPE}                        & \multirow{2}{*}{}       & \multicolumn{3}{c|}{MAPE}                        & \multicolumn{3}{c}{MSPE}                         \\ \cline{3-8} \cline{10-15} 
                        &                         & $h=1$          & $h=2$          & $h=3$          & $h=1$          & $h=2$          & $h=3$          &                         & $h=1$          & $h=2$          & $h=3$          & $h=1$          & $h=2$          & $h=3$          \\ \hline
SegV                    & \multirow{8}{*}{$p=40$} & 1.639          & 1.748          & 1.793          & 0.047          & 0.053          & 0.054          & \multirow{8}{*}{$p=80$} & \textbf{1.996} & 2.058          & 2.071          & \textbf{0.070} & 0.075          & 0.075          \\
Under.SegV              &                         & 1.669          & 1.766          & 1.794          & 0.048          & 0.054          & 0.054          &                         & 2.025          & 2.092          & 2.104          & 0.072          & 0.077          & 0.077          \\
Uni.SegV               &                         & 1.709          & 1.873          & 1.964          & 0.049          & 0.058          & 0.062          &                         & 2.022          & 2.132          & 2.187          & 0.070          & 0.078          & 0.081          \\
FTSegV                  &                         & \textbf{1.637} & \textbf{1.747} & \textbf{1.791} & \textbf{0.047} & \textbf{0.053} & \textbf{0.054} &                         & 2.012          & \textbf{2.055} & \textbf{2.070} & 0.071          & \textbf{0.074} & \textbf{0.074} \\
Under.FTSegV            &                         & 1.669          & 1.766          & 1.793          & 0.048          & 0.054          & 0.054          &                         & 2.040          & 2.087          & 2.104          & 0.073          & 0.076          & 0.077          \\
Uni.FTSegV             &                         & 1.708          & 1.872          & 1.963          & 0.049          & 0.058          & 0.062          &                         & 2.045          & 2.138          & 2.190          & 0.072          & 0.078          & 0.081          \\
UniV                    &                         & 1.867          & 2.009          & 2.109          & 0.058          & 0.067          & 0.072          &                         & 2.221          & 2.362          & 2.463          & 0.083          & 0.093          & 0.100          \\
GSY                     &                         & 2.142          & 2.264          & 2.320          & 0.099          & 0.110          & 0.119          &                         & 2.833          & 2.826          & 2.781          & 0.159          & 0.159          & 0.159          \\ \hline
\end{tabular}
		}	
	\end{center}
	\vspace{-0.5cm}
\end{table}

\appendix

\spacingset{1}\selectfont
\bibliographystyle{dcu}
\bibliography{paperbib}

\newpage
\spacingset{1.7}\selectfont
\setlength{\abovedisplayskip}{0.2\baselineskip}
\setlength{\belowdisplayskip}{0.2\baselineskip}
\setlength{\abovedisplayshortskip}{0.2\baselineskip}
\setlength{\belowdisplayshortskip}{0.2\baselineskip}
\begin{center}
	{\noindent \bf \large Supplementary material  to ``On the modelling and prediction of high-dimensional functional time series"}\\
\end{center}
\begin{center}
	{\noindent Jinyuan Chang, Qin Fang, Xinghao Qiao and Qiwei Yao}
\end{center}
\bigskip

\setcounter{page}{1}
\setcounter{section}{0}
\renewcommand\thesection{\Alph{section}}
\setcounter{lemma}{0}
\renewcommand{\thelemma}{\Alph{section}\arabic{lemma}}
\setcounter{equation}{0}
\renewcommand{\theequation}{S.\arabic{equation}}

\setcounter{table}{0}
\setcounter{figure}{0}
\renewcommand{\thefigure}{S\arabic{figure}}
\renewcommand{\thetable}{S\arabic{table}}

This supplementary material contains all technical proofs supporting Section~\ref{sec.th} in Sections~\ref{supp.lemma}--\ref{supp.lemma.prove}  and additional empirical results in Section~\ref{supp.sec_emp}.
We begin by introducing some notation. For $x, y \in \mathbb{R},$ we use $x \vee y = \max(x,y).$  For a vector $\bb \in \mathbb{R}^p,$ we denote its  $\ell_2$  norm  by $\|\bb\|_2 = (\sum_{j = 1}^p |b_j|^2)^{1/2}.$ 
For any $\bbf = (f_1,\dots,f_p )^{\T}$ and $ \bg = (g_1, \dots, g_p)^{\T}\in \mathbb{H},$ we define the inner product as
$
\inner{\bbf}{\bg}=\int_{\calU}\bbf(u)^{\T}\bg(u)\,{\rm d}u= \sum_{j=1}^p \int_{\calU}f_j(u)g_j(u)\,{\rm d}u
$
with the induced norm $\|\cdot\| =\inner{\cdot}{\cdot}^{1/2},$ and denote by $\bbf \otimes \bg^\T = (f_i\otimes g_j)_{{i,j}\in [p]}.$  
We further denote by $\calL = \calL(\mathbb{H}, \mathbb{H})$ the space of continuous linear operators from $\mathbb{H}$ to $\mathbb{H}$.
For $\bB=(B_{ij})_{p \times p}$ with each $B_{ij} \in \mathbb{S}$, 
we write   $\|\bB\|_{\calS,\tF} = (\sum_{i= 1}^p\sum_{j =1}^p \|B_{ij}\|_\calS^2)^{1/2}$, 
 $\|\bB\|_{\calS,1} = \max_{j\in [p]} \sum_{i=1}^p \|B_{ij}\|_{\calS}$, $ \|\bB\|_{\calS,\infty} = \max_{i\in [p]} \sum_{j = 1}^p \|B_{ij}\|_{\calS}$
 and $\|\bB\|_{\calL}={\text{sup}}_{\|\bbf\| \leq 1, \bbf \in \cH} \|\bB(\bbf)\|$. We define the image space of $\bB$ as $\text{Im}(\bB) = \{\bg \in \mathbb{H}:  \bg = \bB(\bbf),  \bbf \in\mathbb{H}  \}$.
 For two positive sequences $\{a_n\}$ and $\{b_n\}$, we write $a_n\lesssim b_n$ or $b_n\gtrsim a_n$ if there exist a positive constant $c$ such that $a_n/b_n \leq c$ and 
 write $a_n \asymp b_n$ if and only if $a_n \lesssim b_n$ and $b_n\lesssim a_n$ hold simultaneously.   We further write $a_n\ll b_n$ or $b_n\gg a_n$ if  $\lim \sup_{n \to \infty} a_n/b_n = 0$. 
Throughout, we use $c, c_0$ to denote generic positive finite constants that may be different in different uses.

\section{Auxiliary lemmas}
\label{supp.lemma}
To prove Theorems~\ref{thm.seg.msc}--\ref{thm.dr.space}, we need the following inequalities, equality and  auxiliary lemmas, the proofs of which are deferred to Section~\ref{supp.lemma.prove}.

\begin{inequality}
\label{lemma_operator_inequality}
Let $\bB_1=(B_{1,ij})_{p \times p}$ and $\bB_2=(B_{2,ij})_{p \times p}$ with $B_{1,ij}, B_{2,ij}\in \mathbb{S}$ for any $i,j\in[p]$. It holds that {\rm(i)}
$
    \|\int \int\bB_1(u,v)\bB_2(u,v)^{\T}\,{\rm d}u{\rm d}v\|_2 \leq \|\bB_1\|_{\calS,\infty}^{1/2}\|\bB_1\|_{\calS,1}^{1/2}\|\bB_2\|_{\calS,\infty}^{1/2}\|\bB_2\|_{\calS,1}^{1/2}$,  
{\rm (ii)
} $
    \|\int\bB_1(\cdot,w)\bB_2(\cdot,w)^{\T}\,{\rm d}w\|_{\calS,{\rm F}} \leq \|\bB_1\|_{\calS,{ \rm F}}\|\bB_2\|_{\calS,{\rm F}}$, and {\rm (iii)} $\|\bB_1+ \bB_2\|_{\calS,{\rm F}} \leq \|\bB_1\|_{\calS,{\rm F}}+ \| \bB_2\|_{\calS,{\rm F}}$.
\end{inequality}

\begin{inequality} \label{lemma_hs_inequality}
Let $\bB=(B_{ij})_{p \times p}$ with each $B_{ij}\in \mathbb{S}$, $\bb_1\in \mathbb{R}^p$ and $\bb_2\in \mathbb{R}^p.$ Then
$
\|\bb_1 ^\T \bB \bb_2\|_\calS \leq  \|\bb_1\|_2\|\bb_2\|_2 \|\bB\|_{\calS,\infty}^{1/2}\|\bB\|_{\calS,1}^{1/2}
$.
\end{inequality}

\begin{inequality} \label{ineq_F}
Let $\bB=(B_{ij})_{p \times p}$ with each $B_{ij}\in \mathbb{S}$ and  $\bbf \in \mathbb{H}$. Then $\|\int \bB(\cdot,v) \bbf(v)\,{\rm d}v \| \leq \|\bB\|_{\calS, {\rm F}} \|\bbf\|$ and $\|\bB\|_\calL \leq \|\bB\|_{\calS, {\rm F}}$.
\end{inequality}

\begin{equality} \label{lemma_f_inequality}
For any $\bbf $ and $ \bg \in \mathbb{H},$ it holds that
$
        \|\bbf \otimes \bg^\T\|_{\calS,{\rm F}} = \|\bbf\|\| \bg\|.
$
\end{equality}

\begin{lemma} \label{lemma_dev_y}
Let $\{\bY_t(\cdot)\}$ satisfy Conditions {\rm\ref{con_subG}} and {\rm\ref{con_stability}}. There exists some universal constant $\tilde c>0$ such that 
$
\mathbb{P}\{\| {\widehat\Sigma{}^{\ttS}_{y,k,ij}  - \Sigma_{y,k,ij}}\|_\calS > \mathcal{M}_y \eta\} \leq 8\exp\{-\tilde c n\min(\eta^2,\eta)\}
$
for any $\eta>0$, $|k|\leq k_0 \vee m $ and $i,j\in[p]$. 
\end{lemma}

\begin{lemma}
\label{lemma.Sigma.k}
Suppose Condition~{\rm\ref{cond.sA}} holds. Then $\max_{|k|\leq k_0 \vee m }\sum_{i = 1}^p \| \Sigma_{y,k,ij}\|_\calS^\alpha =O(\Xi)=\max_{|k|\leq k_0 \vee m }\sum_{j = 1}^p \| \Sigma_{y,k,ij}\|_\calS^\alpha$ and 
 $\max_{|k|\leq k_0 \vee m }\|\bSigma_{y,k}\|_{\calS,
   1}=O(\Xi p_{\dagger}^{1-\alpha})=\max_{|k|\leq k_0 \vee m }\|\bSigma_{y,k}\|_{\calS,
   \infty}$, where $\Xi =  s_1s_2(2p_{\dagger} +1)$ with $p_{\dagger} = \max_{l \in [q]} p_l$.
\end{lemma}

\begin{lemma} \label{lemma_th_y}
Let Conditions~{\rm\ref{con_subG}--\ref{cond.sA}}  hold. For each $|k|\leq k_0 \vee m $, select  $\omega_k = c_k \mathcal{M}_y (n^{-1}{\log p})^{1/2}$ for some sufficiently large constant $c_k>0$.  If $\log p=o(n)$, then  
$
\max_{|k|\leq k_0 \vee m }\| \calT_{\omega_k}(\widehat \bSigma{}^{\ttS}_{y,k}) - \bSigma_{y,k}\|_{\calS,1} = O_{\rm p}\{\Xi \mathcal{M}_y^{1-\alpha}(n^{-1}\log p)^{(1-\alpha)/{2}}\}=\max_{|k|\leq k_0 \vee m }\| \calT_{\omega_k}(\widehat \bSigma{}^{\ttS}_{y,k}) - \bSigma_{y,k}\|_{\calS,\infty}
$, where $\Xi =  s_1s_2(2p_{\dagger} +1)$ with $p_{\dagger} = \max_{l \in [q]} p_l$. Moreover, if $ p_{\dagger}^{-2}\mathcal{M}_y ^2\log p=o(n)$ is also satisfied, then 
$
	\|\int \int \{ \calT_{\omega_k}(\widehat \bSigma{}^{\ttS}_{y,k}) (u,v)^{\otimes2}-\bSigma_{y,k}(u,v)^{\otimes2} \}\,{\rm d}u{\rm d}v\|_2 = O_{\rm p}\{\Xi^2p_{\dagger}^{1-\alpha} \mathcal{M}_y^{1-\alpha}(n^{-1}\log p)^{(1-\alpha)/2} \}$ for each $|k|\leq k_0 \vee m$.
\end{lemma}

\begin{lemma}[Theorem 8.1.10 of \textcolor{blue}{ Golub and Van Loan (1996)}]\label{la:space}
	 Suppose $\bB$ and $\bB+\bE$ are $m\times m$ symmetric matrices and $\bQ=(\bQ_1,\bQ_2)$, with $\bQ_1 \in \mathbb{R}^{m\times l}$ and $\bQ_2 \in \mathbb{R}^{m\times (m-l)}$,
  is an orthogonal matrix such that $\mathcal{C}(\bQ_1)$ is an invariant subspace for $\bB$, that is, $\bB\cdot\mathcal{C}(\bQ_1)\subset\mathcal{C}(\bQ_1)$. Partition the matrices $\bQ^\T\bB\bQ$ and $\bQ^\T\bE\bQ$ as follows:
	\[
	\bQ^\T\bB\bQ=\left(
	\begin{array}{cc}
	\bD_1 & \bzero \\
	\bzero & \bD_2 \\
	\end{array}
	\right)~~\textrm{and}~~\bQ^\T\bE\bQ=\left(
	\begin{array}{cc}
	\bE_{11} & \bE_{21}^\T \\
	\bE_{21} & \bE_{22} \\
	\end{array}
	\right).
	\]
	If ${\rm sep}(\bD_1,\bD_2)=\min_{\mu_1\in\Lambda(\bD_1),\mu_2\in\Lambda(\bD_2)}|\mu_1-\mu_2|>0$, where $\Lambda(\bM)$ denotes the set of eigenvalues of the matrix $\bM$, and $\|\bE\|_2\leq{\rm sep}(\bD_1,\bD_2)/5$, then there exists a matrix $\bP\in\mathbb{R}^{(m-l)\times l}$ with $\|\bP\|_2\leq 4\|\bE_{21}\|_2/{\rm sep}(\bD_1,\bD_2)$ such that the columns of ${\bQ}_1^\star=(\bQ_1+\bQ_2\bP)(\bI+\bP^\T\bP)^{-1/2}$ define an orthonormal basis for a subspace that is invariant for $\bB+\bE$.
\end{lemma}

From Lemma \ref{la:space}, we have
\begin{equation} \nonumber 
\begin{split}
\|\bQ^{\star}_1-{\bf Q}_1\|_2=&~\|\{{\bf Q}_1+{\bf Q}_2{\bf P}-{\bf Q}_1({\bf I}+{\bf P}^\T{\bf P})^{1/2}\}({\bf I}+{\bf P}^\T{\bf P})^{-1/2}\|_2\\
\leq&~\|{\bf Q}_1\{{\bf I}-({\bf I}+{\bf P}^\T{\bf P})^{1/2}\}\|_2+\|{\bf Q}_2{\bf P}\|_2\\
\leq&~2\|{\bf P}\|_2\leq\frac{8}{{\rm sep}({\bf D}_1,{\bf D}_2)}\|{\bf
	E}_{21}\|_2\leq\frac{8}{{\rm sep}({\bf D}_1,{\bf D}_2)}\|{\bf E}\|_2\,.
\end{split}
\end{equation}

\begin{lemma} \label{lemma_bosq}
Let $\{\theta_j, \bphi_j(\cdot)\}_{j \geq 1}$ and $\{\hat \theta_j, \wh\bphi_j(\cdot)\}_{j \geq 1}$ be the eigenvalue/eigenfunction pairs of $\bQ(\cdot,\cdot)$  and $\wh\bQ(\cdot,\cdot)$ respectively, with the corresponding nonzero eigenvalues sorted in decreasing order.
Then we have
{\rm(i)} $\sup_{j \geq 1}|\hat \theta_{j} -  \theta_{j}| \leq \|\wh \bQ -  \bQ\|_{\calS,{\rm F}}$, and  
{\rm(ii)} $\sup_{j \geq 1} \Delta_j\|\wh \bphi_{j} - \bphi_{j} \|\leq 2\sqrt{2} \|\wh \bQ -  \bQ\|_{\calS,{\rm F}},$ where $\Delta_j = \min_{k \in [j]}(\theta_k - \theta_{k+1}).$
\end{lemma}


\section{Proof of Proposition \ref{lm.space}}
Let $\bOmega_k^{(l)} = (\Omega_{k,ij}^{(l)})_{r_l \times r_l}$. By the decomposition \eqref{kl-decomp}, we write 
\begin{align}
    \bM_k^{(l)} = \sum_{i=1}^{r_l} \sum_{j=1}^{r_l} \Omega_{k,ij}^{(l)} \bvarphi_i^{(l)} \otimes \{\bvarphi_j^{(l)}\}^{\T}. \label{eq.decom1}
\end{align}
Hence, $\text{Im}(\bM_k^{(l)}) \subset {\rm span}\{\bvarphi_1^{(l)}(\cdot), \dots, \bvarphi_{r_l}^{(l)}(\cdot)\}$. Define $\tilde \lambda_{k,i}^{(l)} = \|\sum_{j=1}^{r_l} \Omega_{k,ij}^{(l)} \bvarphi_j^{(l)}\|$ and $ \bphi_{k,i}^{(l)}(\cdot) = \sum_{j=1}^{r_l} \Omega_{k,ij}^{(l)} \bvarphi_j^{(l)}(\cdot)/ \|\sum_{j=1}^{r_l} \Omega_{k,ij}^{(l)} \bvarphi_j^{(l)}\|$, we then rewrite \eqref{eq.decom1} as
\begin{align}
    \bM_k^{(l)} =  \sum_{i = 1}^{r_l} \tilde \lambda_{k,i}^{(l)} \bvarphi_i^{(l)}
 \otimes \{\bphi_{k,i}^{(l)} \}^{\T}. \label{eq.decom}
\end{align}
We next show that the set $\{\bphi_{k,1}^{(l)}(\cdot),\dots, \bphi_{k,r_l}^{(l)}(\cdot)\}$ is linearly independent
for some $k\in [k_0]$.
Let $\bbeta=(\beta_1,\dots, \beta_{r_l})^{\T}$ denote an arbitrary vector in $\mathbb{R}^{r_l}$ and $\bvarphi^{(l)}(\cdot) = \{\bvarphi_1^{(l)}(\cdot)^{\T},\dots, \bvarphi_{r_l}^{(l)}(\cdot)^{\T}\}^{\T}$. Since the set $\{\bvarphi_1^{(l)}(\cdot),\dots, \bvarphi_{r_l}^{(l)}(\cdot)\}$ is linearly independent and $\bOmega_k^{(l)}$ is of full rank for some $k\in [k_0]$, the only solution of
\begin{align}
    \sum_{i=1}^{r_l} \beta_i \bigg\|\sum_{j=1}^{r_l} \Omega_{k,ij}^{(l)} \bvarphi_j^{(l)}\bigg\|\bphi_{k,i}^{(l)}(\cdot) = \bbeta^{\T} \bOmega_k^{(l)}\bvarphi^{(l)}(\cdot) =0 \notag
\end{align}
is $\bbeta=0$ for such $k$. Hence,  the set $\{\bphi_{k,1}^{(l)}(\cdot),\dots, \bphi_{k,r_l}^{(l)}(\cdot)\}$ in \eqref{eq.decom} is linearly independent
for some $k\in [k_0]$. Together with the decomposition \eqref{eq.decom} and the fact that any linearly independent set of $r_l$ elements in a $r_l$-dimensional space forms a basis for that space, it implies that  
$\text{Im}(\bM_k^{(l)}) = {\rm span}\{\bvarphi_1^{(l)}(\cdot), \dots, \bvarphi_{r_l}^{(l)}(\cdot)\}$ for some $k\in [k_0]$.

By the definition of the image space, we further have  
\begin{align}
    &\text{Im}\Big\{\int_\calU \bM_k^{(l)}(\cdot, w) \bM_k^{(l)}(\cdot, w)^\T
\,{\rm d}w\Big\}  \notag
\\ &~~~~~~= \Big\{\bg \in \mathbb{H}:  \bg = \int_\calU\int_\calU \bM_k^{(l)}(u, w) \bM_k^{(l)}(v, w)^\T
\,{\rm d}w\bbf(v)\, {\rm d}v, \, \bbf \in\mathbb{H}  \Big\} \notag
\\ &~~~~~~= \Big\{\bg \in \mathbb{H}:  \bg = \int_\calU \bM_k^{(l)}(u, w) \int_\calU\bM_k^{(l)}(v, w)^\T
\bbf(v)\, {\rm d}v \,{\rm d}w, \, \bbf \in\mathbb{H}  \Big\} \notag
\\ &~~~~~~= \Big\{\bg \in \mathbb{H}:  \bg = \int_\calU \bM_k^{(l)}(u, w) \widetilde
\bbf(w) \,{\rm d}w, \, \widetilde \bbf \in\mathbb{H}  \Big\} = \text{Im}\{\bM_k^{(l)}\}.  \notag
\end{align}
Due to the nonnegativity of $\bK^{(l)}(\cdot,\cdot),$ we have that
$\int_\calU \bK^{(l)}(u,v) \bvartheta(v) {\rm d}v = 0$ if and only if $\int_\calU \int_\calU \bM_k^{(l)}(u, w) \bM_k^{(l)}(v, w)^\T
\,{\rm d}w \bvartheta(v) {\rm d}v = 0$ for all $k \in [k_0]$. This further leads to
$\text{Im}(\bK^{(l)}) = \bigcup_{k \in [k_0]}\text{Im}\{\int_\calU \bM_k^{(l)}(\cdot, w) \bM_k^{(l)}(\cdot, w)^\T
\,{\rm d}w\}= {\rm span}\{\bvarphi_1^{(l)}(\cdot), \dots, \bvarphi_{r_l}^{(l)}(\cdot)\}$. Hence, we complete the proof of part (ii). Furthermore, since $\text{dim}[\text{Im}\{\bK^{(l)}\}]=r_l,$ part (i) follows.
$\hfill\Box$

\section{Proof of Theorem \ref{thm.seg.msc}}
Let $\nu_n = \Xi^2p_{\dagger}^{1-\alpha}\mathcal{M}_y^{1-\alpha}(n^{-1}\log p)^{(1-\alpha)/{2}}$, where $\Xi =  s_1s_2(2p_{\dagger} +1)$ with $p_{\dagger} = \max_{l \in [q]} p_l$.
Recall $\bW_y$ in (\ref{c3}) and $\wh \bW_y$ in (\ref{Wy.th.est}). Since $\rho^{-1}\nu_n \to 0$ implies that $p_{\dagger}^{-2}\mathcal{M}_y ^2\log p =o(n)$,
it follows from Lemma~\ref{lemma_th_y} and fixed $k_0$  that
\begin{align} 
\|\wh{\bW}_y-\bW_y\|_2 \leq\sum_{k=0}^{k_0}\bigg\|\int\int  \big\{\calT_{\omega_k}(\widehat \bSigma{}^{\ttS}_{y,k}) (u,v)^{\otimes2}-\bSigma_{y,k}(u,v)^{\otimes2}\big\}\,{\rm d}u{\rm d}v\bigg\|_2 
= O_{\rm p}(\nu_n)\,.
\label{Wy.rate}
\end{align}
Recall that $ \wh \bGamma_y \bR  = (\wh \bPi_1, \dots, \wh \bPi_q) = (\wh \bgamma_1,\dots, \wh \bgamma_p)$ and $\bGamma_y \bH = (\bA_1 \bGamma_{z,1} \bH_1, \dots, \bA_q \bGamma_{z,q} \bH_q)= (\bgamma_1,\dots, \bgamma_p)$. 
By Lemma \ref{la:space}, 
for each $l\in[q]$,  
we have that
\begin{equation} \label{A.con}
    \|\widehat{\bPi}_l-\bA_l \bGamma_{z,l} \bH_l\|_2\leq 8 \rho^{-1}\|\widehat{\bW}_y-\bW_y\|_2\,.
\end{equation}
Combining (\ref{Wy.rate}) and (\ref{A.con}), it is  immediate to see that 
\begin{equation} \label{eta.con}
    \max_{j \in [p]} \|\wh \bgamma_j - \bgamma_j\|_2 = O_{\rm p} (\rho^{-1}\nu_n)\,.
\end{equation}
Recall that $ \wh \bgamma_{i} =\wh \bfeta_{\pi(i)} $, $ \wh T_{ij}=\max_{ |k|\leq m }\|\widehat{\bfeta}_i^\T \calT_{\omega_k}(\widehat \bSigma{}^{\ttS}_{y,k})\widehat{\bfeta}_j\|_\calS$ and $T_{ij}
 =\max_{|k| \leq m}\|{\bgamma}_{i}^\T  \bSigma_{y,k}{\bgamma}_{j}\|_{\cal S}$. 
Notice that 
\begin{equation} \label{eq.split}
    \widehat{\bfeta}_{\pi(i)}^\T \calT_{\omega_k}(\widehat \bSigma{}^{\ttS}_{y,k})\widehat{\bfeta}_{\pi(j)} - {\bgamma}_{i}^\T  \bSigma_{y,k}{\bgamma}_{j} =  I_1 +I_2+I_3+I_4+I_5\,, 
\end{equation}
where $I_1 = ( \wh \bgamma_{i}-\bgamma_{i})^\T \{ \calT_{\omega_k}(\widehat \bSigma{}^{\ttS}_{y,k})-\bSigma_{y,k}\}  \wh \bgamma_{j}  ,$ $I_2 =( \wh \bgamma_{i}-\bgamma_{i})^\T \bSigma_{y,k} ( \wh \bgamma_{j}-\bgamma_{j}) ,$ $I_3 =( \wh \bgamma_{i}-\bgamma_{i})^\T \bSigma_{y,k} \bgamma_{j},$ $I_4 =\bgamma_{i}^\T\{ \calT_{\omega_k}(\widehat \bSigma{}^{\ttS}_{y,k})-\bSigma_{y,k}\} \wh \bgamma_{j}$ and $I_5 =\bgamma_{i}^\T \bSigma_{y,k} ( \wh \bgamma_{j}-\bgamma_{j}).$ Let $\omega_n = \Xi\mathcal{M}_y^{1-\alpha}(n^{-1}\log p)^{(1-\alpha)/{2}}$. Hence $\omega_n \to 0$ as implied by $\rho^{-1}\nu_n \to 0$.
By (\ref{eta.con}), the orthonormality of $\wh \bgamma_{j}$ and $\bgamma_{j}$,  Lemma~\ref{lemma_th_y} and Inequality~\ref{lemma_hs_inequality}, we obtain that
\begin{equation} \nonumber
    \begin{split}
         \max_{i,j\in[p]} \|I_1\|_\calS = O_{\rm p}( \rho^{-1}\nu_n\omega_n)\,,~~~  \max_{i,j\in[p]} \|I_2\|_\calS = O_{\rm p}( \rho^{-2}
         \Xi p_{\dagger}^{1-\alpha}\nu_n^2)\,,~\,
         \\ 
         \max_{i,j\in[p]} \|I_4\|_\calS=O_{\rm p}( \omega_n)\,,~~~ \max_{i,j\in[p]}\big(\|I_3\|_\calS+\|I_5\|_\calS\big)= O_{\rm p}(\rho^{-1}\Xi p_{\dagger}^{1-\alpha}\nu_n)\,.
    \end{split}
\end{equation}
Together with  $\rho^{-1}\nu_n \to 0$, $\omega_n \to 0$ and $\omega_n = o(\nu_n)$,  it holds that
\begin{align}
        \max_{i,j\in [p]} |\wh T_{\pi(i)\pi(j)} - T_{ij}| \leq \,&
        \max_{i,j\in [p],|k|\leq m} \big |\|\widehat{\bfeta}_{\pi(i)}^\T \calT_{\omega_k}(\widehat \bSigma{}^{\ttS}_{y,k})\widehat{\bfeta}_{\pi(j)}\|_\calS - \|{\bgamma}_{i}^\T  \bSigma_{y,k}{\bgamma}_{j}\|_{\cal S}\big| \label{eq.T.rate} \\ \leq \,&\max_{i,j\in [p],|k|\leq m} \|\widehat{\bfeta}_{\pi(i)}^\T \calT_{\omega_k}(\widehat \bSigma{}^{\ttS}_{y,k})\widehat{\bfeta}_{\pi(j)}-{\bgamma}_{i}^\T  \bSigma_{y,k}{\bgamma}_{j}\|_{\cal S} = 
        O_{\rm p}(\rho^{-1}\Xi p_{\dagger}^{1-\alpha}\nu_n)\,. \notag
\end{align}

We now show that $\hat \varrho $ in (\ref{eq.r}) is a consistent estimate for $\varrho.$  For $k \in [\aleph]$, without loss of generality, we write $T_{(k)} = T_{i_k j_k}$ with $i_k, j_k \in [p]$.  
Since $\rho^{-1}
\Xi p_{\dagger}^{1-\alpha}\nu_n/\delta_n \to 0$ and $\delta_n/\varsigma \to 0$, we can  find some $h_n$ such that  $\rho^{-1}
\Xi p_{\dagger}^{1-\alpha}\nu_n \ll  h_n \ll \delta_n \ll \varsigma$. 
Let $\widetilde\Omega = \{\max_{i,j\in [p]} |\wh T_{\pi(i)\pi(j)} - T_{ij}|   \leq  h_n  \leq \varsigma /2\}$. It is immediate to see that under the event $\widetilde\Omega$ we have $\varsigma/2 \leq \wh T_{\pi(i_k)\pi(j_k)} \leq  \varsigma/2 + T_{(1)}$ for $k \in [\varrho]$ and $0  \leq \wh T_{\pi(i_k)\pi(j_k)} \leq  h_n$ for $k > \varrho$. 
Due to the definition of $\hat \varrho$, we have that 
\begin{align}
\frac{\wh T_{(\hat \varrho)} +\delta_n}{\wh T_{(\hat \varrho+1)} +\delta_n} \geq \frac{\wh T_{ (\varrho)} +\delta_n}{\wh T_{ (\varrho+1)} +\delta_n}\geq \frac{\varsigma/2 + \delta_n}{ h_n + \delta_n}\asymp \frac{\varsigma}{\delta_n} \label{eq.rho.low}
\end{align}
under $\widetilde\Omega$.
If $\hat \varrho < \varrho$, under $\widetilde\Omega$, 
\begin{align}\label{eq:rho.low1}
\frac{\wh T_{(\hat \varrho)} +\delta_n}{\wh T_{(\hat \varrho+1)} +\delta_n} \leq \frac{\varsigma/2 + T_{(1)} + \delta_n}{\varsigma/2 + \delta_n}\asymp\frac{T_{(1)}}{\varsigma}\,. 
\end{align}
Since
$\delta_n T_{(1)}/ \varsigma^2 \to 0$, \eqref{eq.rho.low} and \eqref{eq:rho.low1} imply $\mathbb{P}(\hat \varrho < \varrho \,| \, \widetilde\Omega) \to 0$.
Similarly, if $\hat \varrho > \varrho$,  under $\widetilde\Omega$,
$$
\frac{\wh T_{(\hat \varrho)} +\delta_n}{\wh T_{(\hat \varrho+1)} +\delta_n} \leq \frac{ h_n + \delta_n}{0+ \delta_n} \to 1\, .
$$
This together with (\ref{eq.rho.low}) and $\delta_n/\varsigma \to 0$ yields that $\mathbb{P}(\hat \varrho > \varrho\, | \, \widetilde\Omega)\to 0$. Hence,  $\mathbb{P}(\hat \varrho = \varrho\, | \, \widetilde\Omega)\to 1$. By (\ref{eq.T.rate}), $\mathbb{P}( \widetilde\Omega) \to 1$. Combining the above results, we have
$\mathbb{P}(\hat \varrho = \varrho )\to 1$. Recall $E = \{(i,j): T_{ij} \geq T_{(\varrho)}, 1 \leq i < j \leq p\}$ and $
    \widetilde E = \{(i,j): \wh T_{ij} \geq \wh T_{( \hat \varrho)}, 1 \leq i < j \leq p\}
$. Under the event $\{\hat{\varrho}=\varrho\}$, the permutation $\pi:[p]\to[p]$ actually provides a bijective mapping from the graph $([p],E)$ to $([p],\widetilde{E})$ in the sense that $\{k,(i,j)\}\in [p]\times E \rightarrow \{\pi(k),(\pi(i),\pi(j))\}\in [p]\times \widetilde{E}$. Hence we complete the proof of Theorem~\ref{thm.seg.msc}.
$\hfill\Box$

\section{Proof of Theorem \ref{thm.seg.distance}} 
Let $\nu_n = \Xi^2p_{\dagger}^{1-\alpha}\mathcal{M}_y^{1-\alpha}(n^{-1}\log p)^{(1-\alpha)/{2}}$, where $\Xi =  s_1s_2(2p_{\dagger} +1)$ with $p_{\dagger} = \max_{l \in [q]} p_l$. Since  $\rho^{-1}\nu_n \to 0$, the result in (\ref{Wy.rate}) holds. This, together with (\ref{A.con}) and  the remark for Lemma~1 of \textcolor{blue}{Chang et al. (2018)}, yields that
\begin{align}\label{eq:convspace}
\max_{l \in [q]}  D\{\calC(\wh \bPi_l), \calC(\bA_l)\} \lesssim \rho^{-1}\|\widehat{\bW}_y-\bW_y\|_2 = O_{\rm p}(\rho^{-1} \nu_n)\,.
\end{align}
Recall that $\wh \bPi_l = (\wh \bgamma_i)_{i \in  G_l}$ and $\wh \bA_l = (\wh \bgamma_i)_{i \in \wh G_l}$. Theorem~\ref{thm.seg.msc} implies that
there exists a permutation $\tilde\pi:[q] \to [q]$ such that 
$\mathbb{P}[\bigcap_{l= 1}^q \{\widehat{G}_{\tilde\pi(l)} =  G_l\}\, ,\, \hat q 
 = q ]\rightarrow 1$.  
Let $\Omega_l = \{\wh G_{\tilde \pi(l)} = G_{l}\, ,\, \hat q 
 = q \}$ and $d_n = \rho^{-1}\nu_n $. 
For any $\epsilon>0$, by \eqref{eq:convspace}, there exists a constant $C>0$ such that
\[
\mathbb{P}\bigg[\max_{l \in [q]} d_n^{-1}D\{\calC(\wh \bPi_l), \calC(\bA_l)\} > C \bigg]<\epsilon\,,
\]
which implies
\begin{equation} \nonumber
    \begin{split}
        & \mathbb{P}\bigg[\max_{l \in [q]} \min_{j\in[\hat q]}  d_n^{-1} D\{\calC(\wh \bA_j), \calC(\bA_l)\} >C\bigg]
        \\ & ~~~~~~\leq \mathbb{P}\bigg[\max_{l \in [q]} \min_{j\in[\hat q]}
 d_n^{-1} D\{\calC(\wh \bA_j), \calC(\bA_l)\} > C , \bigcap_{l = 1}^q \Omega_l \bigg] 
         +\mathbb{P}\bigg(
        \bigcup_{l = 1}^q \Omega_l^c \bigg) 
        \\ & ~~~~~~\leq \mathbb{P}\bigg[\max_{l \in [q]} d_n^{-1}D\{\calC(\wh \bPi_l), \calC(\bA_l)\} > C \bigg] + o(1)<\epsilon+o(1)\,.
    \end{split}
\end{equation}
Hence, $\max_{l \in [q]} \min_{j\in[\hat q]}  D\{\calC(\wh \bA_j), \calC(\bA_l)\}=O_{\rm p}(d_n)$. We complete the proof  of Theorem \ref{thm.seg.distance}.
$\hfill\Box$

\section{Proof of Theorem~\ref{thm.dr.dim}}
Recall that $\bPi_{l} = \bA_l \bGamma_{z,l} \bH_l= ( \bgamma_i)_{i \in  G_l}$ and $\wh \bA_l = (\wh \bgamma_i)_{i \in  G_l}$. 
Write $\bM_{k}^{(l)}(u,v)= \bPi_{l}^\T \bSigma_{y,k}(u,v) \bPi_{l}\equiv \{ M_{k,ij}^{(l)}(u,v)\}_{i,j \in [p_l]}$  and $\wh \bM_{k}^{(l)}(u,v) = \wh \bA_l^{\T}  \calT_{\omega_k}(\widehat \bSigma{}^{\ttS}_{y,k})(u,v)\wh \bA_l\equiv \{\wh M_{k,ij}^{(l)}(u,v)\}_{i,j \in [p_l]}.$
Let $\nu_n = \Xi^2p_{\dagger}^{1-\alpha}\mathcal{M}_y^{1-\alpha}(n^{-1}\log p)^{(1-\alpha)/{2}}$, where $\Xi =  s_1s_2(2p_{\dagger} +1)$ with $p_{\dagger} = \max_{l \in [q]} p_l$.
By a similar decomposition to (\ref{eq.split}) and $\rho^{-1}\Xi p_{\dagger}^{1-\alpha}\nu_n \to 0$, we obtain that 
$
        \max_{i,j \in [p_l],l \in [q]} \| \wh M_{k,ij}^{(l)} -   M_{k,ij}^{(l)}\|_\calS = O_{\rm p}(\rho^{-1} \Xi p_{\dagger}^{1-\alpha}\nu_n)$.
Hence,
$$\max_{l \in [q]}\|\wh \bM_{k}^{(l)} -\bM_{k}^{(l)}\|_{\calS,\tF} =\max_{l \in [q]} \bigg(\sum_{i,j \in [p_l]} \| \wh M_{k,ij}^{(l)} -   M_{k,ij}^{(l)}\|_\calS^2\bigg)^{1/2}=  O_{\rm p}(\rho^{-1}\Xi  p_{\dagger}^{2-\alpha}\nu_n)\,.$$
Write $\bZ_{t}^{(l)}(\cdot) = \{Z_{t,1}^{(l)}(\cdot),\dots, Z_{t,p_l}^{(l)}(\cdot)\}^{\T}.$
It follows from Cauchy--Schwartz inequality that
\begin{equation} \nonumber
    \begin{split}
        \max_{l \in [q]}\|\bM_{k}^{(l)}\|_{\calS,\tF}^2 & =\max_{l \in [q]} \sum_{i,j=1}^{p_l} \int \int \{M_{k,ij}^{(l)}(u,v) \}^2 \,{\rm d}u {\rm d}v \\& \leq \max_{ l \in [q] }\sum_{i =1}^{p_l} \int \mathbb{E}[\{Z_{t,i}^{(l)}(u)\}^2]\, {\rm d}u\cdot \max_{ l \in [q] }\sum_{j =1}^{p_l} \int \mathbb{E}[\{Z_{t+k,j}^{(l)}(u)\}^2]\, {\rm d}u
 =O(p_{\dagger}^2)\,.
    \end{split}
\end{equation}
Observe that
$\wh \bK^{(l)}(u,v) - \bK^{(l)}(u,v) =\sum_{k = 1}^{k_0} \int_\calU  \bM_{k}^{(l)}(u,w)\{\wh \bM_{k}^{(l)}(v,w) - \bM_{k}^{(l)}(v,w)\}^{\T}\, {\rm d}w +\sum_{k = 1}^{k_0} \int_\calU \{\wh \bM_{k}^{(l)}(u,w) - \bM_{k}^{(l)}(u,w)\} \bM_{k}^{(l)}(v,w)^{\T}\, {\rm d}w + \sum_{k = 1}^{k_0} \int_\calU \{\wh \bM_{k}^{(l)}(u,w) - \bM_{k}^{(l)}(u,w)\}\{\wh \bM_{k}^{(l)}(v,w) - \bM_{k}^{(l)}(v,w)\}^{\T}\, {\rm d}w$. Together  with    Inequality~\ref{lemma_operator_inequality}, $\rho^{-1}\Xi p_{\dagger}^{1-\alpha}\nu_n \to 0$ and fixed $k_0$, it holds that
\begin{eqnarray} \nonumber
    \begin{split}
       \max_{l \in [q]} \|\wh \bK^{(l)} - \bK^{(l)}\|_{\calS,\tF} 
&\leq \max_{l \in [q]} \sum_{k = 1}^{k_0} \|\wh \bM_{k}^{(l)} - \bM_{k}^{(l)}\|_{\calS,\tF}^2 + 2\max_{l \in [q]}\sum_{k = 1}^{k_0}\|\bM_{k}^{(l)}\|_{\calS,\tF} \|\wh \bM_{k}^{(l)} - \bM_{k}^{(l)}\|_{\calS,\tF} \\ \label{Kest.rate}
&= O_{\rm p}(\rho^{-2}\Xi^2  p_{\dagger}^{4-2\alpha}\nu_n^2) + O_{\rm p}(\rho^{-1}\Xi p_{\dagger}^{3-\alpha}\nu_n) = O_{\rm p}(\rho^{-1}\Xi p_{\dagger}^{3-\alpha}\nu_n)\,.
    \end{split}
\end{eqnarray}
This, together with Lemma~\ref{lemma_bosq}, implies that
\begin{equation}
    \label{eigen.est.rate}
    \max_{l \in [q], j \in [r_l]}|\hat\theta_{j}^{(l)}-\theta_{j}^{(l)}| = O_{\rm p}(\rho^{-1}\Xi p_{\dagger}^{3-\alpha}\nu_n)\,, ~~\max_{l \in [q], j \in [r_l]} \|\wh \bpsi_{j}^{(l)} -  \bpsi_{j}^{(l)}\| = O_{\rm p}(\Delta^{-1}\rho^{-1}\Xi p_{\dagger}^{3-\alpha}\nu_n)\,,
\end{equation}
where $\Delta = \min_{l \in [q], j \in [r_l]}\{\theta_{j}^{(l)} - \theta_{j+1}^{(l)}\}$.

Recall that 
\begin{equation}  \nonumber
    \hat r_l = \arg\max_{j\in [n-k_0]}
    \frac{\hat \theta_{j}^{(l)}+\tilde \delta_{n}}{\hat \theta_{j+1}^{(l)}+\tilde \delta_{n}}\,.
\end{equation}
Note that the condition $\tilde \delta_{n}\max_{l \in [q]}\theta_{1}^{(l)}/\min_{l \in [q]}\{\theta_{r_l}^{(l)}\}^2 \to 0$ 
implies that ${\tilde \delta_{n}}=o(\min_{l \in [q]}\theta_{r_l}^{(l)}).$
By $\rho^{-1}\Xi p_{\dagger}^{3-\alpha}\nu_n/  \tilde \delta_{n} \to 0$ and $\tilde \delta_n/\min_{l \in [q]}\theta_{r_l}^{(l)} \to 0$, we can  find some $\tilde h_n$ such that  $\rho^{-1}\Xi p_{\dagger}^{3-\alpha}\nu_n \ll \tilde h_n \ll \tilde \delta_n \ll \min_{l \in [q]}\theta_{r_l}^{(l)}$. 
Let $\widecheck \Omega = \{\max_{l \in [q], j \in [r_l]} |\hat\theta_{j}^{(l)}-\theta_{j}^{(l)}|   \leq \tilde h_n  \leq \min_{l \in [q]}\theta_{r_l}^{(l)} /2\}$. Under the event $\widecheck \Omega$, we thus  have $\min_{l \in [q]}\theta_{r_l}^{(l)}/2 \leq \hat \theta_{j}^{(l)} \leq  \min_{l \in [q]}\theta_{r_l}^{(l)}/2 + \max_{l \in [q]}\theta_{1}^{(l)}$ if $j \in [r_l]$ and $0  \leq  \hat \theta_{j}^{(l)} \leq \tilde h_n$ if $j > r_l$, for each $l \in [q]$. 
Due to the definition of $\hat r_l$, for each $l \in [q]$, we have that 
\begin{align}
\frac{\hat \theta_{\hat r_l}^{(l)}+\tilde \delta_n}{\hat \theta_{\hat r_l+1}^{(l)}+\tilde \delta_n} \geq \frac{\hat \theta_{r_l}^{(l)} +\tilde \delta_n}{\hat \theta_{r_l+1}^{(l)} +\tilde \delta_n}\geq \frac{\min_{l \in [q]}\theta_{r_l}^{(l)}/2 + \tilde \delta_n}{\tilde h_n + \tilde \delta_n}\asymp \frac{\min_{l \in [q]}\theta_{r_l}^{(l)}}{\tilde \delta_n} \label{eq.r.low}
\end{align}
 under $\widecheck\Omega$. For each $l \in [q]$, if $\hat r_l < r_l$,  under $\widecheck\Omega$, 
\begin{align}\label{eq:r.low1}
\frac{\hat \theta_{\hat r_l}^{(l)}+\tilde \delta_n}{\hat \theta_{\hat r_l+1}^{(l)}+\tilde \delta_n} \leq \frac{\min_{l \in [q]}\theta_{r_l}^{(l)}/2 + \max_{l \in [q]}\theta_{1}^{(l)} + \tilde \delta_n}{\min_{l \in [q]}\theta_{r_l}^{(l)}/2 + \tilde \delta_n}\asymp\frac{\max_{l \in [q]}\theta_{1}^{(l)}}{\min_{l \in [q]}\theta_{r_l}^{(l)}}\,. 
\end{align}
Since
$\tilde \delta_n \max_{l \in [q]}\theta_{1}^{(l)}/ \min_{l \in [q]}\{\theta_{r_l}^{(l)}\}^2 \to 0$, \eqref{eq.r.low} and \eqref{eq:r.low1} imply $\mathbb{P}[\bigcup_{i = 1}^q\{\hat r_l < r_l\}\,| \, \widecheck\Omega] \to 0$.
Similarly, for each $l \in [q]$, if $\hat r_l > r_l$,  under $\widecheck\Omega$,
$$
\frac{\hat \theta_{\hat r_l}^{(l)}+\tilde \delta_n}{\hat \theta_{\hat r_l+1}^{(l)}+\tilde \delta_n} \leq \frac{\tilde h_n + \tilde \delta_n}{0+ \tilde \delta_n} \to 1\, .
$$
This together with (\ref{eq.r.low}) and $\tilde \delta_n/\min_{l \in [q]}\theta_{r_l}^{(l)} \to 0$ yields that $\mathbb{P}[\bigcup_{i = 1}^q\{\hat r_l > r_l\}\, | \, \widecheck\Omega]\to 0$. Thus  $\mathbb{P}[\bigcap_{i = 1}^q\{\hat r_l = r_l\}\, | \, \widecheck\Omega]\to 1$. By (\ref{eigen.est.rate}), $\mathbb{P}( \widecheck\Omega) \to 1$. We complete the proof of Theorem~\ref{thm.dr.dim}. $\hfill\Box$


\section{Proof of Theorem \ref{thm.dr.space}} 
Let $\bpsi_{j}^{(l)}(\cdot)=\{\psi_{j1}^{(l)}(\cdot), \dots, \psi_{jp_l}^{(l)}(\cdot)\}^{\T}$ and $\wh \bpsi_{j}^{(l)}(\cdot)=\{\hat \psi_{j1}^{(l)}(\cdot), \dots, \hat \psi_{jp_l}^{(l)}(\cdot)\}^{\T}$.
Due to the orthonormality of $\bpsi_{j}^{(l)}(\cdot)$ and $\wh \bpsi_{j}^{(l)}(\cdot),$ we obtain that
\begin{align}
    &\bigg\|\sum_{j=1}^{r_l} \big[\wh \bpsi_{j}^{(l)} \otimes \{\wh \bpsi_{j}^{(l)}\}^\T - \bpsi_{j}^{(l)}\otimes \{\bpsi_{j}^{(l)}\}^\T\big] \bigg\|_{\calS,\tF}^2 \notag
    \\ & ~~~~~~= \int \int\sum_{k= 1}^{p_l} \sum_{m= 1}^{p_l}\bigg[\sum_{j = 1}^{r_l}\big \{\hat \psi_{jk}^{(l)}(u) \hat \psi_{jm}^{(l)}(v) - \psi_{jk}^{(l)}(u) \psi_{jm}^{(l)}(v)\big\}
    \bigg]^2{\rm d}u {\rm d}v \notag
    \\&~~~~~~= \sum_{i=1}^{r_l} \sum_{j=1}^{r_l}\int \int \sum_{k=1}^{p_l} \sum_{m=1}^{p_l} \hat \psi_{ik}^{(l)}(u) \hat \psi_{im}^{(l)}(v)\hat \psi_{jk}^{(l)}(u) \hat \psi_{jm}^{(l)}(v)\, {\rm d}u {\rm d}v \notag 
    \\&~~~~~~~~~+ \sum_{i=1}^{r_l} \sum_{j=1}^{r_l} \int \int\sum_{k=1}^{p_l} \sum_{m=1}^{p_l}\psi_{ik}^{(l)}(u) \psi_{im}^{(l)}(v) \psi_{jk}^{(l)}(u)  \psi_{jm}^{(l)}(v)\,{\rm d}u {\rm d}v\notag
    \\ &~~~~~~~~~-2 \sum_{i=1}^{r_l} \sum_{j=1}^{r_l}\int \int\sum_{k=1}^{p_l} \sum_{m=1}^{p_l} \psi_{ik}^{(l)}(u) \psi_{im}^{(l)}(v)\hat \psi_{jk}^{(l)}(u) \hat \psi_{jm}^{(l)}(v)\,{\rm d}u {\rm d}v\notag
    \\& ~~~~~~= \sum_{i=1}^{r_l} \sum_{j=1}^{r_l} \langle \wh \bpsi_{i}^{(l)}, \wh \bpsi_{j}^{(l)} \rangle^2  +  \sum_{i=1}^{r_l} \sum_{j=1}^{r_l}\langle  \bpsi_{i}^{(l)},  \bpsi_{j}^{(l)} \rangle^2 - 2  \sum_{i=1}^{r_l} \sum_{j=1}^{r_l}\langle \wh \bpsi_{i}^{(l)},  \bpsi_{j}^{(l)} \rangle^2 \notag 
    \\ & ~~~~~~= 2r_l-  2\sum_{i=1}^{r_l} \sum_{j=1}^{r_l}\langle \wh \bpsi_{i}^{(l)},  \bpsi_{j}^{(l)} \rangle^2 \, . \notag
\end{align}
Denote
 by $\widetilde\calC_l = \text{span}\{\wh \bpsi_{1}^{(l)}(\cdot),\dots, \wh \bpsi_{r_l}^{(l)}(\cdot)\}$  the dynamic space spanned by $ r_l$ estimated eigenfunctions.
By the  definition of $\widetilde D(\widetilde\calC_l, \calC_l)$, we thus have $\sqrt{2r_l}\widetilde D(\widetilde\calC_l, \calC_l) =  \|\sum_{j=1}^{r_l} [\wh \bpsi_{j}^{(l)} \otimes\{\wh \bpsi_{j}^{(l)}\}^\T - \bpsi_{j}^{(l)}\otimes \{\bpsi_{j}^{(l)}\}^\T]  \|_{\calS,\tF}.$ Let $\nu_n = \Xi^2p_{\dagger}^{1-\alpha}\mathcal{M}_y^{1-\alpha}(n^{-1}\log p)^{(1-\alpha)/{2}}$, where $\Xi =  s_1s_2(2p_{\dagger} +1)$ with $p_{\dagger} = \max_{l \in [q]} p_l$.
Since $\Delta^{-1}\rho^{-1} \Xi p_{\dagger}^{3-\alpha}\nu_n\to 0$, the result in (\ref{eigen.est.rate}) holds. 
This, together with Equality~\ref{lemma_f_inequality}, fixed $r_l$ and  the orthonormality of $\bpsi_{j}^{(l)}(\cdot)$, leads to  
\begin{equation} \nonumber
    \begin{split}
        \max_{l \in [q]} \sqrt{2r_l}\widetilde D(\widetilde\calC_l, \calC_l) =\, & \max_{l \in [q]} \bigg\|\sum_{j=1}^{r_l} \left[\wh \bpsi_{j}^{(l)} \otimes\{\wh \bpsi_{j}^{(l)}\}^\T - \bpsi_{j}^{(l)}\otimes \{\bpsi_{j}^{(l)}\}^\T\right] \bigg\|_{\calS,\tF}
        \\ \leq \,&\max_{l \in [q]}  \sum_{j=1}^{r_l}\|\wh \bpsi_{j}^{(l)} -  \bpsi_{j}^{(l)}\|^2 + 2\max_{l \in [q]} \sum_{j=1}^{r_l}\|\bpsi_{j}^{(l)}\|\|\wh \bpsi_{j}^{(l)} -  \bpsi_{j}^{(l)}\| \\ 
        = \,&  O_{\rm p}(\Delta^{-2}\rho^{-2}\Xi^2 p_{\dagger}^{6-2\alpha}\nu_n^2) + O_{\rm p}(\Delta^{-1}\rho^{-1}\Xi p_{\dagger}^{3-\alpha}\nu_n)  
        \\=\,& O_{\rm p}(\Delta^{-1}\rho^{-1}\Xi p_{\dagger}^{3-\alpha}\nu_n)\,.\\
    \end{split}
\end{equation}
Let $e_n = \Delta^{-1}\rho^{-1}\Xi p_{\dagger}^{3-\alpha}\nu_n$. Theorem~\ref{thm.dr.dim} implies that  $
    \mathbb{P}[\bigcap_{l= 1}^q \{\hat r_l = r_l\}] \to 1$. Thus, for any $\epsilon>0$, there exists a constant $C>0$ such that
\[
\mathbb{P}\bigg\{\max_{l \in [q]} e_n^{-1}\widetilde D(\widetilde\calC_l, \calC_l) > C \bigg\}<\epsilon\,,
\]
which implies
\begin{equation} \nonumber
    \begin{split}
        & \mathbb{P}\bigg\{\max_{l \in [q]} e_n^{-1}\widetilde D(\wh\calC_l, \calC_l) >C\bigg\}
        \\ & ~~~~~~\leq \mathbb{P}\bigg[\max_{l \in [q]} e_n^{-1}\widetilde D(\wh\calC_l, \calC_l) >C ,\, \bigcap_{l= 1}^q \{\hat r_l = r_l\} \bigg] 
         +\mathbb{P}\bigg\{
        \bigcup_{l = 1}^q  [\hat r_l \neq r_l\} \bigg] 
        \\ & ~~~~~~\leq \mathbb{P}\bigg\{\max_{l \in [q]} e_n^{-1}\widetilde D(\widetilde\calC_l, \calC_l) >C \bigg\} + o(1)<\epsilon+o(1)\,.
    \end{split}
\end{equation}
Hence, $\max_{l \in [q]} \widetilde D(\wh \calC_l,\calC_l)= O_{\rm p }(e_n )$. We complete the proof  of Theorem \ref{thm.dr.space}.
 $\hfill\Box$

\section{Proofs of auxiliary lemmas}
\label{supp.lemma.prove}

\subsection{Proof of Inequality \ref{lemma_operator_inequality}}
By  Cauchy--Schwartz inequality, we notice that
        \begin{align}\label{lem:int-mat-pf1}
                \bigg\|\int \int\bB_1(u,v)\bB_2(u,v)^{\T}\,{\rm d}u{\rm d}v\bigg\|_1&= \max\limits_{j \in [p]}\sum_{i=1}^p\bigg|\int \int \sum_{k=1}^p B_{1,ik}(u,v)B_{2,jk}(u,v) \,{\rm d}u{\rm d }v\bigg|\notag
                \\
                &\leq \max\limits_{j \in [p]}\sum_{i=1}^p\sum_{k=1}^p\|B_{1,ik}\|_{\calS}\|B_{2,jk}\|_{\calS}\\
                &\leq \max\limits_{ k\in [p]}\sum_{i=1}^p\|B_{1,ik}\|_{\calS}\cdot\max\limits_{j \in [p]}\sum_{k=1}^p\|B_{2,jk}\|_{\calS} = \|\bB_1\|_{\calS,1}  \|\bB_2\|_{\calS,\infty}\,.\notag
        \end{align}
        By similar argument, we obtain that
        \begin{equation}\label{lem:int-mat-pf2}
        \bigg\|\int\int\bB_1(u,v)\bB_2(u,v)^{\T}{\rm d}u{\rm d}v\bigg\|_{\infty}\leq \|\bB_1\|_{\calS,\infty}  \|\bB_2\|_{\calS,1}\,.
        \end{equation}
Combining  (\ref{lem:int-mat-pf1}) and  (\ref{lem:int-mat-pf2}) and applying the inequality $\|\bE\|^2\leq \|\bE\|_{\infty}\|\bE\|_1$ for any matrix $\bE \in \mathbb{R}^{p \times p},$ we complete the proof of part (i).

Let $\eulbC(u,v) = \int \bB_1(u,w)\bB_2(v,w)^{\T}{\rm d}w= \{\eulC_{ij}(u,v)\}_{i,j \in [p]}$.
It then follows from Cauchy--Schwartz inequality that
\begin{align}\nonumber
        \|\eulbC\|_{\calS, \tF}^2 = &\, \sum_{i,j =1}^p \int \int \eulC_{ij}^2(u,v)\,{\rm d}u{\rm d}v = \sum_{i,j =1}^p \int\int \bigg\{\sum_{k =1}^p \int B_{1,ik}(u,w)B_{2,jk}(v,w)\,{\rm d}w\bigg\}^2{\rm d}u{\rm d}v
        \\ \leq &\,\sum_{i,j =1}^p \int\int \bigg\{\sum_{k =1}^p \int B_{1,ik}^2(u,w)\,{\rm d}w \cdot \sum_{k =1}^p \int B_{2,jk}^2(u,w)\,{\rm d}w\bigg\}\,{\rm d}u{\rm d}v 
        =  \|\bB_1\|_{\calS,{ \rm \tF}}^2\|\bB_2\|_{\calS,{\rm \tF}}^2  \,. \notag
\end{align}
Hence, we complete the proof of part (ii).

By  Cauchy--Schwartz inequality, we further obtain that
\begin{align}
    \|\bB_1+ \bB_2\|_{\calS,{\rm F}}^2 = \,& \sum_{i,j =1}^p \int\int \{B_{1,ij}(u,v) +B_{2,ij}(u,v)\}^2\,{\rm d}u{\rm d}v \notag
    \\= \, & 2 \sum_{i,j =1}^p \int\int B_{1,ij}^2(u,v) B_{2,ij}^2(u,v)\,{\rm d}u{\rm d}v + \|\bB_1\|_{\calS,{\rm F}}^2 + \|\bB_2\|_{\calS,{\rm F}}^2 \notag
    \\ \leq \, & 2 \|\bB_1\|_{\calS,{\rm F}}\|\bB_2\|_{\calS,{\rm F}}+ \|\bB_1\|_{\calS,{\rm F}}^2 + \|\bB_2\|_{\calS,{\rm F}}^2 = (\|\bB_1\|_{\calS,{\rm F}} + \|\bB_2\|_{\calS,{\rm F}})^2\,.
 \notag
\end{align}
Hence, we complete the proof of part (iii).
$\hfill\Box$

\subsection{Proof of Inequality \ref{lemma_hs_inequality}}
By elementary calculations and Inequality~\ref{lemma_operator_inequality}, we obtain that
\begin{align*} 
    \|\bb_1 ^\T \bB \bb_2\|_\calS^2 
     =&\,  \int \int \bb_1^\T \bB(u,v) \bb_2 \bb_2^\T\bB(u,v)^\T \bb_1\, {\rm d}u{\rm d}v
     \\ \leq &\, \int \int \|\bb_2 \bb_2^{\T}\|_2 \|\bB(u,v)^\T \bb_1\|_2^2  \,\, {\rm d}u{\rm d}v \\
     \leq&\, \|\bb_2\|_2^2 \int \int \bb_1^\T \bB(u,v)\bB(u,v)^\T \bb_1 \,{\rm d}u{\rm d}v\leq  \|\bb_1\|_2^2\|\bb_2\|_2^2 \|\bB\|_{\calS,\infty}\|\bB\|_{\calS,1}\,,
\end{align*}
which completes our proof.
$\hfill\Box$

\subsection{Proof of Inequality \ref{ineq_F}}
Let $\bbf(u) = \{f_1(u),\dots,f_p(u) \}^{\T}$ and $\bg(u) = \int \bB(u,v)\bbf(v)\, {\rm d}v = \{g_1(u), \dots, g_p(u)\}^{\T}$.  By Cauchy--Schwartz inequality, we obtain that
\begin{equation} \nonumber
    \begin{split}
        \|\bg\|^2 = \,& \sum_{i =1}^p \int g_{i}(u)^2\,{\rm d}u = \sum_{i=1}^p \int \bigg\{\sum_{k = 1}^p \int B_{ik}(u,v)f_{k}(v)\, {\rm d}v\bigg\}^2\, {\rm d}u
        \\ \leq \, &\sum_{i = 1}^p \int \bigg\{\sum_{k = 1}^p \int B_{ik}^2(u,v)\, {\rm d}v \sum_{k = 1}^p \int f_k^2(v)\, {\rm d}v \bigg\} \, {\rm d}u
        = \|\bB\|_{\calS, {\rm F}}^2 \|\bbf\|^2 \, ,
    \end{split}
\end{equation}
This further leads to
$\|\bB\|_{\calL}={\text{sup}}_{\|\bbf\| \leq 1, \bbf \in \cH} \|\bB(\bbf)\| \leq {\text{sup}}_{\|\bbf\| \leq 1, \bbf \in \cH} \|\bB\|_{\calS, {\rm F}} \|\bbf\| \leq  \|\bB\|_{\calS, {\rm F}}$, which completes our proof.
$\hfill\Box$

\subsection{Proof of Equality \ref{lemma_f_inequality}}
For $\bbf = (f_1,\dots,f_p )^{\T}$ and $ \bg = (g_1, \dots, g_p)^{\T}\in \mathbb{H},$ it holds that
$$
        \|\bbf \otimes \bg^\T\|_{\calS,{\rm \tF}}^2= \sum_{i,j =1}^p \int \int f_i^2(u)g_j^2(v)\,{\rm d}u{\rm d}v =\bigg\{\sum_{i=1}^p \int f_i^2(u)\,{\rm d}u \bigg\}\bigg\{\sum_{j=1}^p \int g_j^2(v)\,{\rm d}v\bigg\}= \|\bbf\|^2\| \bg\|^2\,.
$$
Hence, we  complete our proof.
$\hfill\Box$

\subsection{Proof of Lemma \ref{lemma_dev_y}}
This lemma follows directly from Theorem~1 of \textcolor{blue}{Fang et al. (2022)} and Theorem~2 of \textcolor{blue}{Guo
and Qiao (2023)}and hence the proof is omitted here.
$\hfill\Box$

\subsection{Proof of Lemma \ref{lemma.Sigma.k}}
Recall that $\max_{i \in [p]}\int \mathbb{E}\{Z_{ti}^2(u)\}\,{\rm d}u = O(1).$
Hence,  
\begin{align} \label{eq.z.var}
        { \max_{i,j \in [p], |k|\leq k_0 \vee m}\| \Sigma_{z,k,ij}\|_\calS^2} &= \max_{i,j \in [p], |k|\leq k_0 \vee m}\int\int[\mathbb{E}\{Z_{ti}(u) Z_{(t+k)j}(v)\}]^2\,{\rm d}u{\rm d}v \notag \\
        &\leq \max_{i \in [p]}\int \mathbb{E}\{Z_{ti}^2(u)\}\,{\rm d}u \cdot \max_{j \in [p]} \int \mathbb{E}\{Z_{(t+k)j}^2(v)\}\,{\rm d}v =  O(1)\,.
\end{align}
Let $p_{\dagger} = \max_{l \in [q]} p_l$. Since $\bSigma_{y,k}(u, v) =\bA \bSigma_{z,k}(u, v)\bA^{\T}$ and $\Sigma_{z,k,lm}=0$ for $|l-m|>p_\dagger$, then  
$
\Sigma_{y,k,ij}(u,v) = \sum_{l,m=1}^p A_{il} \Sigma_{z,k,lm}(u,v)A_{jm} =  \sum_{|l-m|\leq p_{\dagger}} A_{il} \Sigma_{z,k,lm}(u,v)A_{jm}
$. 
 By the inequality $(a+b)^\alpha \leq a^\alpha + b^\alpha$ for $a,b \geq 0$ and  $\alpha \in [0,1),$ we obtain that
\begin{align*}
        \sum_{i = 1}^p \| \Sigma_{y,k,ij}\|_\calS^\alpha 
        & =  \sum_{i = 1}^p \bigg\| \sum_{|l-m|\leq p_{\dagger}} A_{il} \Sigma_{z,k,lm}A_{jm}\bigg\|_\calS^\alpha \leq \sum_{i = 1}^p \bigg(\sum_{|l-m|\leq p_{\dagger}}\big\|  A_{il} \Sigma_{z,k,lm}A_{jm}\big\|_\calS\bigg)^\alpha 
\\&\leq \sum_{i = 1}^p\sum_{|l-m|\leq p_{\dagger}} |A_{il}|^\alpha|A_{jm}|^\alpha \| \Sigma_{z,k,lm}\|_\calS^\alpha 
        \\&\leq \max_{l,m \in [p],|k|\leq k_0 \vee m}{ \| \Sigma_{z,k,lm}\|_\calS^\alpha} \cdot\max_{l \in [p]}\sum_{i=1}^p|A_{il}|^\alpha \cdot \sum_{|l-m|\leq p_{\dagger}}|A_{jm}|^\alpha
        \\&\leq \max_{l,m \in [p],|k|\leq k_0 \vee m}{ \| \Sigma_{z,k,lm}\|_\calS^\alpha}\cdot s_2 \cdot (2p_{\dagger} +1)\sum_{m = 1}^p|A_{jm}|^\alpha
         \\&= O(\Xi)\,,
\end{align*}
where $\Xi =  s_1s_2(2p_{\dagger} +1)$.
By (\ref{eq.z.var}) and the block structure of $\bSigma_{z,k}(u,v)$, we further obtain that
$\|\bSigma_{z,k}\|_{\calS,1}= \max_{j \in [p]} \sum_{i = 1}^p \|\Sigma_{z,k,ij}\|_\calS  = O(p_{\dagger})$.
Similarly, we also have $\|\bSigma_{z,k}\|_{\calS,\infty}= O(p_{\dagger}).$
Together with  Inequality \ref{lemma_hs_inequality} and  the orthonormality of the rows in $\bA$, it holds that 
\begin{equation}\nonumber
\|\bSigma_{y,k}\|_{\calS,1} =   \max_{j \in [p]} \sum_{i = 1}^p \|\Sigma_{y,k,ij}\|_\calS \leq  \max_{i,j\in [p]} \|\Sigma_{y,k,ij}\|_\calS^{1-\alpha} \cdot \max_{j\in [p]} \sum_{i = 1}^p \| \Sigma_{y,k,ij}\|_\calS^\alpha   =O(\Xi p_{\dagger}^{1-\alpha})\,.
\end{equation}
Recall $k_0$ and $m$ are fixed integers. Similarly, we can prove the rest of this lemma.
$\hfill\Box$

\subsection{Proof of Lemma \ref{lemma_th_y}}
Denote by $\calT_{\omega_k}(\wh \Sigma{}^{\ttS}_{y,k,ij})$ the $(i,j)$-th component of $ \calT_{\omega_k}(\widehat \bSigma{}^{\ttS}_{y,k}).$ Due to the fact that $\calT_{\omega_k}(\wh \Sigma{}^{\ttS}_{y,k,ij}) = \wh{\Sigma}^{\ttS}_{y,k,ij}(u,v) I\{\|\wh{\Sigma}^{\ttS}_{y,k,ij}\|_{\calS}\geq \omega_k\}$, we have
 $\|\calT_{\omega_k}(\wh \Sigma{}^{\ttS}_{y,k,ij}) - \wh{\Sigma}^{\ttS}_{y,k,ij}\|_\calS \leq \omega_k$.
Under the event $\Omega = \{\max_{i,j\in [p]} \| {\widehat\Sigma{}^{\ttS}_{y,k,ij}  - \Sigma_{y,k,ij}}\|_\calS \le \tilde \theta \omega_k\}$ for $\tilde \theta \in (0,1)$ and  $\omega_k = c_k\mathcal{M}_y (n^{-1}\log p)^{1/2}$, we have 
    \begin{align*}
        &\max_{j \in [p]} \sum_{i = 1}^p\|\calT_{\omega_k}(\wh \Sigma{}^{\ttS}_{y,k,ij}) - \Sigma_{y,k,ij}\|_\calS\\
        &~~~~~~=\max_{j \in [p]} \sum_{i = 1}^p\|\calT_{\omega_k}(\wh \Sigma{}^{\ttS}_{y,k,ij}) - \Sigma_{y,k,ij}\|_\calS I\{\|\wh \Sigma{}^{\ttS}_{y,k,ij}\|_\calS\geq \omega_k\} 
        \\&~~~~~~~~~+\max_{j \in [p]}  \sum_{i = 1}^p\|\calT_{\omega_k}(\wh \Sigma{}^{\ttS}_{y,k,ij}) - \Sigma_{y,k,ij}\|_\calS I\{\|\wh \Sigma{}^{\ttS}_{y,k,ij}\|_\calS < \omega_k\}
        \\ &~~~~~~\leq  \max_{j \in [p]} \sum_{i = 1}^p \|\calT_{\omega_k}(\wh \Sigma{}^{\ttS}_{y,k,ij}) - \wh{\Sigma}^{\ttS}_{y,k,ij}\|_\calS I\{\|\wh \Sigma{}^{\ttS}_{y,k,ij}\|_\calS\geq \omega_k,\| \Sigma_{y,k,ij}\|_\calS\geq \omega_k\}\\
        &~~~~~~~~~+ \max_{j \in [p]} \sum_{i = 1}^p\|\wh{\Sigma}^{\ttS}_{y,k,ij} - \Sigma_{y,k,ij}\|_\calS I\{\|\wh \Sigma{}^{\ttS}_{y,k,ij}\|_\calS\geq \omega_k,\| \Sigma_{y,k,ij}\|_\calS\geq \omega_k\}
        \\ &~~~~~~~~~ + \max_{j \in [p]} \sum_{i = 1}^p\|\calT_{\omega_k}(\wh \Sigma{}^{\ttS}_{y,k,ij}) - \Sigma_{y,k,ij}\|_\calS I\{\|\wh \Sigma{}^{\ttS}_{y,k,ij}\|_\calS\geq \omega_k, \| \Sigma_{y,k,ij}\|_\calS< \omega_k\} 
        \\&~~~~~~~~~+\max_{j \in [p]} \sum_{i = 1}^p\| \Sigma_{y,k,ij}\|_\calS I\{\|\wh \Sigma{}^{\ttS}_{y,k,ij}\|_\calS < \omega_k\}
        \\ &~~~~~~\leq   \underbrace{\omega_k \sum_{i = 1}^p I\{\| \Sigma_{y,k,ij}\|_\calS\geq \omega_k\}}_{Q_1}+\underbrace{\max_{j \in [p]} \sum_{i = 1}^p\| \Sigma_{y,k,ij}\|_\calS I\{\| \Sigma_{y,k,ij}\|_\calS < 2 \omega_k\}}_{Q_2}\\
        &~~~~~~~~~+\underbrace{\max_{j \in [p]} \sum_{i = 1}^p\|\wh \Sigma{}^{\ttS}_{y,k,ij} - \Sigma_{y,k,ij}\|_\calS I\{\|\wh \Sigma{}^{\ttS}_{y,k,ij}\|_\calS\geq \omega_k, \| \Sigma_{y,k,ij}\|_\calS< \omega_k\}}_{Q_3}\,.
    \end{align*}
By Lemma~\ref{lemma.Sigma.k}, we have
 $
     Q_1 +Q_2 \lesssim   \omega_k^{1-\alpha} \sum_{i=1}^p \| \Sigma_{y,k,ij}\|_\calS^\alpha   \lesssim   \omega_k^{1-\alpha}\Xi$ under the event $\Omega$.
Also, 
\begin{equation}\nonumber
    \begin{split}
        Q_3 \leq\, &\max_{j \in [p]} \sum_{i = 1}^p\|\wh \Sigma{}^{\ttS}_{y,k,ij} - \Sigma_{y,k,ij}\|_\calS I\{\|\wh \Sigma{}^{\ttS}_{y,k,ij}\|_\calS\geq \omega_k, \| \Sigma_{y,k,ij}\|_\calS<  (1-\tilde \theta) \omega_k\}  
        \\& + \max_{j \in [p]} \sum_{i = 1}^p\|\wh \Sigma{}^{\ttS}_{y,k,ij} - \Sigma_{y,k,ij}\|_\calS I\{\|\wh \Sigma{}^{\ttS}_{y,k,ij}\|_\calS\geq \omega_k, (1-\tilde \theta) \omega_k \leq \| \Sigma_{y,k,ij}\|_\calS< \omega_k\} 
        \\ \leq \, &  \omega_k \max_{j \in [p]} \sum_{i = 1}^p I\{\|\wh \Sigma{}^{\ttS}_{y,k,ij} - \Sigma_{y,k,ij}\|_\calS > \tilde \theta \omega_k \} + \omega_k  \max_{j \in [p]} \sum_{i = 1}^p I\{\| \Sigma_{y,k,ij}\|_\calS\geq(1-\tilde \theta) \omega_k \}
        \\ = \, &  \omega_k  \max_{j \in [p]} \sum_{i = 1}^p I\{\| \Sigma_{y,k,ij}\|_\calS\geq(1-\tilde \theta) \omega_k \}\lesssim \omega_k^{1-\alpha}  \Xi
    \end{split}
\end{equation}
under the event $\Omega$.
By Lemma~\ref{lemma_dev_y}, if $n \gtrsim \log p$ and $\tilde c {\tilde \theta}^2 c_k^2>2$, then
$
             \mathbb{P}(\Omega^c) 
            \leq 8p^{2-\tilde c {\tilde \theta}^2 c_k^2}\to 0$. 
Combining the above results, we thus have
 $$
\max_{j \in [p]} \sum_{i = 1}^p\|\calT_{\omega_k}(\wh \Sigma{}^{\ttS}_{y,k,ij}) - \Sigma_{y,k,ij}\|_\calS = O_{\rm p}\bigg\{\Xi\mathcal{M}_y^{1-\alpha}\bigg(\frac{\log p}{n}\bigg)^{(1-\alpha)/{2}} \bigg\}\,.
 $$
Recall $k_0$ and $m$ are fixed integers. We have the first result. The second result can be proved in the similar manner. Due to
$\calT_{\omega_k}(\widehat \bSigma{}^{\ttS}_{y,k}) (u,v)^{\otimes2}-\bSigma_{y,k}(u,v)^{\otimes2}
= \{\calT_{\omega_k}(\widehat \bSigma{}^{\ttS}_{y,k}) (u,v)-\bSigma_{y,k}(u,v)\}^{\otimes2}+ \bSigma_{y,k}(u,v)\{\calT_{\omega_k}(\widehat \bSigma{}^{\ttS}_{y,k}) (u,v)-\bSigma_{y,k}(u,v)\}^{\T} + \{\calT_{\omega_k}(\widehat \bSigma{}^{\ttS}_{y,k}) (u,v)-\bSigma_{y,k}(u,v)\}\bSigma_{y,k}(u,v)^{\T}$,
it follows from Inequality~\ref{lemma_operator_inequality} and Lemma~\ref{lemma.Sigma.k} that 
\begin{equation} \nonumber
    \begin{split}
        &\bigg\|\int \int \big\{ \calT_{\omega_k}(\widehat \bSigma{}^{\ttS}_{y,k}) (u,v)^{\otimes2}-\bSigma_{y,k}(u,v)^{\otimes2}\big\}\, {\rm d}u{\rm d}v\bigg\|_2
        \\ &~~~~~~\leq  2\|\bSigma_{y,k}\|_{\calS,1}^{1/2}\|\bSigma_{y,k}\|_{\calS,\infty}^{1/2} \| \calT_{\omega_k}(\widehat \bSigma{}^{\ttS}_{y,k}) - \bSigma_{y,k}\|_{\calS,1}^{1/2}\| \calT_{\omega_k}(\widehat \bSigma{}^{\ttS}_{y,k}) - \bSigma_{y,k}\|_{\calS,\infty}^{1/2}
        \\  &~~~~~~~~~ +\| \calT_{\omega_k}(\widehat \bSigma{}^{\ttS}_{y,k}) - \bSigma_{y,k}\|_{\calS,1}\| \calT_{\omega_k}(\widehat \bSigma{}^{\ttS}_{y,k}) - \bSigma_{y,k}\|_{\calS,\infty}
        \\&~~~~~~ = O_{\rm p}\bigg\{\Xi^2 p_{\dagger}^{1-\alpha}\mathcal{M}_y^{1-\alpha}\bigg(\frac{\log p}{n}\bigg)^{(1-\alpha)/{2}}\bigg\} + O_{\rm p}\bigg\{\Xi^2\mathcal{M}_y^{2-2\alpha}\bigg(\frac{\log p}{n}\bigg)^{1-\alpha}\bigg\}
        \,.
    \end{split}
\end{equation}
Since $ p_{\dagger}^{-2}\mathcal{M}_y ^2\log p=o(n)$, we have the third result. $\hfill\Box$ 

\subsection{Proof of Lemma \ref{lemma_bosq}}
 By the definition of spectral decomposition, we have that $\theta_j = \min_{\bB \in \calL_{j-1}} \| \bQ -\bB\|_{\calL}$ and $\hat \theta_j = \min_{\bB \in \calL_{j-1}} \| \wh \bQ -\bB \|_{\calL},$ where $\calL_{j-1} = \{\bB: \bB \in \calL, \, \text{dim}(\text{Im}(\bB)) \leq j-1\}$. Thus, $\theta_j = \min_{\bB \in \calL_{j-1}} \| \bQ-\bB \|_{\calL} \leq \| \bQ - \wh \bQ\|_{\calL} +\min_{\bB \in \calL_{j-1}} \| \wh \bQ -\bB\|_{\calL} =\|\wh \bQ - \bQ\|_{\calL} + \hat \theta_j$. Similarly, we have $\hat \theta_j = \min_{\bB \in \calL_{j-1}} \| \wh \bQ -\bB \|_{\calL} \leq \|\wh \bQ - \bQ\|_{\calL} +\min_{\bB \in \calL_{j-1}} \| \bQ-\bB \|_{\calL} =\|\wh \bQ - \bQ\|_{\calL} +  \theta_j$. Combining the two above results with Inequality \ref{ineq_F}, we obtain that 
\begin{equation} \label{lm5.1}
    |\hat \theta_{j} -  \theta_{j}|\leq \|\wh \bQ -  \bQ\|_{\calL} \leq  \|\wh \bQ -  \bQ\|_{\calS,\tF}
\end{equation}
holds for all $j \geq 1,$ which completes our proof of part (i).

 Without loss of generality, we assume that $\langle \wh \bphi_{j}, \bphi_{j} \rangle \geq 0$. Since $\sum_{l = 1}^{\infty} \inner{\wh\bphi_j}{\bphi_l}^2 = \|\wh \bphi_j\|^2 =  1$ and $ 0 \leq \langle \wh \bphi_{j}, \bphi_{j} \rangle \leq 1$, it holds that 
\begin{align}
\|\wh \bphi_{j} - \bphi_{j} \|^2 =\,&\sum_{l = 1}^\infty \big(\inner{\wh \bphi_j}{ \bphi_l} - \inner{\bphi_j}{\bphi_l}\big)^2  = \big( \inner{\wh \bphi_j}{ \bphi_j } - 1\big)^2 + \sum_{l \neq j}\inner{\wh \bphi_j}{ \bphi_l}^2 \notag
\\ = \,& \inner{\wh \bphi_j}{ \bphi_j }^2 - 2\inner{\wh \bphi_j}{ \bphi_j }+\sum_{l = 1}^{\infty} \inner{\wh\bphi_j}{\bphi_l}^2 + \sum_{l \neq j}\inner{\wh \bphi_j}{ \bphi_l}^2 \notag
\\ = \,& 2 \sum_{l \neq j}\inner{\wh \bphi_j}{ \bphi_l}^2 + 2(\inner{\wh \bphi_j}{ \bphi_j }^2 - \inner{\wh \bphi_j}{ \bphi_j }) \leq 2 \sum_{l \neq j}\inner{\wh \bphi_j}{ \bphi_l}^2\,. \label{lm5.2}
\end{align}
Observe that $
\bQ( \wh \bphi_{j})(u)- \theta_j \wh \bphi_{j}(u) = (\bQ - \wh \bQ)(\wh \bphi_{j}) (u) + (\hat \theta_j - \theta_j) \wh \bphi_{j}(u)
$. This together with Inequality \ref{ineq_F}, \eqref{lm5.1} and the orthonormality of $\wh \bphi_{j}$ implies that 
\begin{equation} \label{lm5.3}
    \|\bQ( \wh \bphi_{j})- \theta_j \wh \bphi_{j}\| \leq 2\|\wh \bQ -\bQ\|_{\calS,\tF}\,.
\end{equation}
We further write 
\begin{align}
    \|\bQ( \wh \bphi_{j})- \theta_j \wh \bphi_{j}\|^2 = \, & \sum_{l = 1}^{\infty}\big( \inner{\wh \bphi_j}{\bQ(  \bphi_{l})} - \inner{\theta_j \wh \bphi_{j}}{\bphi_l}\big)^2 \notag
    \\ = \, &\sum_{l \neq j}(\theta_l -\theta_j)^2\inner{\wh\bphi_{j}}{\bphi_l}^2 \geq \Delta_j^2\sum_{l \neq j}\inner{\wh\bphi_{j}}{\bphi_l}^2\,. \label{lm5.4}
\end{align}
Combining \eqref{lm5.2}--\eqref{lm5.4}, we complete the proof of part (ii). 
$\hfill\Box$ 

\section{Additional empirical results}
\label{supp.sec_emp}
\subsection{Simulated data with VFAR model}
\label{supp.sec_nogroup}
In Section~\ref{sec.sim.nogroup}, we illustrate the usefulness of decorrelation transformation with an example of a vector functional autoregressive (VFAR) model. 
In this section, we present the corresponding  details of the data generating process, the prediction procedure of the VFAR method and some additional simulation results. 

In each simulated scenario, we generate functional variables by $Y_{tj}(u)=\bs(u)^{\T}\btheta_{tj}$ for $t \in [n], j \in [p]$ and $u \in \calU=[0,1],$ where $\bs(u)$ is a $5$-dimensional Fourier basis function and  $\btheta_{t}=(\btheta_{t1}^{\T},\dots,\btheta_{tp}^{\T} )^{\T} \in \mathbb{R}^{5p}$ are 
generated from a stationary vector autoregressive (VAR) model, $\btheta_t = \bB \btheta_{t-1} + \boldsymbol{e}_t$, with  transition matrix $\bB \in \mathbb{R}^{5p \times 5p}$, whose entries are randomly sampled from ${\cal N}(0,1)$ and innovations $\boldsymbol e_t =(\boldsymbol e_{t1}^{\T},\dots,\boldsymbol e_{tp}^{\T} )^{\T} \in \mathbb{R}^{5p}$ being independently sampled from ${\cal N}(\boldsymbol 0, \bI_{5p}).$ To guarantee the stationarity of $\bY_t(\cdot)$, we rescale  $\bB$ by $\iota\bB/\rho(\bB)$ with $\iota \sim\text{Uniform}[0.5,1]$. 
Write $\bQ(u,v) = \{Q_{ij}(u,v)\}_{i,j \in [p]}$ and $\bB = (\bB_{ij})_{i,j \in [p]}$ with $\bB_{ij} \in \mathbb{R}^{5 \times 5}$.
According to Section F.3 of the supplementary material in \textcolor{blue}{Guo
and Qiao (2023)}, $\bY_t(\cdot)$ follows from a VFAR model of order 1 as in \eqref{VFAR1}, where $\epsilon_{tj}(u) = \bs(u)^{\T}\boldsymbol e_{tj}$ and the $(i,j)$-th entry of $\bQ(u,v)$ is given by 
$Q_{ij}(u,v) =\bs(u)^{\T}\bB_{ij} \bs(u)$.

We next implement the standard three-step estimation procedure in the VFAR method to estimate $\bQ(u,v)$.
\begin{enumerate}
\item[Step~1.] Perform  FPCA on  $\{Y_{tj}(\cdot)\}_{t \in [200]}$ thus obtaining the estimated eigenfunctions $\widehat\bs_j(\cdot) = \{\hat s_{j1}(\cdot), \dots,\hat s_{j5}(\cdot) \}^{\T}$ and the corresponding estimated principal component scores $\widehat \btheta_{tj}= (\hat \theta_{t1},\dots, \hat \theta_{t5})^{\T}$ with $\hat \theta_{tl} = \langle Y_{tj}, \hat s_{jl}\rangle$ for each $j$. 

\item[Step~2.] Write $\bTheta_1 = (\widehat \btheta_{tj})_{t\in [199], j \in [p] } \in \mathbb{R}^{199 \times 5p}$ and  $\bTheta_2 = (\widehat \btheta_{tj})_{t\in [200]\setminus[1], j \in [p] } \in \mathbb{R}^{199 \times 5p}$.  Obtain the least-squares estimator of $\bB$ as
$\widehat \bB = \{(\bTheta_1^{\T}\bTheta_1)^{-1} \bTheta_1^{\T}\bTheta_2\}^{\T}\equiv (\widehat \bB_{ij})_{i,j \in [p]}.$

\item[Step~3.] Recover the functional coefficient by $\widehat Q_{ij}(u,v) =\widehat \bs_i(u)^{\T}\widehat \bB_{ij}\widehat  \bs_j(u)$.
\end{enumerate}
Let $\widehat Q_{ij}(\cdot,\cdot)$ be the $(i,j)$-th entry of $\widehat \bQ(\cdot,\cdot)$. 
We then compute $\int_\calU \widehat \bQ(\cdot,v) \bY_{200}(v) {\rm d} v$  as
the one-step ahead prediction for the original
curves $\bY_{201}(\cdot)$.

Finally, we summarize in Table~\ref{table.nogroup.A} the effect of decorrelation transformation on the identified group numbers  $\mathring q_y$ of Seg+Y and 
 $\mathring q_z$ of Seg+Z. Notably, the transformation step results in the identification of more distinct groups.

\begin{table}[!t]
	\caption{\label{table.nogroup.A} The means of identified group numbers for Seg+Y and Seg+Z over 500 simulation runs.}
	\begin{center}
		\vspace{-0.5cm}
		\resizebox{4in}{!}{
		\begin{tabular}{c|cccccc}
\hline
                 & $p=10$ & $p=15$ & $p=20$ & $p=25$ & $p=30$ & $p=35$ \\ \hline
 $\mathring q_y$                         & 6.038  & 5.642  & 4.195  & 3.094  & 2.557  & 2.158  \\
 $\mathring q_z$                          & 6.126  & 6.394  & 6.737  & 6.612  & 5.687  & 6.712  \\
                                \hline
\end{tabular}
		}	
	\end{center}
	\vspace{-0.2cm}
\end{table}

\subsection{UK annual temperature data}
\label{sec.real.temp}
The dataset, which is available at \url{https://www.metoffice.gov.uk/research/climate/maps-and-data/historic-station-data}, consists of monthly mean temperature collected at $p=22$ measuring stations across Britain from 1959 to 2020 ($n=62$). 
Let $\check Y_{tj}(v_i)$ ($t\in [62],$ $j\in [22],$ $i \in [12]$) be the mean temperature during month $v_i=i$ of year $1958+t$ measured at the $j$-th station.  The observed temperature curves are smoothed using a $10$-dimensional Fourier basis that characterize the periodic pattern over the annual cycle. The post-sample prediction are carried out in an identical way to Section~\ref{sec.real.mortality}, we choose $k_0=m=3$ in our estimation procedure and treat the smoothed curves in the first $n_1=41$ years and the last $n_2=21$ years as the training sample and the test sample, respectively.
The values of MAPE and MSPE for $h\in\{1, 2, 3\}$ defined in (\ref{MSP}) are summarized in Table~\ref{table.temp}. 
 Again it is obvious that SegV and FTSegV perform similarly well and both
provide the highest predictive accuracies among all comparison methods for all $h.$


\begin{table}[!ht]
	\caption{\label{table.temp} 
	MAFEs and MSFEs for eight competing methods on the UK temperature curves for $h\in\{1,2,3\}.$}
	\begin{center}
		\vspace{-0.2cm}
		\resizebox{4in}{!}{
\begin{tabular}{c|ccc|ccc}
\hline
\multirow{2}{*}{Method} & \multicolumn{3}{c|}{MAFE}                        & \multicolumn{3}{c}{MSFE}                         \\ \cline{2-7} 
                        & $h=1$          & $h=2$          & $h=3$          & $h=1$          & $h=2$          & $h=3$          \\ \hline
SegV                    & \textbf{0.786} & 0.806          & \textbf{0.827} & \textbf{1.073} & 1.075          & \textbf{1.155} \\
Under.SegV              & 0.805          & 0.826          & 0.883          & 1.152          & 1.135          & 1.266          \\
Uni.SegV               & 0.797          & 0.821          & 0.845          & 1.101          & 1.126          & 1.174          \\
FTSegV                  & 0.789          & \textbf{0.806} & 0.828          & 1.077          & \textbf{1.073} & 1.158          \\
Under.FTSegV            & 0.791          & 0.820           & 0.872          & 1.105          & 1.112          & 1.250           \\
Uni.FTSegV             & 0.797          & 0.821          & 0.845          & 1.101          & 1.126          & 1.174          \\ 
UniV                    & 0.936          & 0.951          & 0.976          & 1.450           & 1.450           & 1.458          \\
GSY                     & 0.894          & 0.884          & 0.854          & 1.346          & 1.338          & 1.219          \\ \hline
\end{tabular}

		}	
	\end{center}
	\vspace{-0.2cm}
\end{table}

\subsection{Age-specific mortality data}
Table \ref{tab.country} gives a list of $p = 29$
inclusive countries with corresponding ISO Alpha-3 codes under our study.
\begin{table}[!ht]
	\caption{\label{tab.country} 
	List of inclusive countries with corresponding ISO Alpha-3 codes.}
	\begin{center}
		\vspace{-0.2cm}
		\resizebox{5.5in}{!}{
\begin{tabular}{lclclclc}
\hline
Country        & Code & Country   & Code & Country     & Code & Country       & Code \\
\hline
Australia      & AUS  & Finland   & FIN  & Norway      & NOR  & Sweden        & SWE  \\
Austria        & AUT  & France    & FRA  & Portugal    & PRT  & Switzerland   & CHE  \\
Belgium        & BEL  & Hungary   & HUN  & Poland      & POL  & Great Britain & GBR  \\
Belarus        & BLR  & Ireland   & IRE  & Netherlands & NLD  & United States & USA  \\
Bulgaria       & BGR  & Italy     & ITA  & New Zealand & NZL  & Ukraine       & UKR  \\
Canada         & CAN  & Japan     & JPN  & Russia      & RUS  &               &      \\
Denmark        & DEN  & Lithuania & LTU  & Slovakia    & SVK  &               &      \\
Czech Republic & CZE  & Latvia    & LVA  & Spain       & ESP  &               &     \\
\hline
\end{tabular}
		}	
	\end{center}
	\vspace{-0.2cm}
\end{table}

\begin{figure}[tbp]
\centering
\begin{subfigure}{1\linewidth}
\centering\includegraphics[width=16cm,height=6.5cm]{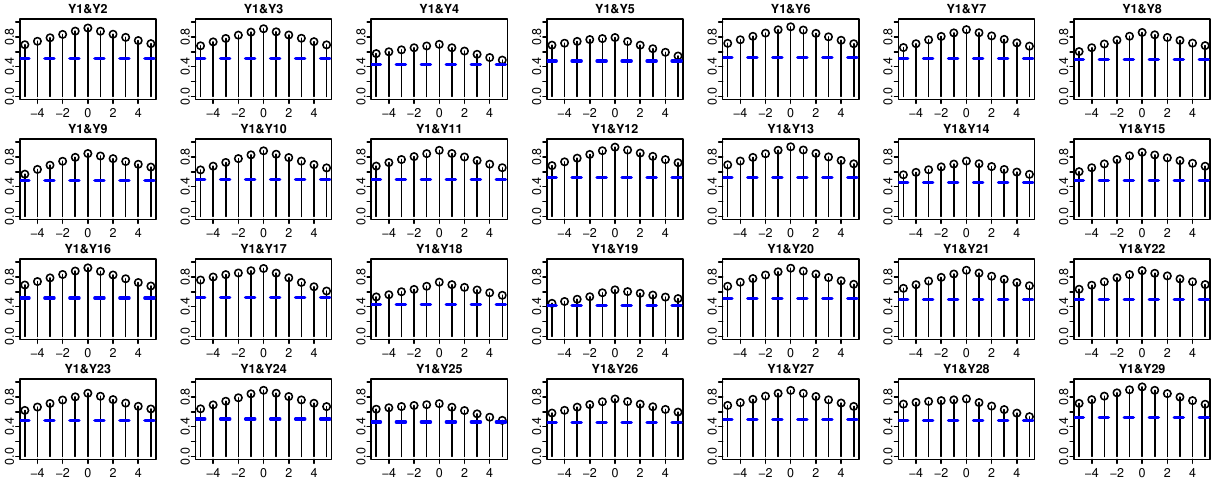}
 \end{subfigure}
  \vspace{-0.2cm} 
\caption{\label{plot1}{Functional cross-autocorrelation of the original female mortality data
}}
\end{figure}

\begin{figure}[tbp]
\centering
\begin{subfigure}{1\linewidth}
  \centering\includegraphics[width=16cm,height=6.5cm]{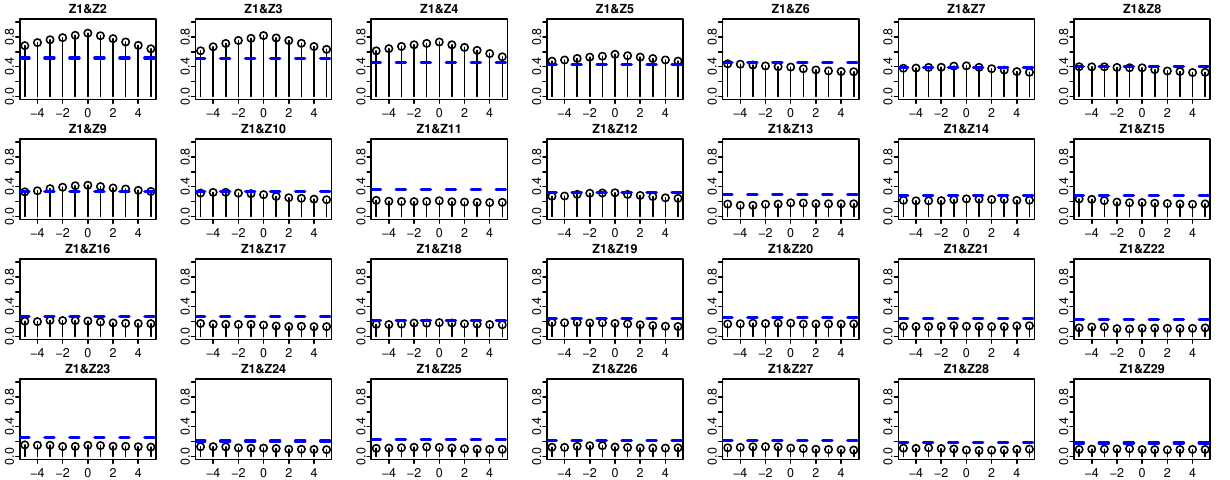}
  \vspace{-0.2cm} 
  \caption{Functional cross-autocorrelation of the 1st component series of the transformed curves} 
 \end{subfigure}

\vspace{0.2cm}
\begin{subfigure}{1\linewidth}
  \centering\includegraphics[width=16cm,height=6.5cm]{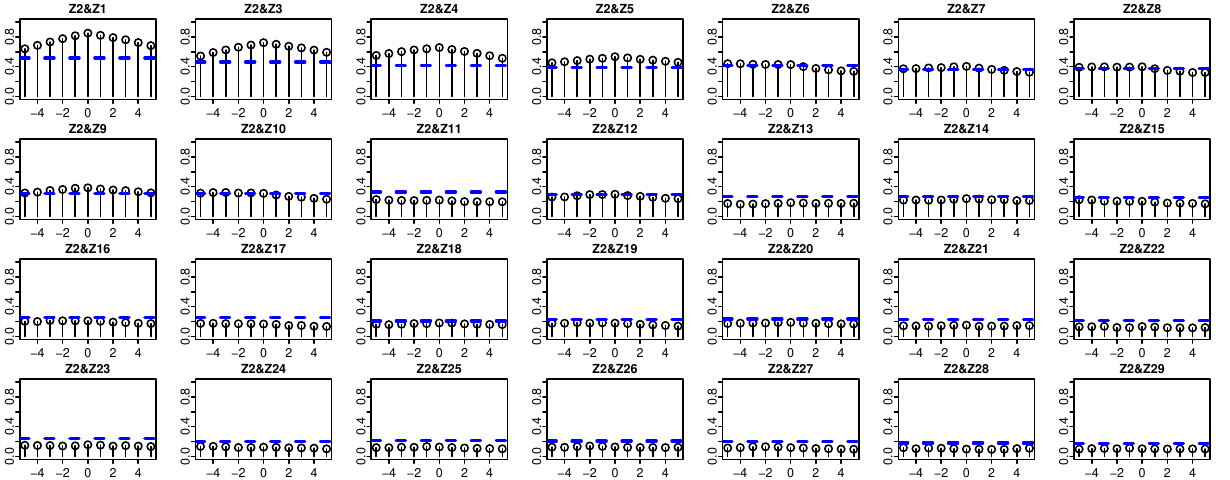}
  \vspace{-0.2cm} 
  \caption{Functional cross-autocorrelation of the 2nd component series of the transformed curves} 
 \end{subfigure}

 \vspace{0.2cm}
\begin{subfigure}{1\linewidth}
  \centering\includegraphics[width=16cm,height=6.5cm]{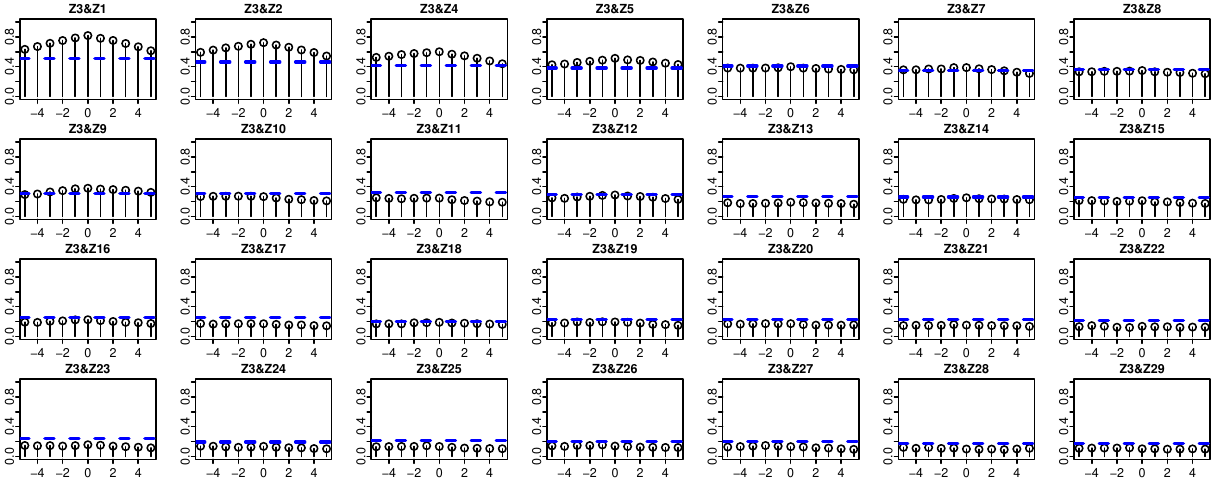}
  \vspace{-0.2cm} 
  \caption{Functional cross-autocorrelation of the 3rd component series of the transformed curves} 
 \end{subfigure}
\centering
\caption{\label{plot21}{Selected functional cross-autocorrelation of the transformed female mortality data
}}
\end{figure}

\begin{figure}[tbp]
\centering
\begin{subfigure}{1\linewidth}
  \centering\includegraphics[width=16cm,height=6.5cm]{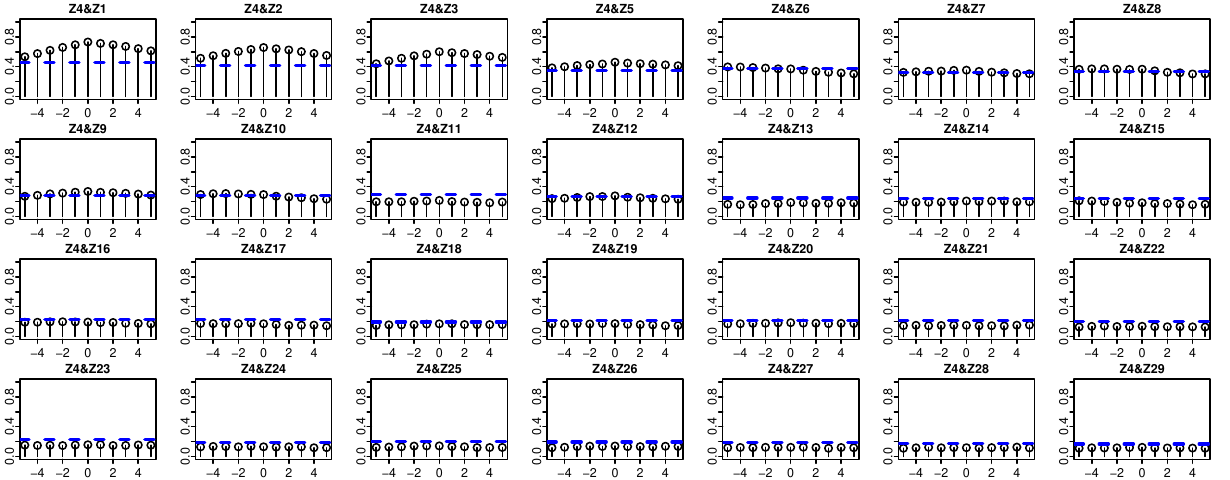}
  \vspace{-0.2cm} 
  \caption{Functional cross-autocorrelation of the 4th component series of the transformed curves} 
 \end{subfigure}

\vspace{0.2cm}
\begin{subfigure}{1\linewidth}
  \centering\includegraphics[width=16cm,height=6.5cm]{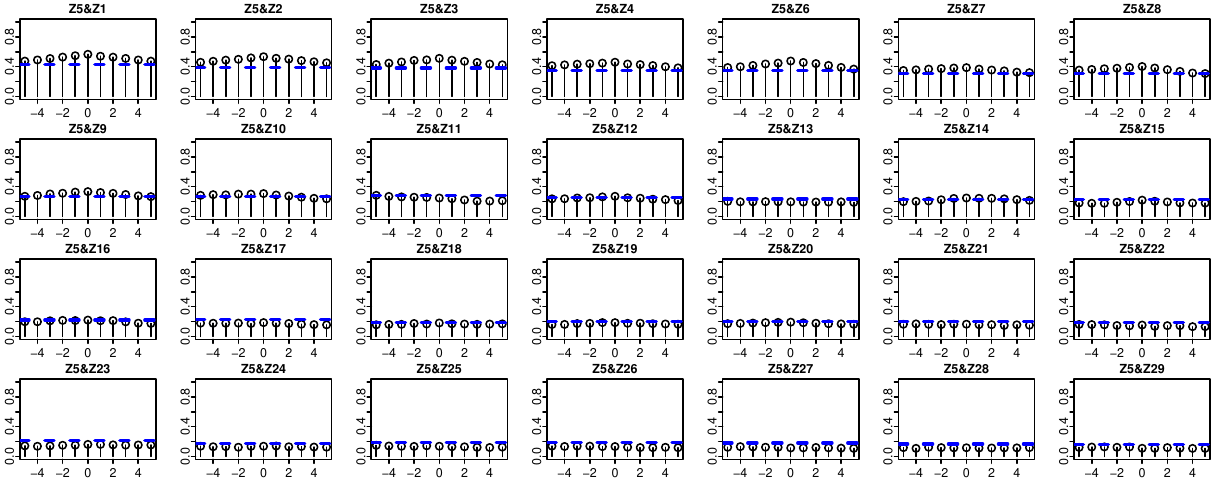}
  \vspace{-0.2cm} 
  \caption{Functional cross-autocorrelation of the 5th component series of the transformed curves} 
 \end{subfigure}

 \vspace{0.2cm}
\begin{subfigure}{1\linewidth}
  \centering\includegraphics[width=16cm,height=6.5cm]{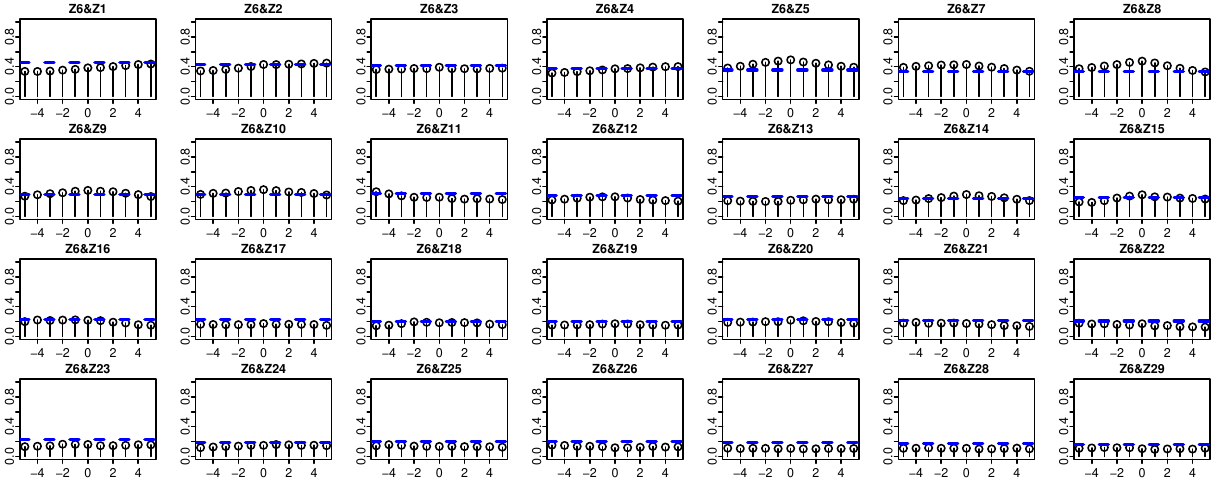}
  \vspace{-0.2cm} 
  \caption{Functional cross-autocorrelation of the 6th component series of the transformed curves} 
 \end{subfigure}

\centering
\caption{{Selected functional cross-autocorrelation of the transformed female mortality data
}}
\end{figure}

\begin{figure}[tbp]
\centering
\begin{subfigure}{1\linewidth}
  \centering\includegraphics[width=16cm,height=6.5cm]{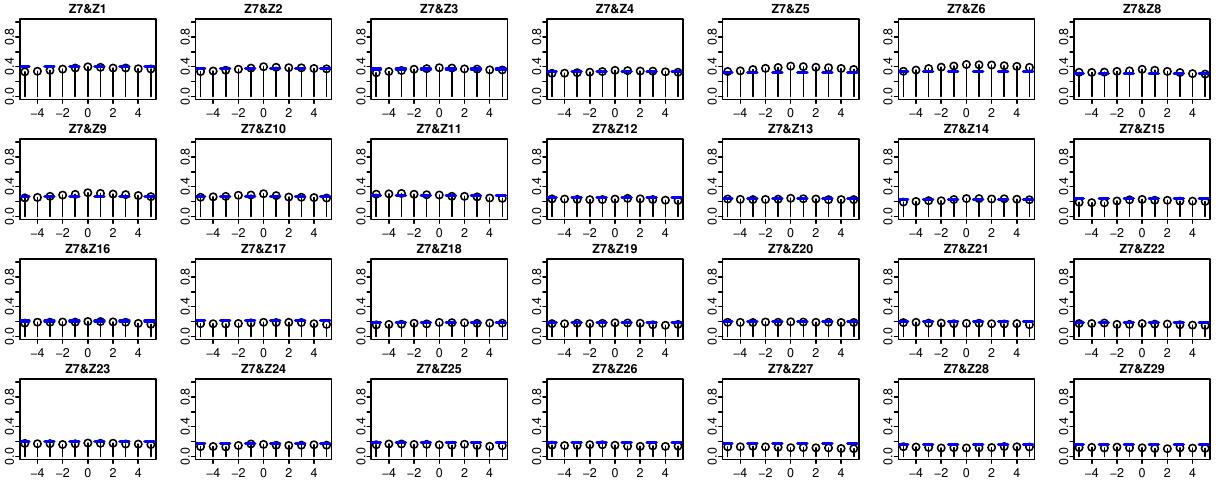}
  \vspace{-0.2cm} 
  \caption{Functional cross-autocorrelation of the 7th component series of the transformed curves} 
 \end{subfigure}

\vspace{0.2cm}
\begin{subfigure}{1\linewidth}
  \centering\includegraphics[width=16cm,height=6.5cm]{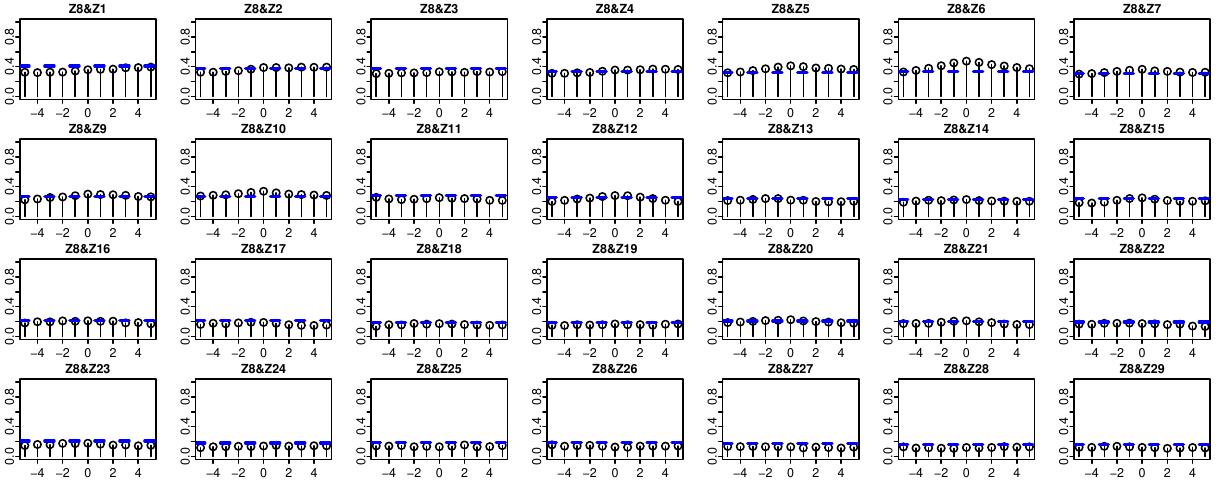}
  \vspace{-0.2cm} 
  \caption{Functional cross-autocorrelation of the 8th component series of the transformed curves} 
 \end{subfigure}

 \vspace{0.2cm}
\begin{subfigure}{1\linewidth}
  \centering\includegraphics[width=16cm,height=6.5cm]{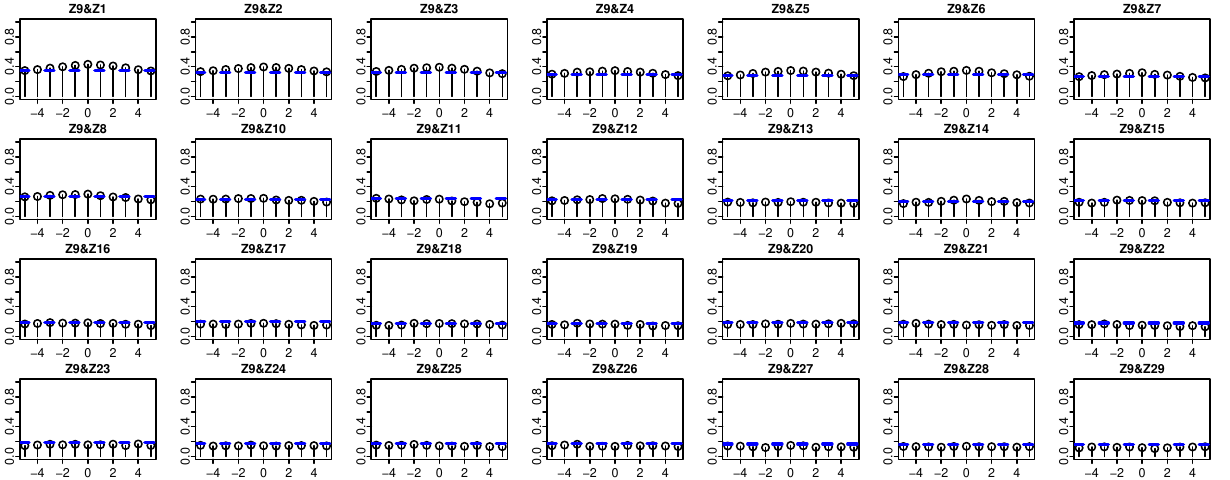}
  \vspace{-0.2cm} 
  \caption{Functional cross-autocorrelation of the 9th component series of the transformed curves} 
 \end{subfigure}

\centering
\caption{\label{plot23}{Selected functional cross-autocorrelation of the transformed female mortality data
}}
\end{figure}

\begin{figure}[tbp]
\centering
\begin{subfigure}{1\linewidth}
  \centering\includegraphics[width=16cm,height=6.5cm]{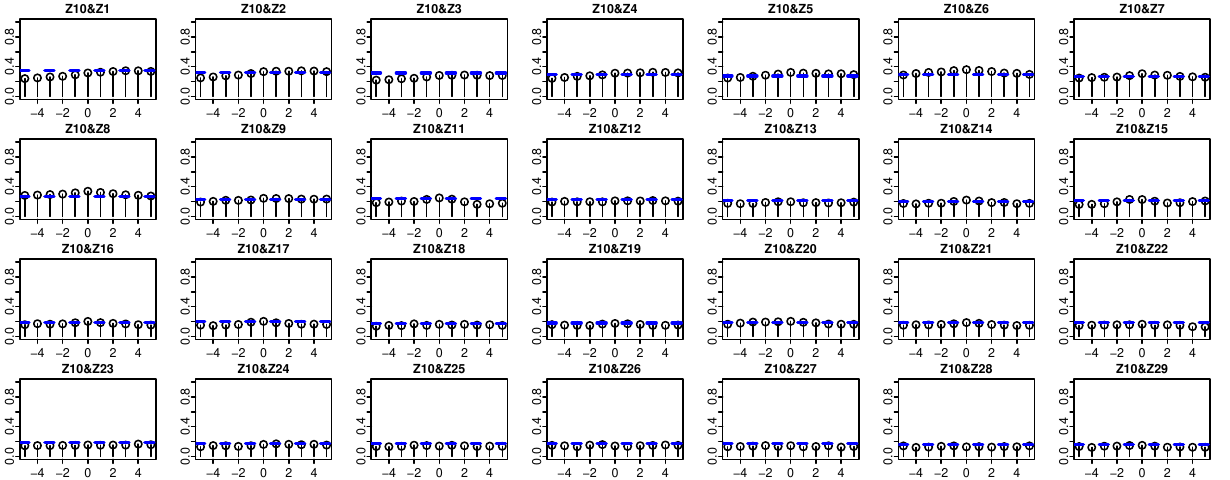}
  \vspace{-0.2cm} 
  \caption{Functional cross-autocorrelation of the 10th component series of the transformed curves} 
 \end{subfigure}

\vspace{0.2cm}
\begin{subfigure}{1\linewidth}
  \centering\includegraphics[width=16cm,height=6.5cm]{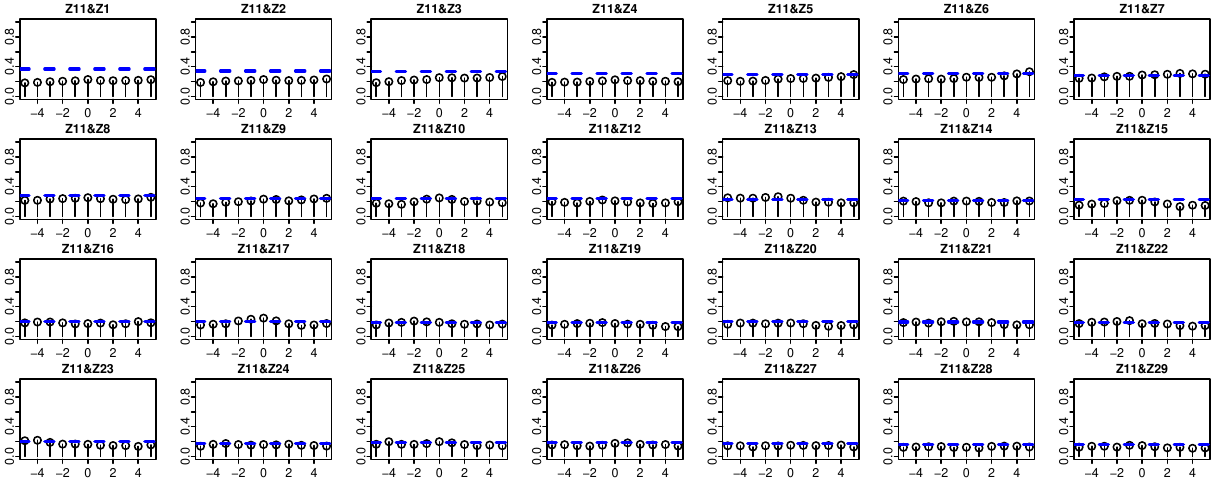}
  \vspace{-0.2cm} 
  \caption{Functional cross-autocorrelation of the 11th component series of the transformed curves} 
 \end{subfigure}

 \vspace{0.2cm}
\begin{subfigure}{1\linewidth}
  \centering\includegraphics[width=16cm,height=6.5cm]{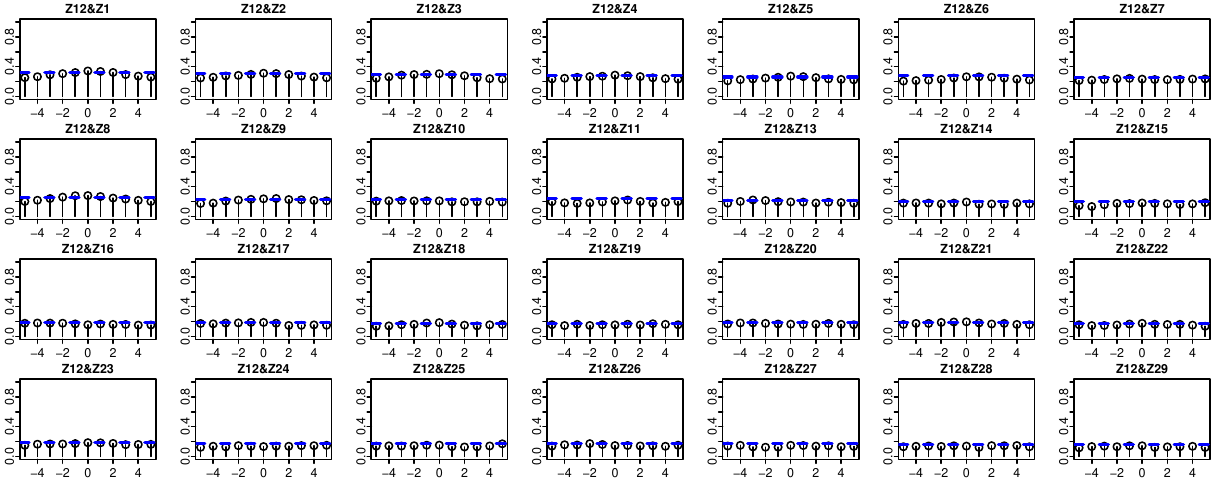}
  \vspace{-0.2cm} 
  \caption{Functional cross-autocorrelation of the 12th component series of the transformed curves} 
 \end{subfigure}

\centering
\caption{{Selected functional cross-autocorrelation of the transformed female mortality data
}}
\end{figure}

\begin{figure}[tbp]
\centering
\begin{subfigure}{1\linewidth}
  \centering\includegraphics[width=16cm,height=6.5cm]{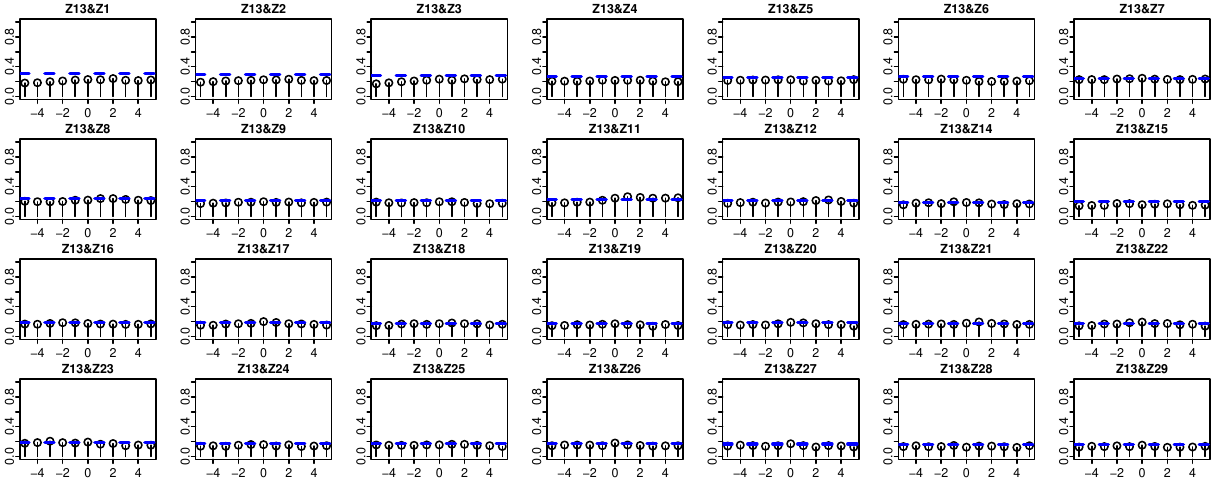}
  \vspace{-0.2cm} 
  \caption{Functional cross-autocorrelation of the 13th component series of the transformed curves} 
 \end{subfigure}

\vspace{0.2cm}
\begin{subfigure}{1\linewidth}
  \centering\includegraphics[width=16cm,height=6.5cm]{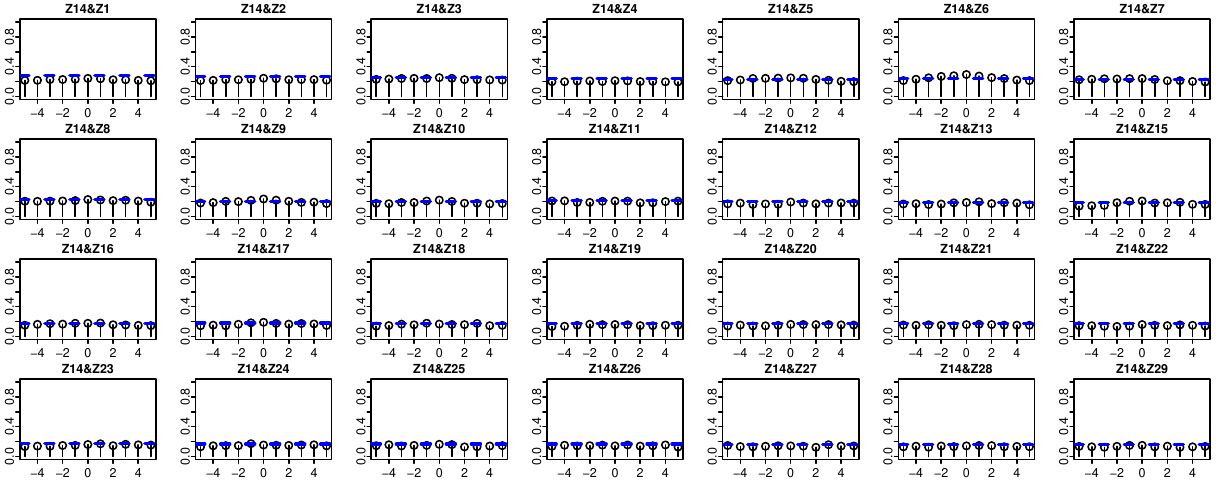}
  \vspace{-0.2cm} 
  \caption{Functional cross-autocorrelation of the 14th component series of the transformed curves} 
 \end{subfigure}

 \vspace{0.2cm}
\begin{subfigure}{1\linewidth}
  \centering\includegraphics[width=16cm,height=6.5cm]{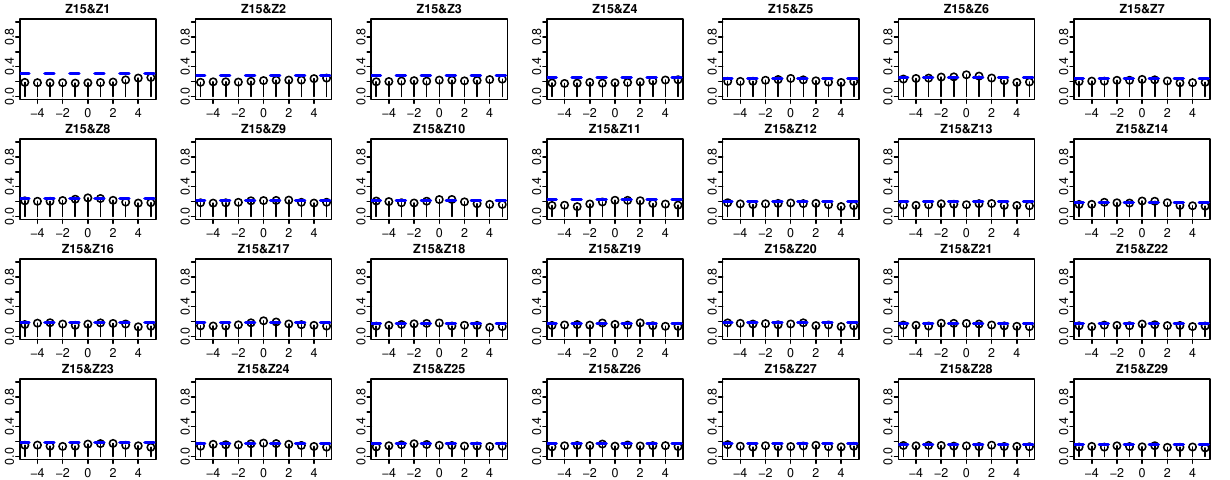}
  \vspace{-0.2cm} 
  \caption{Functional cross-autocorrelation of the 15th component series of the transformed curves} 
 \end{subfigure}
\centering
\caption{{Selected functional cross-autocorrelation of the transformed female mortality data
}}
\end{figure}

\begin{figure}[tbp]
\centering
\begin{subfigure}{1\linewidth}
  \centering\includegraphics[width=16cm,height=6.5cm]{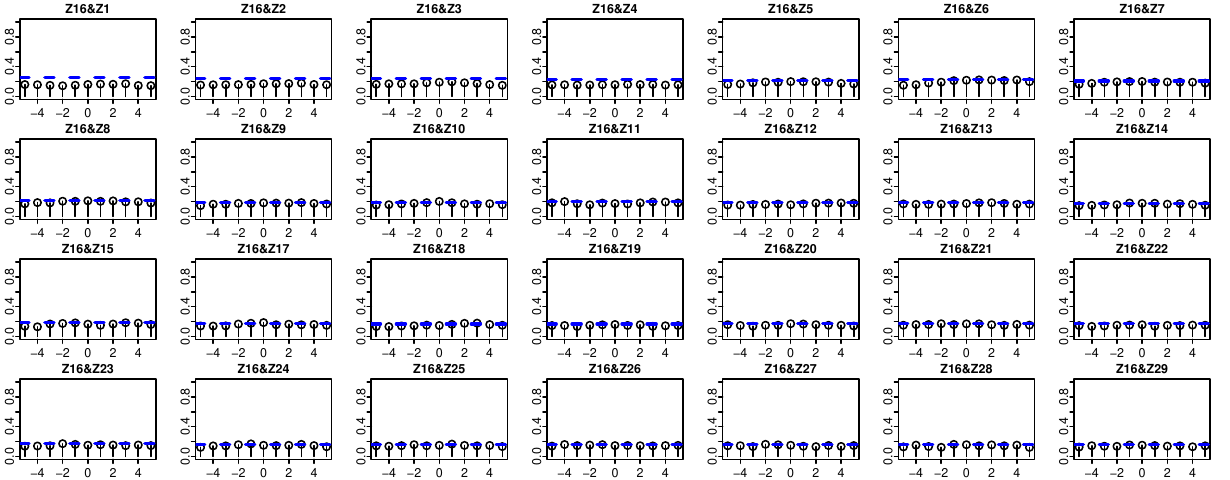}
  \vspace{-0.2cm} 
  \caption{Functional cross-autocorrelation of the 16th component series of the transformed curves} 
 \end{subfigure}

\vspace{0.2cm}
\begin{subfigure}{1\linewidth}
  \centering\includegraphics[width=16cm,height=6.5cm]{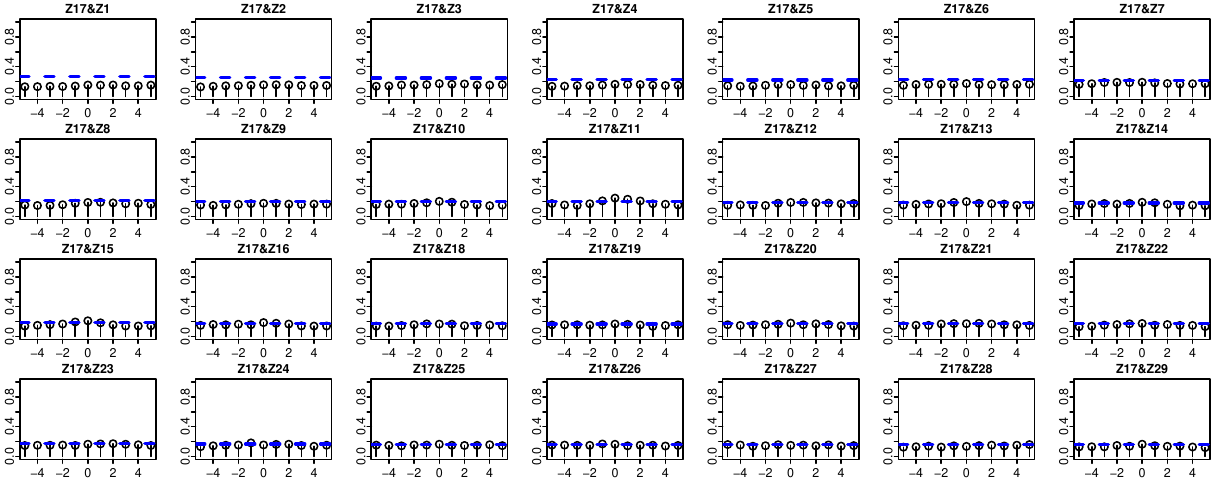}
  \vspace{-0.2cm} 
  \caption{Functional cross-autocorrelation of the 17th component series of the transformed curves} 
 \end{subfigure}

 \vspace{0.2cm}
\begin{subfigure}{1\linewidth}
  \centering\includegraphics[width=16cm,height=6.5cm]{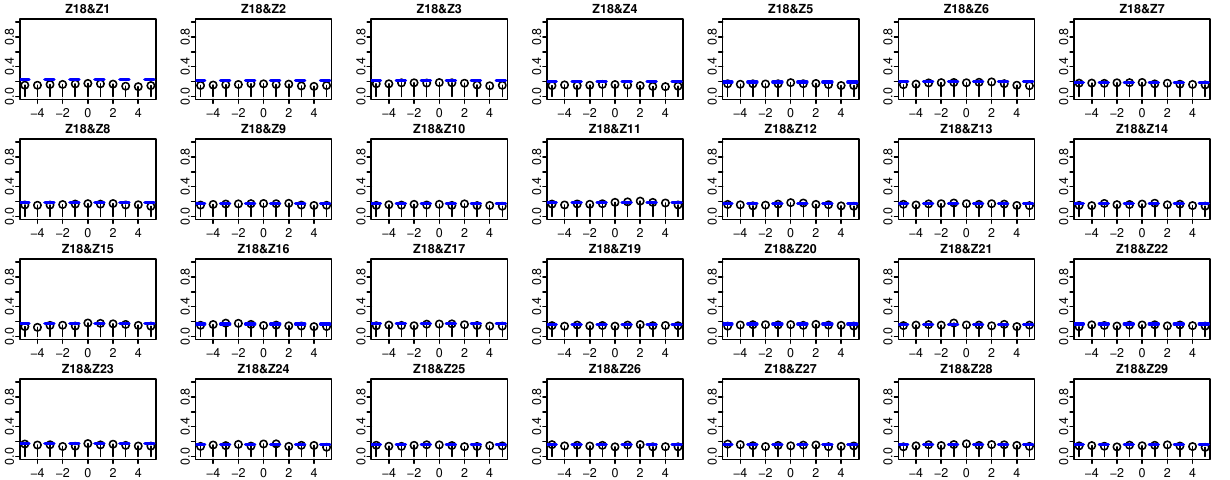}
  \vspace{-0.2cm} 
  \caption{Functional cross-autocorrelation of the 18th component series of the transformed curves} 
 \end{subfigure}
\centering
\caption{{Selected functional cross-autocorrelation of the transformed female mortality data
}}
\end{figure}

\begin{figure}[tbp]
\centering
\begin{subfigure}{1\linewidth}
  \centering\includegraphics[width=16cm,height=6.5cm]{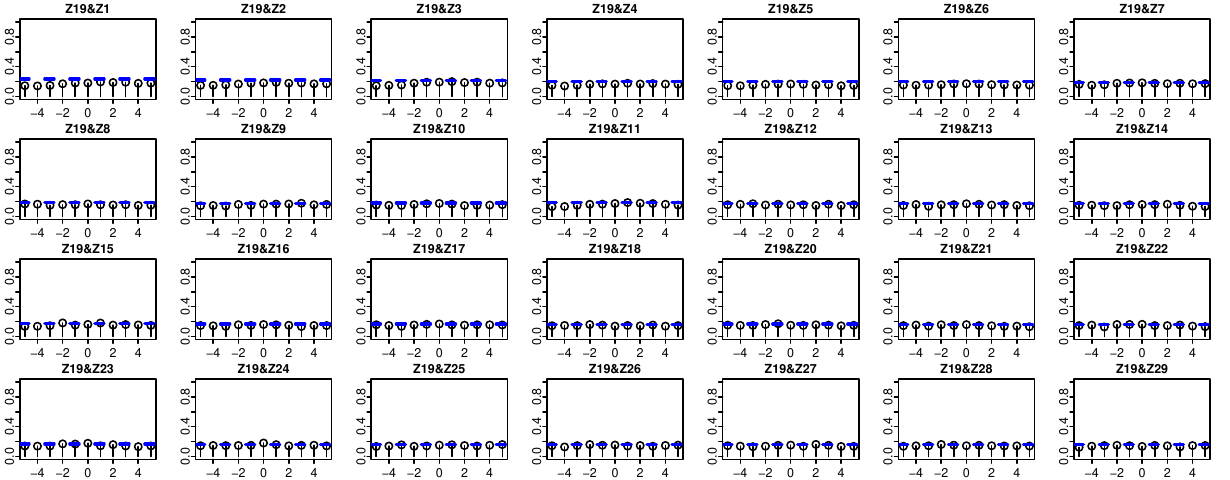}
  \vspace{-0.2cm} 
  \caption{Functional cross-autocorrelation of the 19th component series of the transformed curves} 
 \end{subfigure}

\vspace{0.2cm}
\begin{subfigure}{1\linewidth}
  \centering\includegraphics[width=16cm,height=6.5cm]{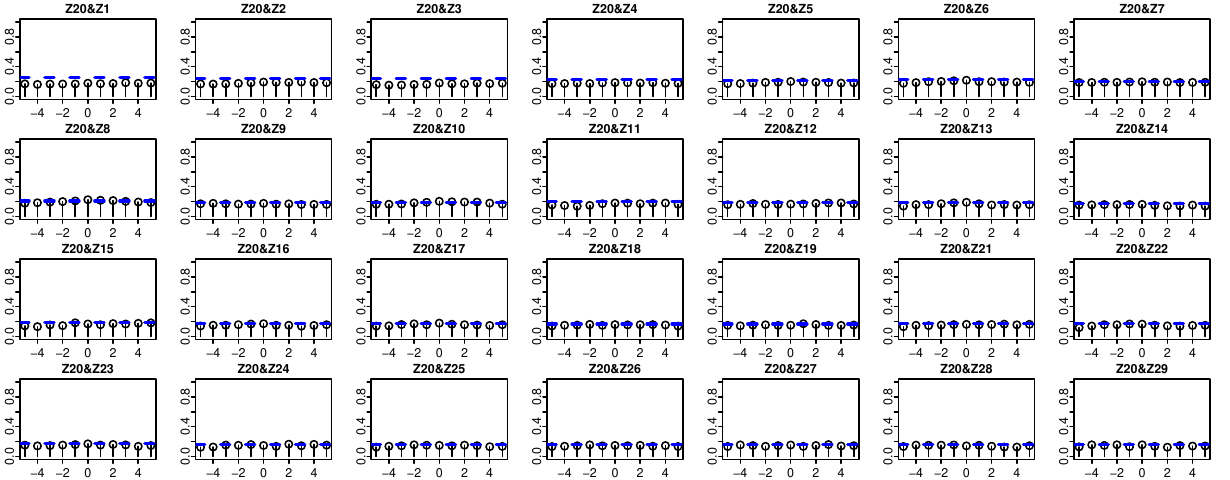}
  \vspace{-0.2cm} 
  \caption{Functional cross-autocorrelation of the 20th component series of the transformed curves} 
 \end{subfigure}

 \vspace{0.2cm}
\begin{subfigure}{1\linewidth}
  \centering\includegraphics[width=16cm,height=6.5cm]{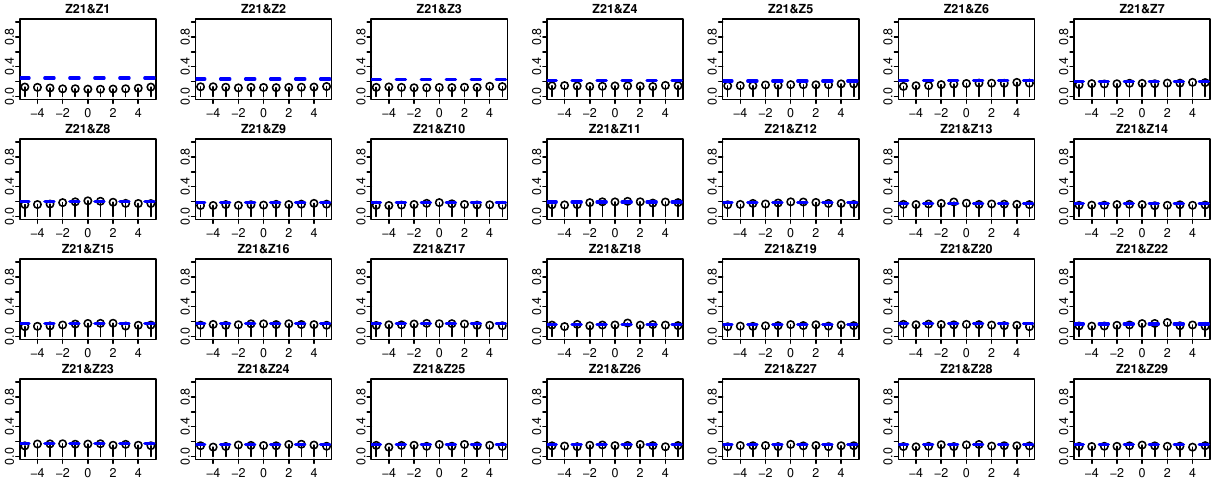}
  \vspace{-0.2cm} 
  \caption{Functional cross-autocorrelation of the 21st component series of the transformed curves} 
 \end{subfigure}
\centering
\caption{{Selected functional cross-autocorrelation of the transformed female mortality data
}}
\end{figure}

\begin{figure}[tbp]
\centering
\begin{subfigure}{1\linewidth}
  \centering\includegraphics[width=16cm,height=6.5cm]{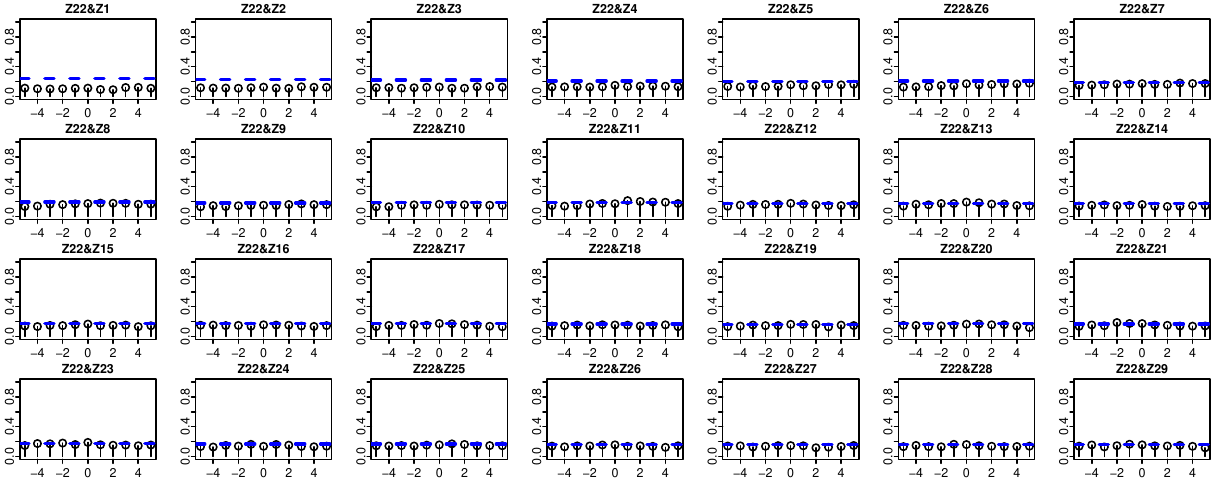}
  \vspace{-0.2cm} 
  \caption{Functional cross-autocorrelation of the 22nd component series of the transformed curves} 
 \end{subfigure}

\vspace{0.2cm}
\begin{subfigure}{1\linewidth}
  \centering\includegraphics[width=16cm,height=6.5cm]{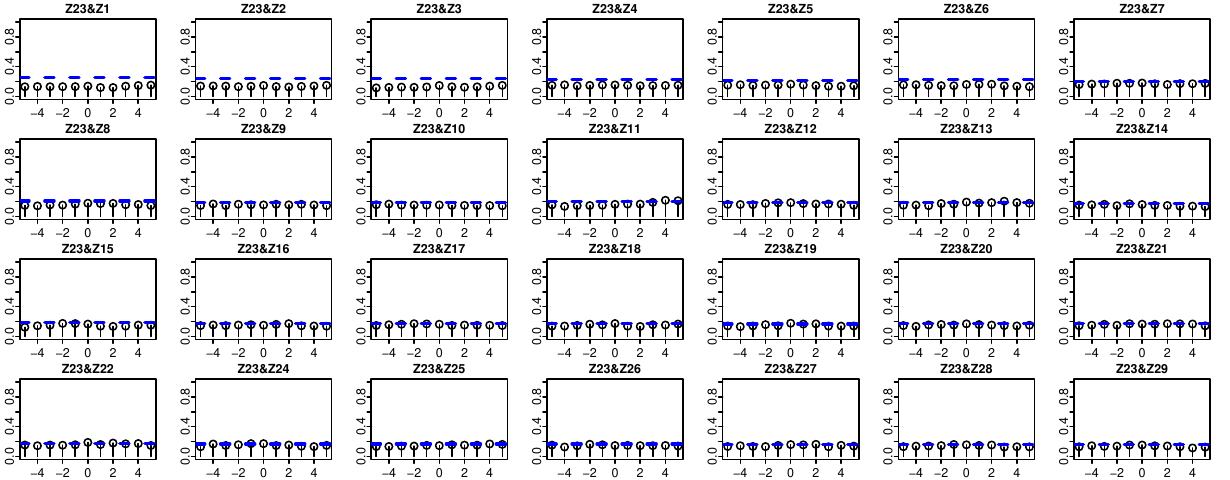}
  \vspace{-0.2cm} 
  \caption{Functional cross-autocorrelation of the 23rd component series of the transformed curves} 
 \end{subfigure}

 \vspace{0.2cm}
\begin{subfigure}{1\linewidth}
  \centering\includegraphics[width=16cm,height=6.5cm]{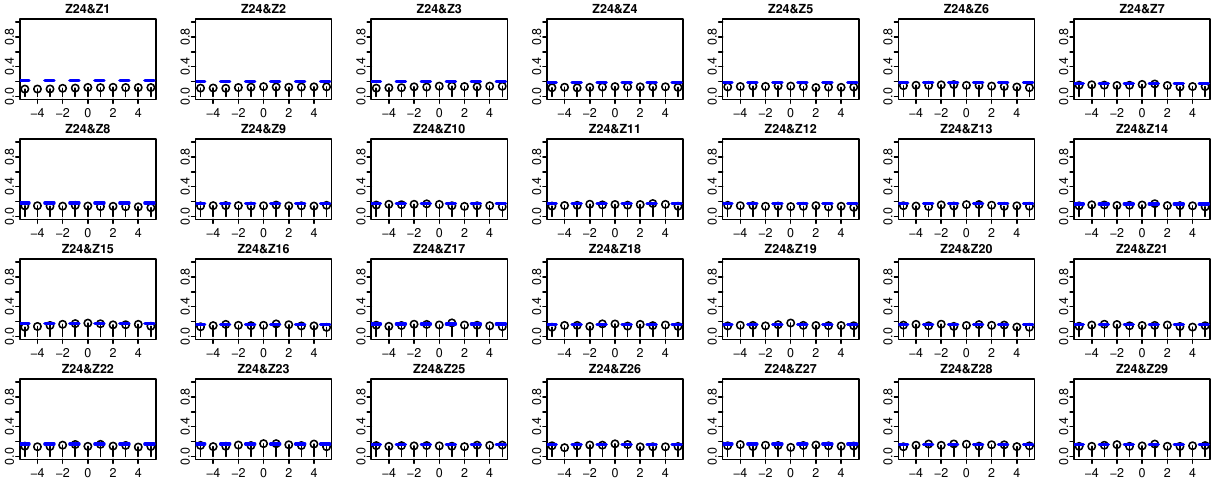}
  \vspace{-0.2cm} 
  \caption{Functional cross-autocorrelation of the 24th component series of the transformed curves} 
 \end{subfigure} 
\centering
\caption{{Selected functional cross-autocorrelation of the transformed female mortality data
}}
\end{figure}

\begin{figure}[tbp]
\centering
\begin{subfigure}{1\linewidth}
  \centering\includegraphics[width=16cm,height=6.5cm]{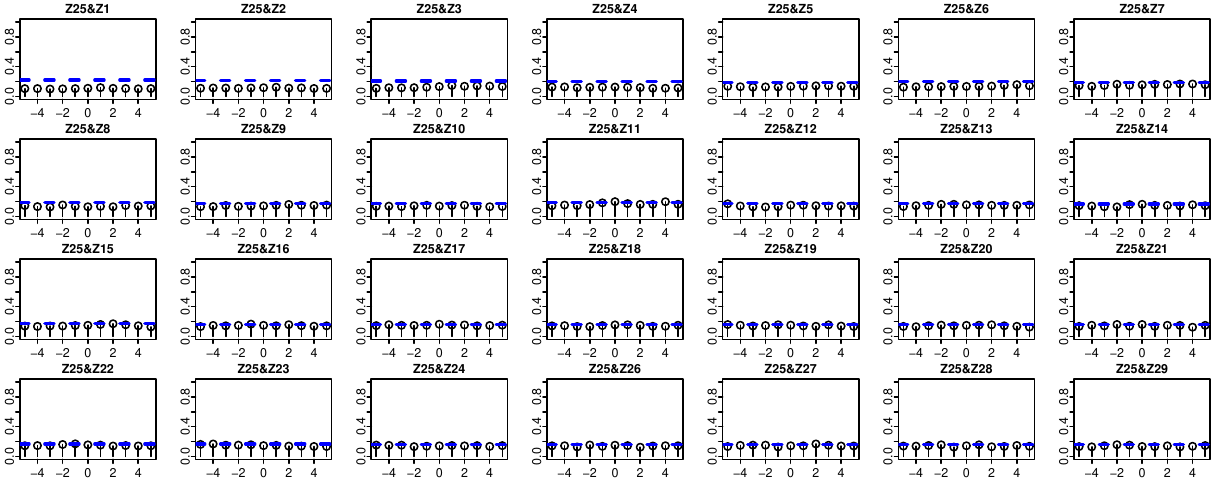}
  \vspace{-0.2cm} 
  \caption{Functional cross-autocorrelation of the 25th component series of the transformed curves} 
 \end{subfigure}

\vspace{0.2cm}
\begin{subfigure}{1\linewidth}
  \centering\includegraphics[width=16cm,height=6.5cm]{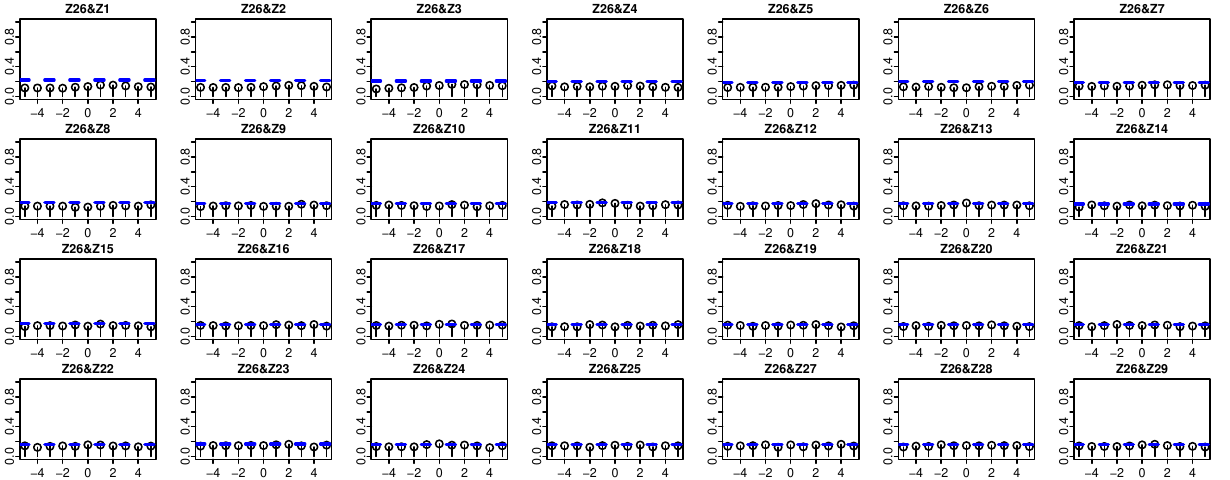}
  \vspace{-0.2cm} 
  \caption{Functional cross-autocorrelation of the 26th component series of the transformed curves} 
 \end{subfigure}

 \vspace{0.2cm}
\begin{subfigure}{1\linewidth}
  \centering\includegraphics[width=16cm,height=6.5cm]{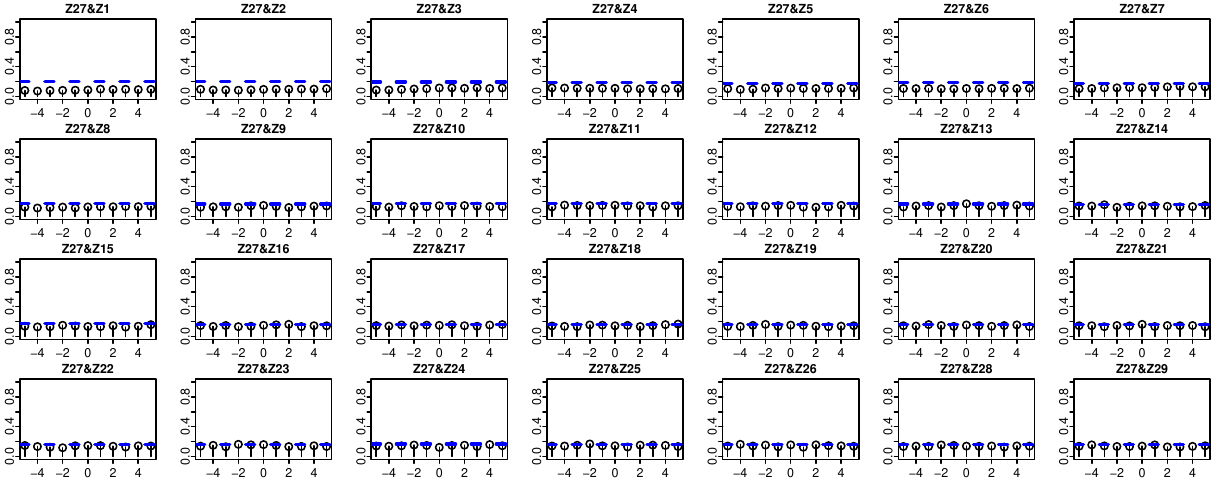}
  \vspace{-0.2cm} 
  \caption{Functional cross-autocorrelation of the 27th component series of the transformed curves} 
 \end{subfigure}
\centering
\caption{{Selected functional cross-autocorrelation of the transformed female mortality data
}}
\end{figure}

\begin{figure}[tbp]
\centering
\begin{subfigure}{1\linewidth}
  \centering\includegraphics[width=16cm,height=6.5cm]{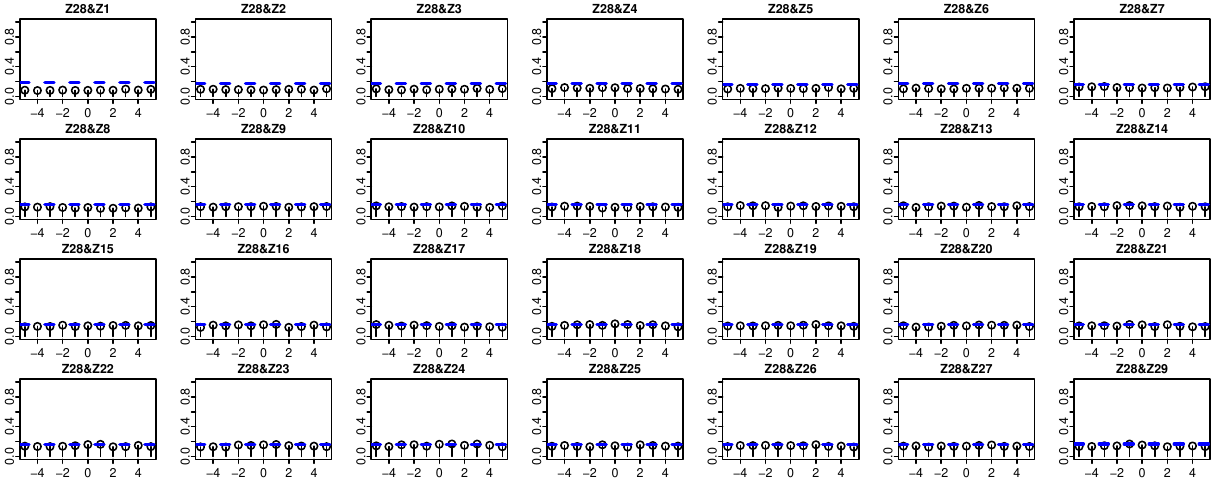}
  \vspace{-0.2cm} 
  \caption{Functional cross-autocorrelation of the 28th component series of the transformed curves} 
 \end{subfigure}

\vspace{0.2cm}
\begin{subfigure}{1\linewidth}
  \centering\includegraphics[width=16cm,height=6.5cm]{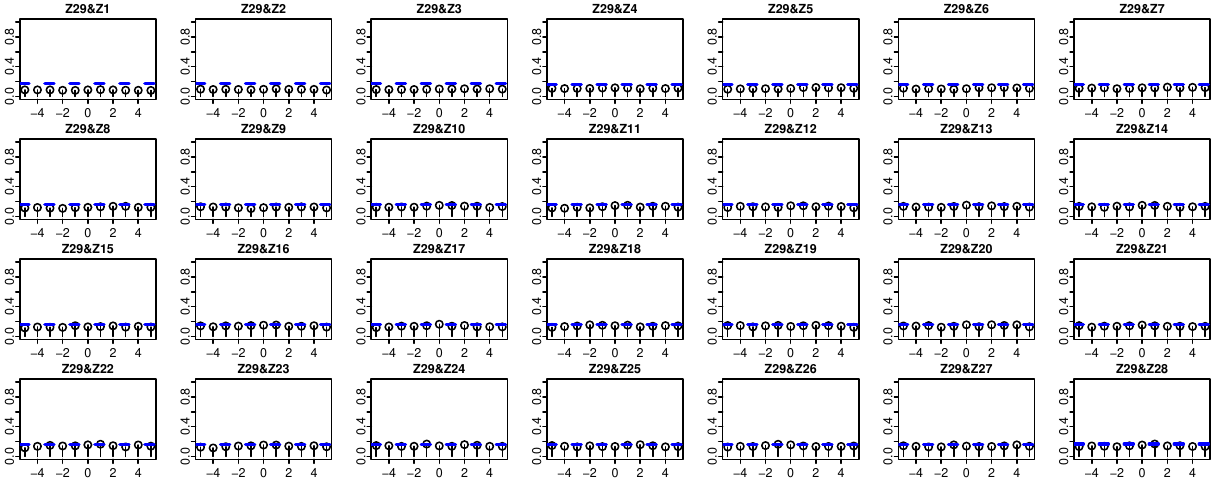}
  \vspace{-0.2cm} 
  \caption{Functional cross-autocorrelation of the 29th component series of the transformed curves} 
 \end{subfigure}

\centering
\caption{\label{plot22}{Selected functional cross-autocorrelation of the transformed female mortality data
}}
\end{figure}

\newpage
\spacingset{1.2}
\section*{References}

\begin{description}
\item Chang, J., Guo, B. and Yao, Q. (2018). Principal component analysis for second-order
stationary vector time series, {\it The Annals of Statistics} {\bf 46}: 2094–2124.
    \item 
    Fang, Q., Guo, S. and Qiao, X. (2022).
	\newblock   Finite sample theory for high-dimensional functional/scalar time series with applications, {\it Electronic Journal of Statistics} {\bf 16}: 527-591.

 \item Golub, G. H. and Van Loan, C. F. (1996). {\it Matrix Computations}, Johns Hopkins Studies
in the Mathematical Sciences, fourth edn, Johns Hopkins University Press, Baltimore,
MD.

\item Guo, S. and Qiao, X. (2023). On consistency and sparsity for high-dimensional functional
time series with application to autoregressions, \textit{Bernoulli} \textbf{29}: 451–472.

\end{description}

\end{document}